\documentclass[singlecolumn,prd,amssymb,amsmath,preprintnumbers,aps,12pt]{revtex4-2}
\usepackage{lineno}
\usepackage{bm,color,xcolor}
\usepackage{slashed} 
\usepackage{graphicx}
\usepackage{booktabs}
\usepackage{makecell}
\usepackage{multirow}
\usepackage{comment}
\usepackage{mathtools}
\usepackage{appendix}
\usepackage[colorlinks=true
,urlcolor=blue
,anchorcolor=blue
,citecolor=blue 
,filecolor=blue
,linkcolor=blue
,menucolor=blue
,linktocpage=true
,pdfproducer=medialab
,pdfa=true
]{hyperref}
\usepackage{lineno}

\newcommand{\beq}{\begin{equation}}
\newcommand{\eeq}{\end{equation}}
\newcommand{\bea}{\begin{eqnarray}}
\newcommand{\eea}{\end{eqnarray}}

\begin{document}

 
\begin{flushright}
\end{flushright}

\title{\bf \Large The Physics potential of the CEPC  \\{\it Prepared for the US Snowmass Community Planning Exercise (Snowmass 2021)}}




\author{CEPC Physics Study Group}




\maketitle
\section*{Contributors}

\begin{itemize}
    \item Huajie Cheng, Department of Applied Physics, Naval University of Engineering, Jiefang Blvd 717, Qiaokou District, Wuhan 430033, China
    \item Wen Han Chiu, Department of Physics, University of Chicago, Chicago, IL 60637. USA
    \item Yaquan Fang, Institute of High Energy Physics, University of Chinese Academy of Science, Beijing, 100049, China
    \item Yu Gao, Key Laboratory of Particle Astrophysics, Institute of High Energy Physics, Chinese Academy of Sciences, Beijing, 100049, China
    \item Jiayin Gu, Department of Physics, Center for Field Theory and Particle Physics, Key Laboratory of Nuclear Physics and Ion-beam Application (MOE), Fudan University, Shanghai 200438, China
    \item Gang Li, Institute of High Energy Physics, University of Chinese Academy of Science, Beijing, 100049, China
    \item Lingfeng Li, Department of Physics, Brown University, Providence, RI 02912, USA
  \item Tianjun Li, CAS Key Laboratory of Theoretical Physics, Institute of Theoretical Physics, Chinese Academy of Sciences, Beijing 100190, China
  \item Zhijun Liang, Institute of High Energy Physics, University of Chinese Academy of Science, Beijing, 100049, China
    \item Bo Liu, Institute of High Energy Physics, University of Chinese Academy of Science, Beijing, 100049, China

        \item Jia Liu, School of Physics and State Key Laboratory of Nuclear Physics and Technology, Peking University, Beijing 100871, China
    \item Zhen Liu, School of Physics and Astronomy, University of Minnesota, Minneapolis, MN 55455, USA
    \item Manqi Ruan, Institute of High Energy Physics, Chinese Academy of Sciences, Beijing, 100049, China
    \item Jing Shu, School of Physics and State Key Laboratory of Nuclear Physics and Technology, Peking University, Beijing 100871, China
    \item Kechen Wang, Department of Physics, School of Science, Wuhan University of Technology, 430070 Wuhan, Hubei, China 
    \item Lian-Tao Wang, Department of Physics, University of Chicago, Chicago, IL 60637. USA
    \item Ke-Pan Xie, Department of Physics and Astronomy, University of Nebraska, Lincoln, NE 68588, USA
    \item Shuo Yang, Department of Physics, Liaoning Normal University, Dalian 116029, China
    \item Jiarong Yuan, School of Physical Sciences, University of Chinese Academy of Sciences, No. 19A Yuquan Road, Beijing 100049, China
    \item Kaili Zhang, Institute of High Energy Physics, University of Chinese Academy of Science, Beijing, 100049, China
    \item Mengchao Zhang, Department of Physics and Siyuan Laboratory, Jinan University, Guangzhou 510632, P.R. China
    \item Yang Zhang, School of Physics, Zhengzhou University, Zhengzhou 450000, China
    \item Xuai Zhuang, Institute of High Energy Physics, University of Chinese Academy of Science, Beijing, 100049, China
\end{itemize}

\newpage
\section*{Abstract}
The Circular Electron Positron Collider (CEPC) is a large-scale collider facility that can serve as a factory of the Higgs, $Z$, and $W$ bosons and is upgradable to run at the $t\bar t $ threshold. 
This document describes the latest  CEPC nominal operation scenario and particle yields and updates the corresponding physics potential. A new detector concept is also briefly described.
This submission is for consideration by the Snowmass process. 

\tableofcontents
\clearpage

\section{Executive Summary}

The Higgs boson, discovered in 2012 by the ATLAS and CMS Collaborations at the Large Hadron Collider (LHC), plays a central role in the Standard Model. Higgs precision measurements will advance our understanding of many critical questions, including the origin of the electroweak scale, the nature of the electroweak phase transition, and the origin of the matter anti-matter asymmetry in the universe. The Higgs boson is also a window for exploring new physics, such as dark matter and its associated dark sector, heavy sterile neutrino, and many others. Compared to the LHC, an electron-positron Higgs factory can provide 
crucial and highly complementary information, significantly enhance our knowledge of this mysterious particle, and help us make progress in answering the critical questions mentioned above. Therefore, it is widely regarded as the highest priority among various proposed future collider facilities. 

The Circular Electron Positron Collider (CEPC)~\cite{CEPC-SPPCStudyGroup:2015csa,CEPC-SPPCStudyGroup:2015esa}, proposed by the Chinese high energy physics community in 2012, is designed to run primarily at a center-of-mass energy of 240 GeV as a Higgs factory. In addition, it will also be operated on the Z-pole as a Z factory, perform a precise $WW$ threshold scan, and be upgraded to a center-of-mass energy of 360 GeV, close to the $t \bar t $ threshold. The CEPC study group published its Conceptual Design Report (CDR)~\cite{CEPCStudyGroup:2018ghi} in Nov. 2018. Since then, many critical technologies R$\&$D have been carried out. Multiple prototypes have been produced and tested, especially those components on the critical accelerator sub-systems. The performances have achieved or surpassed the designed goal. In addition, significant progress has been made on the accelerator design, leading to significantly enhanced instantaneous luminosities compared to those presented in the CDR. Based on the progress, the CEPC study group proposes a set of new beam parameters and a new nominal operation scenario. 

In this new nominal operation scenario, the CEPC is expected to deliver 4 million Higgs bosons, nearly 4 trillion Z bosons, 
over 400 million W-boson pairs (of which $\sim$100 million are near the WW threshold), 
and potentially one million top quarks with the high-energy upgrade. Many Higgs boson couplings can be measured with precision about one order of magnitude better than those achievable at the High Luminosity LHC (HL-LHC). In addition, 
the CEPC is projected to improve the current precision of many of the electroweak observables by about one order of magnitude or more, complementary to the Higgs boson coupling measurements. The CEPC also offers excellent opportunities for searching for rare decays of the Higgs, W, and Z bosons and many other new physics signals. The large quantities of bottom quarks, charm quarks, and tau leptons produced from the decays of the Z bosons are ideal for a suite of important flavor physics measurements. While the results reported here are based on the updated running scenario, we note that there is room for further improvement as a possible expansion to 4 IPs is also under investigation. 

This report briefly introduces the new nominal operation scenario and summarizes many recent physics potential studies. This submission for consideration by the Snowmass is part of our dedicated effort to seek international collaboration and support. Given the importance of the precision Higgs boson measurements and the search for new physics beyond the Standard Model, the CEPC team is actively promoting the CEPC project and is motivated to collaborate on other proposals for electron-positron collider based Higgs factories.
 
\newpage

\section{Introduction}

The Higgs field is at the heart of many mysteries of the Standard Model (SM), such as the origin of the electroweak scale, the nature of the electroweak phase transition, the flavor structure and so on. 
%
Meanwhile, it could 
be deeply connected with many fundamental phenomena beyond the Standard Model, such as the origin of matter and anti-matter asymmetry, the nature of dark matter and dark energy, and the fundamental forces that drove inflation. 
The discovery of the Higgs boson, an excited state of the Higgs field, completes the SM particle spectrum and offers an excellent probe for those fundamental mysteries and phenomena. 
A Higgs factory that can measure the properties of the Higgs boson to an unprecedented precision 
is vital for this exploration. 

The LHC is a powerful Higgs factory. 
Ultimately, the high luminosity run of the LHC (HL-LHC) will produce 100 million Higgs bosons in the coming decades. 
However, due to the large backgrounds and the large theoretical/systematical uncertainties,
the ultimate accuracies of the Higgs property measurements at the HL-LHC are typically limited to a few percent.

Compared to the LHC, the electron-positron colliders have significant advantages for the Higgs property measurements. 
They are free of the QCD background. The ratio of the Higgs signal versus the SM background is 7-8 orders of magnitude higher than the HL-LHC. 
The electron-positron colliders have precisely known and adjustable initial states and can determine the absolute values of the Higgs boson width and couplings. 
Multiple electron-positron Higgs factories have been proposed, including the International Linear Collider (ILC) \cite{ILCInternationalDevelopmentTeam:2022izu}, the Compact Linear Collider (CLIC) \cite{CLIC:2018fvx}, the Future Circular Collider (FCC) \cite{FCC:2018evy}, and
the Circular Electron Positron Collider (CEPC) \cite{CEPC-SPPCStudyGroup:2015csa,CEPC-SPPCStudyGroup:2015esa}. Meanwhile, many new possibilities are also under consideration, such as a 125 GeV Muon collider~\cite{deBlas:2022aow}, C3~\cite{Bai:2021rdg}, ReliC~\cite{Litvinenko:2022qbd} and CERC~\cite{Litvinenko:2022mrt}. 

The CEPC is proposed by the High Energy Physics community right after the Higgs discovery and is expected to be hosted in China. The CEPC working group kicked off in September 2013.
In 2015, the CEPC study achieved its first milestone, the pre-CDR \cite{CEPC-SPPCStudyGroup:2015csa,CEPC-SPPCStudyGroup:2015esa}, in which the study group concluded there is no shower-stopper for this facility. 
Followed by intensive R\&D activities and physics study, the CEPC study group delivered the CEPC Conceptual Design Report (CDR) \cite{CEPCStudyGroup:2018ghi} in Nov. 2018. 

The CEPC has the main ring with a total circumference of 100 km. 
It is designed to operate at around $E_{\rm CM} = 91.2$ GeV as a Z factory, close to W pair production threshold $E_{\rm CM} \simeq 160$ GeV,  at $E_{\rm CM} = 240$ GeV as a Higgs factory. 
The center of mass energy of CEPC can be upgraded to 360 GeV, enabling the $t\bar{t}$ pair production. 
With an eye on future upgrades, the tunnel is designed to be wide enough to accommodate both the CEPC and SPPC \cite{Tang:2022fzs}.

In the CDR, the CEPC is envisioned to operate with two detectors. It has a ten-year nominal operation plan which will deliver total combined integrated luminosities of 16, 2.6, and 5.6 $ab^{-1}$ for the Z, the W, and the Higgs operation, respectively. 
It will produce close to one trillion Z bosons, 100 million W bosons, and over one million Higgs bosons. 
Billions of bottom quarks, charm quarks, and tau-leptons will be produced in the Z boson decays, making it a B-factory and a tau-charm factory.

\begin{figure}[h!]
    \includegraphics[width=0.6\textwidth]{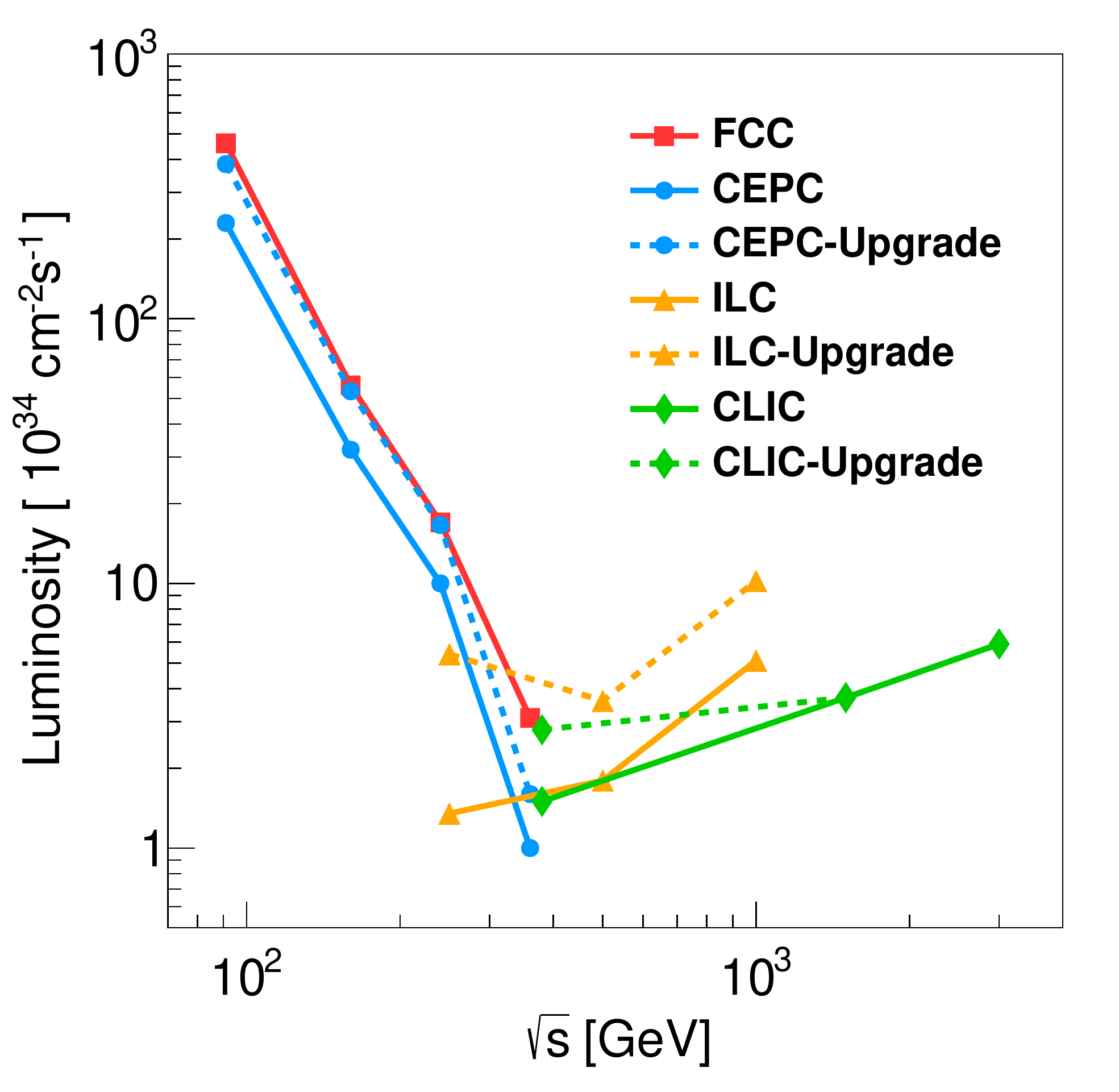}
    \caption{The updated run plan of the CEPC, with the baseline and upgrade shown in solid and dashed blue curves, respectively. The run plans for several other proposals of the $e^+ e^-$ colliders are also shown for comparison. See \cite{Gao:2022lew} for details.}
    \label{fig:lumiplan}
\end{figure}

After the delivery of the CEPC CDR, the CEPC study group continued its physics study and technology R$\&$D.
The CEPC accelerator study entered the Technical Design Report (TDR) phase, endorsed by CEPC International Advisory Committee (IAC). 
Key technologies are developed and validated, especially at the core accelerator sub-systems, including high-quality Superconducting Radio-Frequency (SRF) system, high precision magnets for the booster and collider rings, vacuum system, Machine-Detector Interface (MDI), and so on. 
Multiple prototypes have been produced and tested, validating the design by achieving and surpassing the required performances. All these progress lead to an update of the nominal run plan for the CEPC, shown in \autoref{fig:lumiplan}. Currently, the CEPC Study group is working for getting project approval around 2027.

The CEPC plans to take a staging approach to realize its construction and the targeted performance.
The baseline CEPC will be operating at the center-of-mass energies ranging from the $Z$ mass to 240 GeV, with maximal synchrotron radiation power per beam limited to 30 MW. 
Increasing the power supply and the heat load capability of the cooling system, the maximal synchrotron radiation power per beam can be increased from 30 MW to 50 MW, resulting in a linear increase of the instantaneous luminosity. 
The center-of-mass energy can be raised to 360 GeV by increasing the RF cavities. 

\begin{table}[h!]
	\centering
  	\resizebox{0.8 \textwidth}{!}{
	\begin{tabular}[t]{ccccc}
		\toprule[1pt]
		Operation mode & Z factory  & WW threshold  & Higgs factory & $t\bar{t}$ \\
		\midrule
		$\sqrt{s}$ (GeV) & 91.2 & 160 & 240 & 360 \\
		Run time (year) & 2 & 1 & 10 & 5 \\
		\makecell{Instantaneous luminosity \\ ($10^{34} cm^{-2}s^{-1}$, per IP)} & 191.7 & 26.6 & 8.3 & 0.83 \\
		\makecell{Integrated luminosity \\ ($ab^{-1}$, 2~IPs)} & 100 & 6 & 20 & 1 \\
		Event yields & $3 \times 10^{12}$ & $1 \times 10^{8}$ & $4 \times 10^{6}$ & $5 \times 10^{5}$ \\
		
		\bottomrule[1pt]
	\end{tabular}
	}
	\caption{Nominal CEPC operation scheme, and the physics yield, of four different modes. See \cite{Gao:2022lew} for details. }
	\label{tab:Yield_T1}
\end{table}

A new nominal data-taking scenario is also developed from the upgraded performance of the CEPC accelerator, emphasizing the scientific program as a Higgs and $Z$ factory, summarized in \autoref{tab:Yield_T1}. Detailed description of the updated running scenarios can be found in another contribution to the Snowmass from the CEPC study group \cite{Gao:2022lew}. It aims at ten years of data taking at $E_{\rm CM}=240$ GeV, with two interaction points (IPs), accumulating an integrated luminosity of 20 $ab^{-1}$, and producing 4 million Higgs bosons. 
It also plans to operate for at least two years near the $Z$ pole, which corresponds to an integrated luminosity of 100 $ab^{-1}$, and 3 trillion $e^{+}e^{-}\to Z\to {q \bar q}$ events.
About one year will be devoted to the WW threshold scan, providing MeV level accuracies on the W boson mass and width.
After the high energy upgrade, the CEPC will be operated for at least five years at a 360 GeV center-of-mass energy with a 1 $ab^{-1}$ integrated luminosity. About 500 thousand $t\bar{t}$ events and 150 thousand inclusive Higgs events will be produced. A possible expansion to 4 IPs is also under investigation. Hence, the statistics can potentially be further improved.

The CEPC Physics study groups have continued to explore the physics potential, focusing on a broad range of topics, including  Higgs precision measurements, precise EW measurements, Flavor Physics,  QCD measurements, and direct new physics searches. In addition, the physics studies identified a handful of critical detector requirements, quantified its impact on different physics benchmarks, and set clear performance goals for its detector system. 
These requirements include the separation of final state particles, the precise reconstruction of energy/momentum of different kinds of final state particles, the identification of physics objects in high-multiplicity events, the monitoring, and calibration of beam energy and instant luminosity, and so on. 
The CEPC detector group initiated a series of detector technology R$\&$D programs to satisfy the physics requirements better and leverage the latest detector development technology.

We submit this document as an input to the Snowmass process based on the updated operating scenario of the CEPC. Here, we update the CEPC physics reaches based on recent new studies and describes briefly the detector R$\&$D activities.
The European Strategy for Particle Physics Update in 2020 \cite{ESU} concluded that an electron-positron Higgs factory is the highest-priority next collider. We firmly believe that such a facility is an indispensable step for high-energy physics explorations in the future.

\section{Higgs, EW and top physics}

Higgs, electroweak, and top quark physics make up the primary physics output of the CEPC program. This section reviews the corresponding physics potential with up-to-date simulation results according to the upgraded CEPC run plan. The top quark physics, including top quark mass and width measurements and top quark coupling measurements, enabled by the $t\bar t$ run, are still under development. We will report the corresponding projections in future works.

\subsection{Measurements of the SM Higgs processes}

In addition to the CEPC Higgs white paper~\cite{An_2019} and CDR~\cite{CEPCStudyGroup:2018ghi}, several analyses on the measurements of the Higgs cross-section~\cite{Chen:2016zpw} as well as individual Higgs decaying channels have been documented in~\cite{Yu:2019oal} and~\cite{Bai:2019qwd}. After the CDR, various updates have been implemented in the analysis of the individual channels, and for some of them, substantial improvements on the measurement precision have been achieved. Multivariate analyses (MVA) have been applied on  $H \to ZZ$~\cite{Kiuchi:2021roe} and $H \to \gamma\gamma$, leading to 5-10\% improvements on the precision with respect to the CDR results. Dedicated MVA can also enhance the sensitivity to the Higgs invisible decay measurement ($H\xrightarrow{}inv$)~\cite{Tan_2020}. For the analyses of the $H\rightarrow{}bb/cc/gg$ channels, the Color Singlet Identification (CSI) method is introduced for the $(Z\rightarrow{}q\bar{q})H$ channel, and flavor taggings are introduced for both $(Z\rightarrow{}\nu\bar{\nu}) H$ and $(Z\rightarrow{}q\bar{q})H$ channels, which were previously absent in the CDR analyses.  This results in an improvement of 2\%/52\%/10\% (for $bb/cc/gg$, respectively) in the $\nu\bar{\nu} H$ channel and 35\%/107\%/169\% in the $q\bar{q}H$ channel for the precision of the signal strength measurement~\cite{Zhu:2022lzv}. Furthermore, a global analysis based on Machine Learning (ML) is developed for the Higgs measurement which can potentially increase the corresponding precision by a factor of 2 with respect to the results reported in the  CDR~\cite{LI:2021kjh}. 

At present, the relevant data sets for Higgs measurements are those collected at the 240\,GeV run with a luminosity of  20~${\rm ab}^{-1}$, and the 360\,GeV run with 1~${\rm ab}^{-1}$, as shown in \autoref{tab:Yield_T1}. The cross-section of the ZH production for the 360\,GeV run is 36\% lower than that of the 240\,GeV run, while the cross-sections of the WW and ZZ fusion Higgs production processes are a factor of 3.8 and 4.6 higher, respectively. These channels provide important inputs for the Higgs width determination.  

An important case is the correlation between the 
two different production modes, $ZH\rightarrow bb$ and $\nu\nu H\rightarrow bb$ (from WW fusion).  It is difficult to separate the latter from  $ZH\rightarrow \nu\nu bb$ at 240\,GeV due to their similar kinematics.  
As a result, the two modes have a large correlation of $-46\%$ in the 240\,GeV run.
This is reduced to $-16\%$ in the 360\,GeV run due to a better separation of the recoiling mass of the two $b$ quarks. In addition, the correlations of the $H\rightarrow{}bb/cc/gg $ channels for the 240\,GeV run are also studied in~\cite{Zhu:2022lzv} and are estimated to be  $-15\%$, $-17\%$ and $-26\%$ for $bb/cc$, $bb/gg$ and $cc/gg$, respectively. 

The projections for the measurements of the 360\,GeV run are obtained by rescaling from the ones of the 240\,GeV run 
according to the cross sections of signal and backgrounds, 
while implementing the new developments described above and using the same statistical method as the CDR study. The $\nu\nu H\rightarrow bb$ channel is fully simulated as a cross check.  The projections of the measurement precision for various Higgs 
production and decay channels at the CEPC 240\,GeV and 360\,GeV runs are summarized in \autoref{tab:exclusiveprecision}. 

In comparison with the CDR results, the inclusive Higgs cross-section measurement has been improved from 0.5\% to 0.26\%, 
the measurements of the exclusive channels (in terms of $\sigma \times {\rm BR}$) are also significantly improved, and the precision reach of the Higgs width determination is improved from 2.8\% to 1.65\%,\footnote{The precision of the Higgs width is extracted from the $\kappa$-fit of all Higgs measurements described in \autoref{sec:HiggsCoupling}.} 
for the 240\,GeV run.  
This is mainly due to the higher luminosity ($20{\rm ab}^{-1}$ {\it vs.} $5.6{\rm ab}^{-1}$) for the updated run plan.  
Thanks to the 360\,GeV run, the combined precision of the Higgs width determination with the two runs can reach a remarkable 1.10\%.

\begin{table}[]
\resizebox{0.7 \textwidth}{!}{
\begin{tabular}{|c|cc|ccc|}
\hline
                  & \multicolumn{2}{c|}{240\,GeV, 20 ${\rm ab}^{-1}$}                & \multicolumn{3}{c|}{360\,GeV, 1 ${\rm ab}^{-1}$}                                                                        \\ \hline
                  & \multicolumn{1}{c|}{ZH}                 & \textbf{vvH}      & \multicolumn{1}{c|}{ZH}                        & \multicolumn{1}{c|}{\textbf{vvH}}              & \textbf{eeH}     \\ \hline
inclusive                & \multicolumn{1}{c|}{\textbf{0.26\%}}    &                   & \multicolumn{1}{c|}{\textbf{1.40\%}}           & \multicolumn{1}{c|}{\textbf{\textbackslash{}}} & \textbackslash{} \\ \hline
H→bb               & \multicolumn{1}{c|}{\textbf{0.14\%}}    & \textbf{1.59\%}   & \multicolumn{1}{c|}{\textbf{0.90\%}}           & \multicolumn{1}{c|}{\textbf{1.10\%}}           & \textbf{4.30\%}           \\ \hline
H→cc               & \multicolumn{1}{c|}{\textbf{2.02\%}}    &                   & \multicolumn{1}{c|}{\textbf{8.80\%}}           & \multicolumn{1}{c|}{\textbf{16\%}}             & \textbf{20\%}             \\ \hline
H→gg               & \multicolumn{1}{c|}{\textbf{0.81\%}}    &                   & \multicolumn{1}{c|}{\textbf{3.40\%}}           & \multicolumn{1}{c|}{\textbf{4.50\%}}           & \textbf{12\%}             \\ \hline
H→WW               & \multicolumn{1}{c|}{\textbf{0.53\%}}    &                   & \multicolumn{1}{c|}{\textbf{2.80\%}}           & \multicolumn{1}{c|}{\textbf{4.40\%}}           & \textbf{6.50\%}          \\ \hline
H→ZZ               & \multicolumn{1}{c|}{\textbf{4.17\%}}    &                   & \multicolumn{1}{c|}{\textbf{20\%}}             & \multicolumn{1}{c|}{\textbf{21\%}}             &                  \\ \hline
$H\xrightarrow{}\tau\tau$               & \multicolumn{1}{c|}{\textbf{0.42\%}}    &                   & \multicolumn{1}{c|}{\textbf{2.10\%}}           & \multicolumn{1}{c|}{\textbf{4.20\%}}           & \textbf{7.50\%}           \\ \hline
$H\xrightarrow{}\gamma\gamma$               & \multicolumn{1}{c|}{\textbf{3.02\%}}    &                   & \multicolumn{1}{c|}{\textbf{11\%}}             & \multicolumn{1}{c|}{\textbf{16\%}}             &                  \\ \hline
$H\xrightarrow{}\mu\mu$               & \multicolumn{1}{c|}{\textbf{6.36\%}}    &                   & \multicolumn{1}{c|}{\textbf{41\%}}             & \multicolumn{1}{c|}{\textbf{57\%}}             &                  \\ \hline

$H\xrightarrow{}Z\gamma$     & \multicolumn{1}{c|}{\textbf{8.50\%}}    &                   & \multicolumn{1}{c|}{\textbf{35\%}}             &  \multicolumn{1}{c|}{\textbf{}} &                  \\ \hline
${\rm Br}_{upper} (H\xrightarrow{}inv.)$ & \multicolumn{1}{c|}{\textbf{0.07\%}}    &                   & \multicolumn{1}{c|}{\textbf{}} & \multicolumn{1}{c|}{\textbf{}}  &                  \\ \hline
$\Gamma_H$              & \multicolumn{2}{c|}{{\textbf{1.65\%}}} & \multicolumn{3}{c|}{{\textbf{1.10\%}}}                                                        \\ \hline
\end{tabular}
}
\caption{The projected precision for the measurements of the inclusive Higgs cross section,  cross section times branching ratio and the Higgs total width ($\Gamma_H$) for the 240\,GeV and 360\,GeV runs at the CEPC.  All results are in terms of the one-sigma precision except for the invisible decay branching ratio ${\rm Br}_{upper} (H\xrightarrow{}inv.)$, which shows the 95\% confidence-level upper bound.  The precision of the Higgs total width is obtained from a $\kappa$-fit of all the Higgs measurements.
}
\label{tab:exclusiveprecision}
\end{table}

\subsection{Higgs coupling determination}
\label{sec:HiggsCoupling}

The determination of the Higgs couplings 
is one of the major goal of the CEPC physics program.  In order to compare with the SM predictions, a systematic parameterization of BSM contributions to the Higgs couplings is needed.  The two common frameworks are the so-called $\kappa$ framework and the effective-field-theory (EFT) approach, which are both considered in the CEPC study.  

\begin{figure}
    \includegraphics[width=0.8\textwidth]{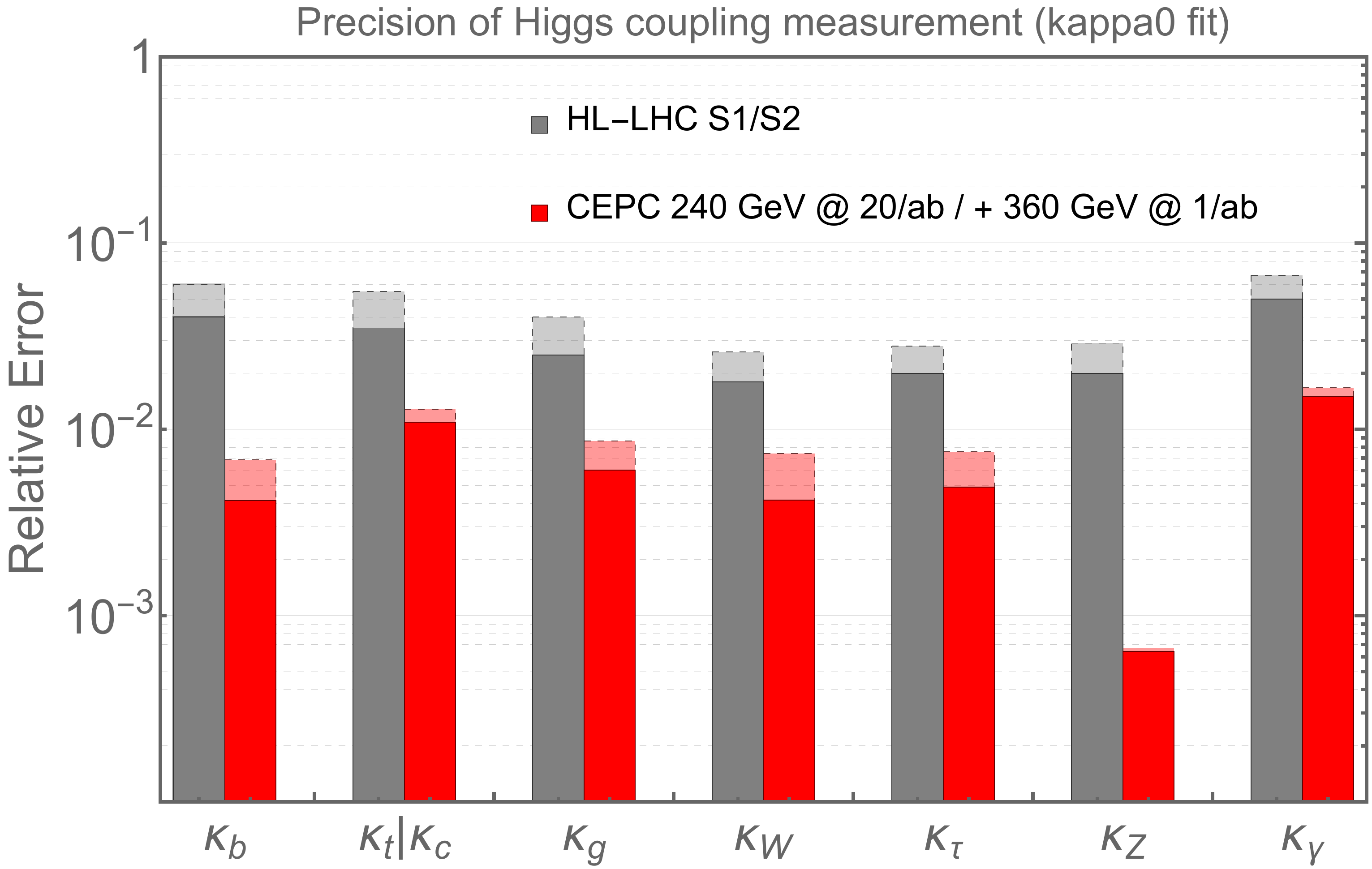}
    \includegraphics[width=0.8\textwidth]{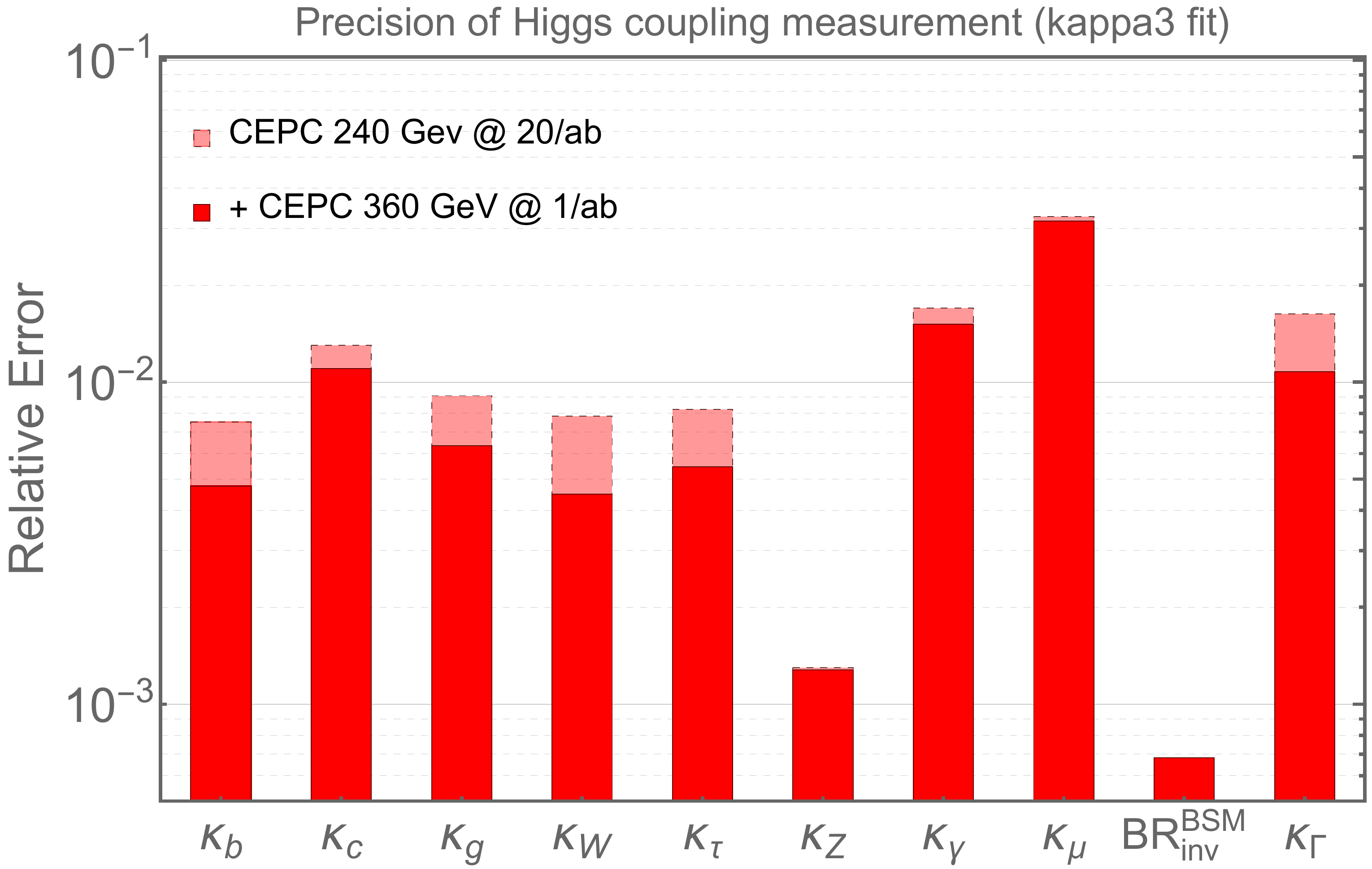}
    \caption{The projections of the precision of the Higgs coupling measurement  at the CEPC in different kappa fit frameworks, shown with red bars. The projections with CEPC 240 GeV run only are shown in dashed opaque bars, and the improved result with CEPC 360 GeV run are shown in solid bars. Note that with the upgrade CEPC run plans, including the HL-LHC in the combination does not lead to any visible changes in the results in the chosen set of free parameters. Hence, we do not consider them separately here anymore. The upper panel has the total width as a derived quantity instead of a free parameter (as in the lower panel), enabling a direct comparison with the HL-LHC sensitivities shown in gray bars. HL-LHC has a large flat direction with total width as a free parameter, and hence its results are not shown in the lower panel.}
    \label{fig:kapparesults}
\end{figure}

The $\kappa$ framework defines $\kappa$s as the rescaling factors of the Higgs couplings with respect to their SM values. The precision projection on $\kappa$ directly reflects the Higgs precision program in a simplified manner. These $\kappa$s can be related to the Wilson coefficients in the Higgs EFT (HEFT) or the Standard Model EFT, often encapsulating several operators' combined direction. The EFT framework becomes more suitable when describing the physics across 
a broad range of energy scales or between different sectors (such as the Higgs and the EW ones).

For the $\kappa$ framework, we make projections in two schemes that have already been adopted in the CEPC pre-CDR~\cite{CEPCpreCDR} and CDR~\cite{CEPCStudyGroup:2018ghi,An:2018dwb}. These two schemes are also equivalent to the so-called ``kappa0" and ``kappa3" schemes 
in the European Strategy Updates~\cite{deBlas:2019rxi}. In the ``kappa0" scheme, the Higgs total width is not a free parameter, and is assumed to be the sum of all the partial widths of the SM Higgs decay channels. Since it is difficult to measure the Higgs width precisely at the LHC, using this scheme allows for  precise extraction of the couplings  at the LHC and enables comparisons between the HL-LHC performances and the CEPC. In the ``kappa3" framework, the Higgs width is a free parameter, containing (inclusively) various exotic decay modes that are not directly searched. The lepton collider Higgs factories have unique capability to determine the Higgs width through inclusive measurements~\cite{Peskin:2012we,Han:2013kya,Dawson:2013bba,Han:2015ofa,CEPCpreCDR,CEPCStudyGroup:2018ghi,An:2018dwb,deBlas:2019rxi}.\footnote{Other means of lineshape scan is also possible but often requires very precise beam and high luminosity~\cite{Han:2012rb,Greco:2016izi,dEnterria:2021xij,deBlas:2022aow}, in particular, for $e^+e^-$ machines~\cite{Greco:2016izi,dEnterria:2021xij}.}

We show the projected Higgs precision in these two kappa fit frameworks in \autoref{fig:kapparesults}. The CEPC projections with the upgraded run plan are shown in red bars, with opaque ones with dashed boundary representing 240 GeV run with $20~{\rm ab}^{-1}$ results alone. The full projections, in combination with a 360 GeV run with an integrated luminosity of $1~{\rm ab}^{-1}$, are shown with solid red bars. 

The upper panel shows the constrained fit result where Higgs width is not a free parameter. The gray bars represent the projections for the HL-LHC, with two scenarios of uncertainty assumption~\cite{Cepeda:2019klc}. In comparison with the HL-LHC, the CEPC improves most coupling precisions by more than a factor of 5. Notably, the Higgs to $ZZ$ coupling can be improved by more than a factor of 30, enabling us to dive deep into new physics which can modify Higgs couplings. 

We show the kappa fit result with the Higgs width as a free parameter in the lower panel. Historically, this scheme is often dubbed the ``model-independent" fit. The Higgs width could contain additional contributions from new decay modes which are hard to search for or tag. HL-LHC has a sizeable flat direction with total width as a free parameter. Consequently, the HL-LHC results are not shown in the lower panel. CEPC is able to determine the Higgs boson total width at the level of 1.1\%. CEPC can also set an upper limit on the branching ratio of BSM invisible decay at 0.07\% at 95\% CL from the exclusive measurement. 
Note that with the upgraded CEPC run plans, including the HL-LHC in the combination does not lead to any visible changes in the results in the projections for the chosen set of free parameters. Hence, we do not consider them separately here anymore.

\begin{figure}
    \includegraphics[width=0.9\textwidth]{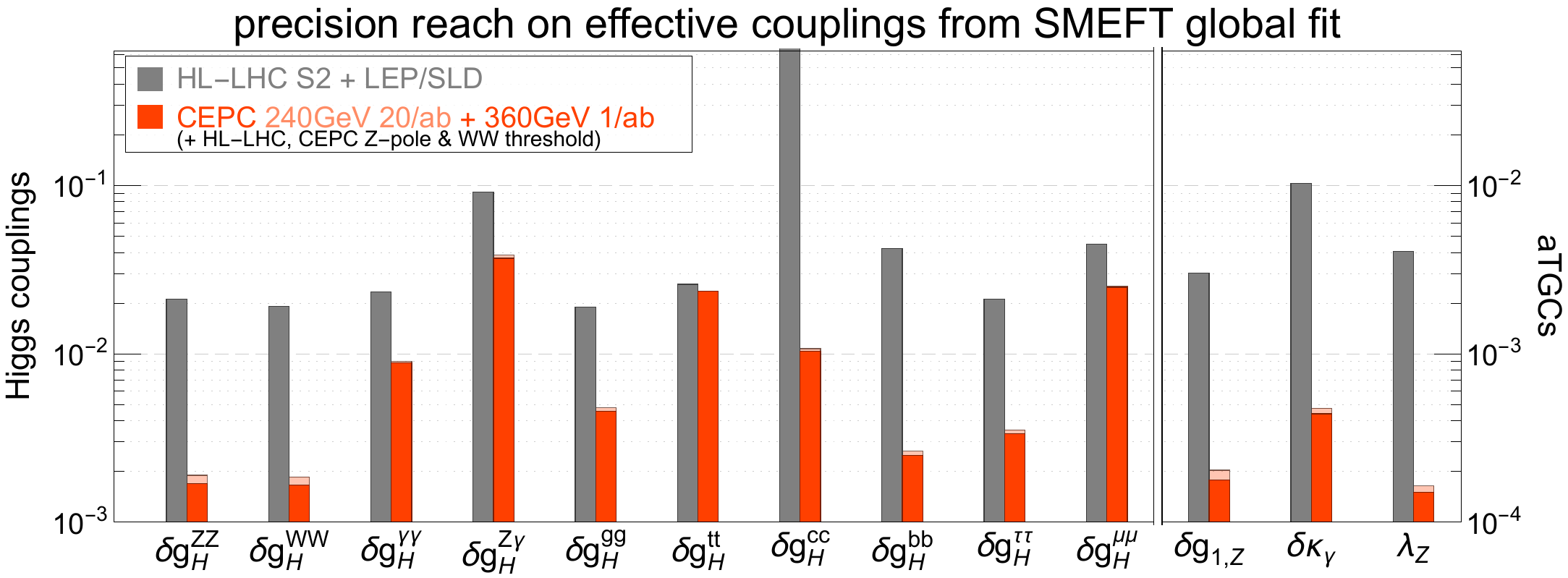}
    \caption{Precision reach on the effective Higgs couplings and aTGCs at the HL-LHC (S2 scenario) and the CEPC, obtained from a global fit of the Higgs and electroweak measurements in the SMEFT framework.  For the CEPC, the light-shaded columns show the results with runs up to 240\,GeV, while the solid columns also include the run at 360\,GeV.}
    \label{fig:eft1}
\end{figure}

In the EFT framework, 
the new particles are assumed to be heavy and can be integrated out, and the new physics effects can be mapped to a series of higher dimensional operators.   
We work in the Standard Model Effective Field Theory (SMEFT) in which the electroweak symmetry breaking is linearly realized.  A more detailed SMEFT analysis of the Higgs and electroweak measurements is presented later in \autoref{sec:smeft}.  Here we project the SMEFT results onto the effective Higgs couplings, which are defined such that they are proportional to the square root of the corresponding Higgs decay~\cite{Barklow:2017suo}\footnote{The only exception is the $ht\bar{t}$ coupling which can be simply defined as the effective Yukawa coupling.}.  The results are presented in \autoref{fig:eft1}. 
The reaches on the anomalous triple gauge couplings (aTGCs) from the same fit are also shown.  For comparison, we also show the results for the HL-LHC S2 scenario with the measurement inputs from \cite{Cepeda:2019klc}, following the treatment in \cite{DeBlas:2019qco}.

In \autoref{fig:eft1}, we again observe a precision reach of $\mathcal{O}(10^{-2}-10^{-3})$ for most of the Higgs couplings at the CEPC, clearly demonstrating its great potential in  precision Higgs physics. 
There are a few important differences to the $\kappa$-framework results that should be noted.  First, 
the reach on the $hWW$ coupling ($\delta g^{WW}_H$) is significantly better in the EFT framework than in the $\kappa$ one.  The former is similar to the reach on $\delta g^{ZZ}_H$ despite the fact that the $hZ$ production channel is much better measured than the $WW$-fusion one.  This is mainly due to the fact that a large deviation between $\delta g^{WW}_H$ and $\delta g^{ZZ}_H$ violates the custodial symmetry, and is therefore tightly constrained by the electroweak measurements at the CEPC.
Similarly, many other operators contribute to both Higgs and electroweak processes. 
Thanks to the outstanding electroweak precision program at the CEPC, these operators are strongly constrained and have very small impacts on the extraction of Higgs couplings~\cite{DeBlas:2019qco}. 
The measurement of the $h \to Z\gamma$ channel also plays a crucial role in the EFT framework as it provides important constraints on the operators (such as $|H|^2 W_{\mu\nu} W^{\mu\nu}$ and $|H|^2 B_{\mu\nu} B^{\mu\nu}$) that also contribute to the $h\gamma\gamma$, $hWW$ and $hZZ$ couplings.  This measurement also constrains the contribution of the $hZ\gamma$ coupling to the $hZ$ production via the $s$-channel photon to a negligible level, which would otherwise have an significant impact on the extraction of the $hZZ$ coupling.

\subsection{CP violation in the Higgs couplings}

The Higgs boson is predicted to be a scalar particle ($J^{P}=0^{++}$) under the SM of particle physics. 
Any observation of charge-parity violation (\textit{CPV}) in the Higgs couplings would be a clear sign of new physics. Such new physics could potentially account for 
(at least some of) the observed baryon asymmetry of the universe.  
At the HL-LHC~\cite{Cepeda:2019klc}, the $CP$-odd $HVV$ couplings are can be probed by the measurement of Higgs decay, and the expected bound on the $CP$-odd parameters $\tilde{c}_{Z \gamma}$ and $\tilde{c}_{Z Z}$ is estimated to be $\pm0.22$ and $\pm0.33$, respectively, at the $68\%$ confidence level. 

CEPC, with good reconstructions of the Higgs production events and much smaller QCD background, provides a unique opportunity to probe the CP-odd $HVV$ couplings. 
Such an analysis was perform in \cite{sha:probing} using the $e^{+} e^{-} \rightarrow Z H \rightarrow \mu^{+} \mu^{-} H(\rightarrow b \bar{b} / c \bar{c} / g g)$ process with the luminosity presented in the CDR. 
A likelihood fit to the binned analysis of the azimuthal angle $\phi$ between the production and the $Z$-decay planes was carried out, and the 68\%CL bound on $\tilde{c}_{Z \gamma}$ and $\tilde{c}_{Z Z}$ is around $\pm 0.37$ and $\pm 0.10$, respectively.  This is further improved to $\pm 0.36$ and $\pm 0.08$ with the use of optimal observables~\cite{Davier:1992nw}.  The reach on $\tilde{c}_{Z Z}$ at CEPC is significantly better than the HL-LHC one, while for $\tilde{c}_{Z \gamma}$ the CEPC bound is weaker due to its suppressed contribution to the $hZ$ production channel.  
With the luminosity updated from $5.6 {\rm \ ab}^{-1}$ to $20 {\rm \ ab}^{-1}$, and more channels included, we expect the reach on the $CP$-odd parameters to be improved by at least a factor of two.

  \subsection{ $W, Z$ electroweak precision measurements at the CEPC \label{w-z-measurement}}
The CEPC offers the possibility of dedicated 
runs at the Z pole and the WW threshold. The expected integrated luminosity for the CEPC Z pole run is 100~$ {\rm ab}^{-1}$, corresponding to $3 \times 10^{12}$ Z bosons. It will also produce over 400 million W boson pairs (of which ~100 million are near the WW threshold with 6~$ {\rm ab}^{-1}$). 
With large integrated luminosity, the CEPC will reach a new level of precision for  
measurements of the properties of the $W$ and $Z$ bosons. Precise measurements of the $W$ and $Z$ boson masses, widths, 
and couplings are critical to test the consistency of the SM. 
In addition, many BSM models predict new couplings of the $W$ and $Z$ bosons to other elementary particles. 
Precise electroweak measurements performed at the CEPC could discover deviations from the SM predictions and reveal the existence of new particles that are beyond the reach of current experiments.

\begin{table*}[h!]
\begin{center}
\begin{tabular}{ccccc}\hline\hline
Observable & current precision   & CEPC  precision (Stat. Unc.)  & CEPC runs  & main systematic          \\\hline
$\Delta{m_Z}$            & 2.1 MeV ~\cite{ALEPH:2005ab,OPAL:2000ufp,DELPHI:2000wje,L3:2000vgx,ALEPH:1999smx}    & 0.1 MeV  (0.005 MeV)        & $Z$ threshold  & $E_{beam}$             \\
$\Delta{\Gamma_Z}$        & 2.3 MeV ~\cite{ALEPH:2005ab,OPAL:2000ufp,DELPHI:2000wje,L3:2000vgx,ALEPH:1999smx}    & 0.025 MeV (0.005 MeV)        & $Z$ threshold    & $E_{beam}$             \\
$\Delta{m_W}$      & 9 MeV  ~\cite{CDF:2022hxs,ATLAS:2017rzl,D0:2013jba,CDF:2012gpf,ALEPH:2013dgf}        & 0.5 MeV (0.35 MeV)           & $WW$ threshold      & $E_{beam}$        \\
$\Delta{\Gamma_W}$       & 49 MeV ~\cite{D0:2009oet,CDF:2007tdb,TevatronElectroweakWorkingGroup:2010mao,ALEPH:2013dgf}         & 2.0 MeV (1.8 MeV)           & $WW$ threshold  & $E_{beam}$         \\
$\Delta{m_t}$ & 0.76 GeV \cite{ATLAS:2014wva} & ${\cal O}(10)$ MeV\footnote{Simulation studies are still on-going to obtain accurate projections of the precision of top quark mass and coupling measurements at the CEPC. } & $t \bar t$ threshold & \\
$\Delta{A_e}$    & $4.9\times10^{-3}$   ~\cite{ALEPH:2005ab,ALEPH:2001uca,DELPHI:1999yne,SLD:2000ujp,OPAL:2001brm,L3:1998oan}        & $1.5\times10^{-5}$ ($1.5\times 10^{-5}$)           & $Z$ pole ($Z \to \tau\tau$)   &   Stat. Unc.    \\
$\Delta{A_\mu}$       & $0.015$  ~\cite{SLD:2000ujp,ALEPH:2005ab}        & $3.5\times10^{-5}$  ($3.0\times 10^{-5}$)         & $Z$ pole ($Z \to \mu\mu$)&   point-to-point Unc.       \\
$\Delta{A_\tau}$     & $4.3\times10^{-3}$   ~\cite{ALEPH:2005ab,ALEPH:2001uca,DELPHI:1999yne,SLD:2000ujp,OPAL:2001brm,L3:1998oan}        & $7.0\times10^{-5}$ ($1.2 \times 10^{-5}$)             & $Z$ pole ($Z \to \tau\tau$)      &tau decay model     \\
$\Delta{A_b}$       & 0.02 ~\cite{ALEPH:2005ab,SLD:2004kjl}        & $20\times10^{-5}$ ($3\times10^{-5}$)          & $Z$ pole    &    QCD effects   \\
$\Delta{A_c}$       & 0.027 ~\cite{ALEPH:2005ab,SLD:2004kjl}         & $30\times10^{-5}$ ($6\times10^{-5}$)         & $Z$ pole     &     QCD effects   \\
$\Delta{\sigma_{had}}$ & 37 pb ~\cite{ALEPH:2005ab,OPAL:2000ufp,DELPHI:2000wje,L3:2000vgx,ALEPH:1999smx}   &     2 pb (0.05 pb)      & $Z$ pole & lumiosity \\
$\delta R^0_b$      & 0.003  ~\cite{ALEPH:2005ab,SLD:2005zyw,L3:1999aer,OPAL:1998kxc,DELPHI:1998cnd,ALEPH:1997xqy}         & 0.0002 ($5\times10^{-6}$)         & $Z$ pole  &      gluon splitting         \\
$\delta R^0_c$      & 0.017  ~\cite{ALEPH:2005ab,SLD:2005zyw,DELPHI:1999slw,ALEPH:1999syy,OPAL:1997edj,OPAL:1996ikk} & 0.001  ($2\times10^{-5}$)        & $Z$ pole    &    gluon splitting         \\
$\delta R^0_e$      & 0.0012  ~\cite{ALEPH:2005ab,OPAL:2000ufp,DELPHI:2000wje,L3:2000vgx,ALEPH:1999smx}         & $2\times10^{-4}$ ($3\times10^{-6}$)         & $Z$ pole  & $E_{beam}$ and t channel          \\
$\delta R^0_\mu$      & 0.002    ~\cite{ALEPH:2005ab,OPAL:2000ufp,DELPHI:2000wje,L3:2000vgx,ALEPH:1999smx}       & $1\times10^{-4}$ ($3\times10^{-6}$)          & $Z$ pole &   $E_{beam}$             \\
$\delta R^0_\tau$      & 0.017 ~\cite{ALEPH:2005ab,OPAL:2000ufp,DELPHI:2000wje,L3:2000vgx,ALEPH:1999smx}           & $1\times10^{-4}$ ($3\times10^{-6}$)          & $Z$ pole  &  $E_{beam}$             \\

$\delta N_\nu$    & 0.0025  ~\cite{ALEPH:2005ab,Janot:2019oyi}         & $2\times10^{-4}$ ($3\times10^{-5}$ )         & $ZH$ run ($\nu \nu \gamma$)     &   Calo energy scale \\
         \hline\hline
\end{tabular}
\end{center}
\caption{The expected precision for a selected set of EW precision measurements at the CEPC and the comparison with the precision from the previous measurements~\cite{ALEPH:2005ab,OPAL:2000ufp,DELPHI:2000wje,L3:2000vgx,ALEPH:1999smx,ATLAS:2017rzl,D0:2013jba,CDF:2012gpf,ALEPH:2013dgf,CDF:2022hxs,ATLAS:2017rzl,D0:2013jba,CDF:2012gpf,ALEPH:2013dgf,ALEPH:2001uca,DELPHI:1999yne,SLD:2000ujp,OPAL:2001brm,L3:1998oan,D0:2009oet,CDF:2007tdb,TevatronElectroweakWorkingGroup:2010mao,ALEPH:2013dgf,SLD:2004kjl,SLD:2005zyw,L3:1999aer,OPAL:1998kxc,DELPHI:1998cnd,ALEPH:1997xqy,DELPHI:1999slw,ALEPH:1999syy,OPAL:1997edj,OPAL:1996ikk,Janot:2019oyi}. The CEPC accelerator running mode and  total integrated luminosity expected for each measurement are also listed. Relative uncertainties are quoted for $\delta R^0_b$, $\delta R^0_c$, $\delta R^0_e$, $\delta R^0_\mu$, $\delta R^0_\tau$ and $\delta N_\nu$ measurements, while absolute uncertainties are quoted for other measurements.}
\label{tab:precision-WZ} 
\end{table*}
Significant improvements are expected from the CEPC measurements for some of these variables. \autoref{tab:precision-WZ} lists the expected precision from CEPC compared to achieved precisions from the previous measurements~\cite{ALEPH:2005ab,OPAL:2000ufp,DELPHI:2000wje,L3:2000vgx,ALEPH:1999smx,ATLAS:2017rzl,D0:2013jba,CDF:2012gpf,ALEPH:2013dgf,CDF:2022hxs,ATLAS:2017rzl,D0:2013jba,CDF:2012gpf,ALEPH:2013dgf,ALEPH:2001uca,DELPHI:1999yne,SLD:2000ujp,OPAL:2001brm,L3:1998oan,D0:2009oet,CDF:2007tdb,TevatronElectroweakWorkingGroup:2010mao,ALEPH:2013dgf,SLD:2004kjl,SLD:2005zyw,L3:1999aer,OPAL:1998kxc,DELPHI:1998cnd,ALEPH:1997xqy,DELPHI:1999slw,ALEPH:1999syy,OPAL:1997edj,OPAL:1996ikk,Janot:2019oyi}. 
These runs allow high precision electroweak measurements of  the $Z$ boson decay  partial widths, {\it i.e.} the parameters $R_b=\Gamma_{ Z\to b\bar{b}}/\Gamma_{\rm had}$, $R_c=\Gamma_{ Z\to c\bar{c}}/\Gamma_{\rm had}$ and $R_\ell =
\Gamma_{\rm had}/\Gamma_{ Z\to \ell\bar{\ell}}$ where $l=e,\,\mu,\,\tau$. 
It would also perform high precision measurements of the forward-backward 
charge asymmetry ($A^f_{\rm FB}$), the effective weak mixing angle ($\sin^2 \theta_W^{\rm eff}$), number of light neutrino 
species ($N_\nu$), and the mass of the $Z$ boson ($m_Z$). 
Even through the polarized beam is not the default option in the CEPC physics program, the asymmetry observables $A_{e}$ and $A_{\tau}$ can still be measured by probing the tau polarization asymmetry in the tau hadronic decay mode using $Z \to \tau\tau$ events at the Z pole runs with unpolarized beams. 
Projections for the other $A_f$ variables are dervied from the measurements of $A_{e}$ and $A^f_{\rm FB}$ using the relation $A^{f}_{\rm FB}=\frac{3}{4}  A_e  A_f$. 
The threshold scan runs are also crucial for the calibrations of leptons and jets. 
It is also possible to perform some measurements with the $Z$ boson without these dedicated low-energy runs near or at the $Z$ pole. 
For example, the direct measurement of the number of light neutrino species can be performed in $ZH$ runs intended for Higgs boson measurements. 

$m_Z$ can be measured at CEPC $Z$ pole runs and $Z$ threshold scans runs. The dominant systematic is expected to come from beam energy measurements. By using resonant depolarization method, the beam energy is expected to be calibrated to a precision of 0.1 MeV~\cite{Koratzinos:2015hya,Arnaudon:1994zq,Assmann:1998qb}, which is a factor of 20 improvement compared to LEP measurements~\cite{ALEPH:2005ab,OPAL:2000ufp,DELPHI:2000wje,L3:2000vgx,ALEPH:1999smx}. 

The W boson mass ($m_W$) can be measured with WW threshold runs at CEPC with unprecedented experimental precision~\cite{Shen:2018afo}. Recent result from CDF collaboration showed that the measured $m_W$ was significantly higher than the Standard Model predicts, with a discrepancy of 7 $\sigma$ standard deviations~\cite{CDF:2022hxs}. CEPC will offer an opportunity to improve the experimental systematic by one order of magnitude, revealing insight on the potential new physics behind the deviations in the $m_W$ measurement. The systematic uncertainty in beam energy measurement is expected to be one of the dominated experimental uncertainties. A new method for high energy beam energy measurement with microwave-beam Compton back-scattering was proposed to calibrate the beam energy at WW threshold runs~\cite{Si:2021nsr,Tang:2020gmv}. 

The weak mixing angle is one of the most important parameter for the Standard Model electroweak sector. The asymmetry between the right-handed and left-handed lepton couplings to the $Z$ boson ($A_\ell$) is a powerful experimental observable to measure the weak mixing angle.  As mentioned above, 
$A_e$ and $A_\tau$ can be measured precisely with the final-state tau polarizations at the CEPC. 
The forward-backward asymmetry ($A^f_{\rm FB}$) measurement at the Z pole can provides a similarly accurate determination of the weak mixing angle. 


\subsection{Measurement of the $e^+e^-\to WW$ process}
\label{sec:ww}

The measurement of the $e^+e^-\to WW$ process provides important constraints on various new physics contributions and is a crucial input for the global SMEFT analysis. 
Conventionally, the results are presented in terms of the bounds on the three CP-even anomalous triple gauge couplings (aTGCs).   Additional modifications are allowed in the global SMEFT framework.  Focusing on dimension-6 CP-even tree-level effects, the new physics contributions in SMEFT can be written in terms of the following 7 parameters~\footnote{ Here we do not consider the contributions to the $W$ branching ratios, which only modify the rates and can be separated from the contributions to the $WW$ production.}, 
\begin{equation}                     
\delta g_{1,Z}\,,~~\delta \kappa_{\gamma}\,,~~\lambda_Z\,,~~\delta g^{ee}_{Z,L}\,,~~\delta g^{ee}_{Z,R}\,,~~\delta g^{e\nu}_{W}\,,~~\delta_{m_W} \,,  \nonumber
\end{equation}   
where $\delta g_{1,Z}$, $\kappa_{\gamma}$ and $\lambda_Z$ are the three aTGCs, $g^{ee}_{Z,L}$, $\delta g^{ee}_{Z,R}$ and $\delta g^{e\nu}_{W}$ are the modifications to the gauge couplings of electron (and neutrino) to the Z, W bosons, and $\delta_{m_W}$ modifes the W-boson mass.

Apart from $\delta_{m_W}$ which is strongly constrained by the W-boson mass measurement, the EFT parameters receive strong energy enhancements and are much better probed at 240\,GeV than around the $WW$ threshold.  The parameters are also sensitive to the differential distributions, which can be parameterized as five angles (the production polar angle and two decay angles for each $W$).  We implement the so-called ``optimal observables''~\cite{Diehl:1993br} and perform a phenomenological analysis to efficiently extract the precision reach of the EFT parameters from the differential distributions.  Our analysis follow closely the one in \cite{DeBlas:2019qco}.  We focus on the semi-leptonic channel. 
We consider only statistical uncertainties, and apply a conservative $50\%$ signal selection efficiency to account for possible effects of detector acceptance, systematic effects and 
signal selection cuts.  The resulting likelihood of the 7 parameters above is combined with the Higgs and other electroweak measurements in the global SMEFT analysis in \autoref{sec:smeft}.

\subsection{SMEFT global fit of Higgs and electroweak processes}
\label{sec:smeft}

\begin{figure}
    \includegraphics[width=\textwidth]{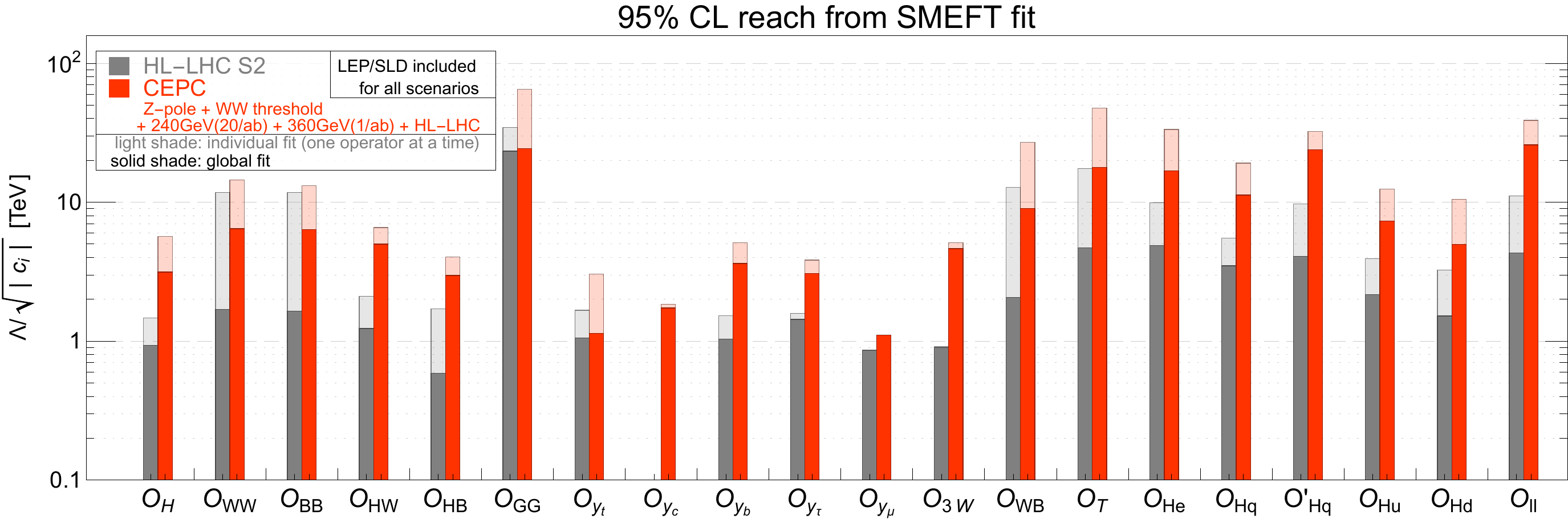}
    \caption{$95\%$ reach on new physics scale ($\Lambda/\sqrt{c_i}$) for various operators at HL-LHC and CEPC.  The solid shaded columns are the results of the global fits, while the light shaded ones are the results of the individual fits, obtained by switching on one operator coefficient at a time.  The operators are listed in Table~\ref{tab:d6}. }
    \label{fig:eft2}
\end{figure}

The Standard Model Effective Field Theory (SMEFT) offers a systematic 
parameterization of the new physics effects in the 
Higgs and electroweak measurements.  Assuming the scale of the new physics is significantly larger than the electroweak scale, and the electroweak symmetry breaking is linearly realized, the SM Lagrangian can be extended with a series of higher dimensional operators~\cite{Buchmuller:1985jz, Grzadkowski:2010es}, 
\begin{equation}
\mathcal{L}_{\rm SMEFT} =
    \mathcal{L}_{\rm SM} + 
    \sum_i \frac{c^{(6)}_i}{\Lambda^2} \mathcal{O}^{(6)}_i + 
    \sum_j \frac{c^{(8)}_j}{\Lambda^4} \mathcal{O}^{(8)}_j +         \cdots   \label{eq:smeft}
\end{equation}
where $\Lambda$ denotes the scale of the new physics.  Only operators of even dimensions are allowed if baryon and lepton numbers are conserved.  For a sufficiently large $\Lambda$,  the new physics effects could be well approximated by the dimension-6 operators.  The general bottom-up approach is to start with a non-redundant basis and perform a global fit with all possible measurements to obtain the bounds on the operator coefficients.  
A more practical approach is to focus on a subset of the measurements, such as the the Higgs and electroweak ones, and obtain bounds on the operator coefficients that contribute to them at the leading order.  This has been done in previous studies~\cite{Falkowski:2015jaa, Ellis:2018gqa, Dawson:2020oco, Ellis:2020unq, Ethier:2021bye, Almeida:2021asy, Barklow:2017suo, deBlas:2019rxi, DeBlas:2019qco}.  Our analysis here follow closely the one in \cite{DeBlas:2019qco}.  For CEPC, the measurement inputs we use are the Higgs rate measurements in \autoref{tab:exclusiveprecision}~\footnote{Note that the bound on the Higgs width is a derived quantity and is not used here as a measurement input.}, the EW precision observables in \autoref{tab:precision-WZ}, and the measurements of $e^+e^-\to WW$ in \autoref{sec:ww}. For comparison, we also perform the same analysis for the HL-LHC S2 scenario, with the Higgs measurement inputs from \cite{Cepeda:2019klc}, the analysis of $pp\to WW/WZ$ in \cite{Grojean:2018dqj}, and the electroweak measurements at the LEP and the SLD in Ref.~\cite{ALEPH:2005ab}. 
The results of our analysis are shown in \autoref{fig:eft2} in terms of the $95\%$ reach on the combination $\Lambda/\sqrt{c_i}$ (which corresponds to the scale of new physics if $c_i=1$), with operators defined in \autoref{tab:d6}. Here we have imposed the flavor universality condition on the operators that contributes to the $Zf\bar{f}$ couplings.  The SMEFT fit results are also projected on the effective Higgs couplings and aTGCs shown previously in \autoref{fig:eft1}. In this case, the flavor universality condition is reduced to a $U(2)$ symmetry of the first two generation quarks.  It can be seen in \autoref{fig:eft2} that the CEPC is able to provide $\mathcal{O}(1$-$10)$ improvements on the reach on the new physics scale of the HL-LHC and LEP/SLD measurements.  
Compared with the SMEFT analysis in the CDR, one important difference is that the one in CDR assumes perfect Z-pole measurements, so that the corresponding EW operators can be eliminated in order to focus on the operators that only contributes to the Higgs and $WW$ processes.  This framework is expanded here to include realistic Z-pole measurements, and the constraints on the corresponding EW operators are also obtained.

\begin{table}
\centering
\begin{tabular}{l|l} \hline\hline
$\mathcal{O}_H = \frac{1}{2} (\partial_\mu |H|^2 )^2$ &  $\mathcal{O}_{GG} =  g_s^2 |H|^2 G^A_{\mu\nu} G^{A,\mu\nu}$  \\ 
$\mathcal{O}_{WW} =  g^2 |H|^2 W^a_{\mu\nu} W^{a,\mu\nu}$  & $\mathcal{O}_{y_u} = y_u |H|^2 \bar{q}_L \tilde{H} u_R \,+\, {\rm h.c.}$ \hspace{0.25cm} {\scriptsize $(u \to t, c)$}  \\
$\mathcal{O}_{BB} =  g'^2 |H|^2 B_{\mu\nu} B^{\mu\nu}$ &  $\mathcal{O}_{y_d} = y_d |H|^2 \bar{q}_L H d_R \,+\, {\rm h.c.}$ \hspace{0.3cm} {\scriptsize $(d \to b)$}  \\
$\mathcal{O}_{HW} =  ig(D^\mu H)^\dagger \sigma^a (D^\nu H) W^a_{\mu\nu}$  &  $\mathcal{O}_{y_e} = y_e |H|^2 \bar{l}_L H e_R \,+\, {\rm h.c.}$ \hspace{0.36cm} {\scriptsize $(e \to \tau, \mu)$}  \\
$\mathcal{O}_{HB} =  ig'(D^\mu H)^\dagger  (D^\nu H) B_{\mu\nu}$ &      $\mathcal{O}_{3W} = \frac{1}{3!} g \epsilon_{abc} W^{a\,\nu}_\mu W^b_{\nu \rho} W^{c\,\rho\mu} $  \\   \hline
$\mathcal{O}_{WB} = gg' H^\dagger \sigma^a H W^a_{\mu\nu} B^{\mu\nu}$  &   
$\mathcal{O}_{\ell\ell} =  (\bar{\ell}_L \gamma^\mu \ell_L) (\bar{\ell}_L \gamma_\mu \ell_L)$  \\
$\mathcal{O}_{T} = \frac{1}{2}(H^\dagger \overleftrightarrow{D_\mu} H)^2 $  &   
$\mathcal{O}_{He} = i H^\dagger \overleftrightarrow{D_\mu} H \bar{e}_R \gamma^\mu e_R $  \\
%
$\mathcal{O}_{Hq} = i  H^\dagger \overleftrightarrow{D_\mu} H \bar{q}_L \gamma^\mu q_L  $ &  $\mathcal{O}_{Hu} = i H^\dagger \overleftrightarrow{D_\mu} H \bar{u}_R \gamma^\mu u_R $ \\
$\mathcal{O}'_{Hq} = i H^\dagger \sigma^a \overleftrightarrow{D_\mu} H \bar{q}_L \sigma^a \gamma^\mu q_L $ &  $\mathcal{O}_{Hd} = i H^\dagger \overleftrightarrow{D_\mu} H \bar{d}_R \gamma^\mu d_R $ \\  \hline\hline
\end{tabular}
\caption{A set of non-redundant dimension-six operators that contributes to the Higgs and EW processes in our analysis.  }
\label{tab:d6}
\end{table}

\section{Flavor Physics}
Flavor physics is an inseparable part of the physics case at CEPC. The rich phenomenology of flavored final states offers efficient ways to understand the SM better and scrutinize BSM physics. The paradigm of flavor physics relies on heavy quark or lepton decays, where the leading SM amplitudes are mediated by the heavy $W$ boson. The decay width of a fermion $f$ is therefore suppressed by the small $m_f^4/m_W^4$, and $f$ becomes long-lived. Even minor theoretical corrections within the SM or BSM effects could reveal themselves on top of the extremely narrow SM widths.

The Tera-$Z$ phase of CEPC stands at a privileged position of studying flavor physics~\cite{CEPCStudyGroup:2018ghi}. By operating at the $Z$-pole, vast amounts of heavy flavor final states will be produced, thanks to the large $\sigma (e^+e^-\to Z \to b\bar{b},~c\bar{c},~\tau^+\tau^-)$ and a high integrated luminosity. About $\mathcal{O}(10^{11})$ $b\bar{b}$, $c\bar{c}$, and $\tau^+\tau^-$ pairs will be produced at the $Z$ factory phase of CEPC. Meanwhile, $m_Z$ is sufficiently larger than $m_{b,c,\tau}$, creating a surplus of energy. Final state particles will thus be boosted even if they are soft in the rest frame of their mother particle decays, enhancing measurement accuracies. The abundant energy from $Z$ decays also generates a variety of hadrons, including many rare ($e.g.$, $B_c$~\cite{Zheng:2019gnb}) and exotic ($e.g.$, tetraquarks~\cite{Qin:2020zlg}) species. Moreover, the cleanness of a lepton collider enables investigating decay modes that are extremely rare or containing neutral/invisible particles. Finally, unexpected discoveries may also come from the phases beyond the $Z$-pole run, such as the $WW$-threshold or Higgs factory phases. We summarize representative studies at CEPC and similar projects in the following.

\subsection{Precise Measurements of Flavor Physics Parameters}
Parameters like the meson masses, CKM matrix elements, and $CP$ phases are the cornerstone of flavor physics. Given the merits described above, we expect CEPC to push precisions of multiple parameters to unprecedented limits. It is also possible that the evidence of BSM physics first appears as discrepancies between precision measurements and the theory. Experimental approaches to these fundamental parameters have been developed in the past decades. Measurements at CEPC can follow those well-established methods, or more likely, by adopting new techniques exploiting the future collider. Examples include measuring final states with neutral/invisible particles or high track multiplicities, where the high particle boosts and low backgrounds make CEPC a powerful tool for their searches.

$CP$ violation measurements at CEPC and other $Z$ factory projects are under active progress. Promising projections are demonstrated when determining several angles like $\gamma$ and $\beta_s$. The Tera-$Z$ precision on the angle $\beta_s$ using the fully charged $B_s\to J/\psi(\to \mu^+\mu^-)\phi(\to K^+K^-)$~\cite{Aleksan:2021gii,Talk_Mingrui} is projected to be $\simeq 0.1^\circ$, comparable to the LHCb extrapolation after the high-luminosity runs. The phenomenology study on angle $\gamma$ via a fully-charged $B_s\to D_s^\pm (\to K^+ K^-\pi^\pm) K^\mp$ mode is also conducted~\cite{Aleksan:2021gii}, resulting a Tera-$Z$ sensitivity about $1^\circ$. Alternatively, the determination of angle $\gamma_s$ is also viable by including $B^\pm\to D^0(\bar{D}^0)K^\pm$ decays~\cite{Aleksan:2021fbx}. Besides, preliminary analysis on $B_{(s)}\to 2\pi^0(\to 4\gamma)$ decays~\cite{Talk_Yuexin} indicates the potential of pushing the $\alpha$ angle uncertainty below $2^\circ$, while stronger constraints can be achieved by measuring $B\to \rho\rho$ decays~\cite{Kou:2018nap}. Other promising directions include observing the $CP$ violations in semileptonic $B$ decays and thus $CP$ violations in $B$ mixings~\cite{Charles:2020dfl}.

For $|V_{ij}|$ measurements, the global picture is less clear. The most remarkable progress known is the determination of $|V_{cb}|$ in $W$ decays, given the abundant $W$ in the $WW$-threshold run and latest flavor tagging developments~\cite{Bedeschi:2022rnj}. It serves as a manifest example of runs beyond the $Z$-pole that contribute to flavor physics. The relative size $|V_{ub}/V_{cb}|$ can also be determined from semileptonic $b\to u(c)\ell\nu$ transitions~\cite{Abada:2019lih}. However, systematic exemplifications of many statements above are not yet available. We hope that such a gap of knowledge will be fully addressed by future studies.

\subsection{(Semi)leptonic and Rare Decays}
$Z$-pole phenomenology of (semi)leptonic and rare flavor changing neutral current (FCNC) decays, especially $b$-hadron decays, is significantly motivated by the recent observation of $B$ anomalies~\cite{Azatov:2018knx,Bordone:2016gaq}. These decays are the direct test of lepton flavor universality (LFU)~\cite{Alguero:2019ptt} and thus also sensitive to BSM physics coupling to the lepton sector~\cite{Buras:2014fpa,Calibbi:2015kma,deMedeirosVarzielas:2015yxm,Capdevila:2017iqn,Bordone:2018nbg}.

For charged-current-induced $b\to c\ell(\tau)\nu$ transitions, the bottleneck of measurement would be the $\tau$ modes as $\tau$ decays lead to extra neutrinos and softer tracks. Breakthroughs in these channels are expected because of the high final state boost and cleanness at the $Z$ pole. The first example is the $B_c \to \tau \nu$ decay, where the constraint from $\Gamma(B_c)$ is weak~\cite{Azatov:2018knx}. The CEPC Tera-$Z$ study~\cite{Zheng:2020emi} uses full simulation and leptonic $\tau$ decays (see also~\cite{Amhis:2021cfy} where the hadronic decay is used). The sensitivities from both methods agree and the number of $B_c\to \tau\nu$ decays is measured $\mathcal{O}(10^{-2})$ level. A work in preparation~\cite{Kwok_FCCC} simulates the measurements of $R_{D_s}$, $R_{D_s^\ast}$, $R_{J/\psi}$, and $R_{\Lambda_c}$ at Tera-$Z$. All four $R$ values can be pinned down with precisions of (sub)percent level with high signal-to-noise ratios. The combined results can push the scale of BSM physics up to the multi-TeV range~\cite{DiLuzio:2017chi}.

Another obstacle emerges when it comes to FCNC rare decays, is their low SM branching ratio $\leq 10^{-5}$. Careful background mitigation and accurate event reconstruction are necessary to probe such modes at CEPC. Based on the same arguments above, CEPC can bring unique sensitivities to di-$\tau$ modes of $b\to s\tau^+\tau^-$~\cite{Kamenik:2017ghi,Li:2020bvr}, where the future sensitivities from Belle II and LHCb cannot cover the small SM predicted BRs$\sim 10^{-7}$~\cite{Kou:2018nap,Bediaga:2018lhg}. The Tera-$Z$ sensitivities to BR($B^0\to K^\ast \tau^+\tau^-$), BR($B_s\to \phi \tau^+\tau^-$), BR($B^\pm\to K^\pm \tau^+\tau^-$), and BR($B_s\to \tau^+\tau^-$) decays are evaluated using 3-prong tau decays~\cite{Li:2020bvr}, reaching $\mathcal{O}(10^{-5})$ for the two-body $B_s\to \tau^+\tau^-$ mode and $\mathcal{O}(10^{-7})$ for other three-body modes. For the baseline CEPC luminosity, such sensitivities can $O(1)$ deviations from the SM. For di-muon modes, the current benchmarks are the $B^0 \to \mu^+\mu^-$ and $B_s \to \mu^+\mu^-$ measurements~\cite{Monteil:2021ith}. While the former is primarily limited by statistics and reaching high sensitivity, the later will also affected by $B^0\to\pi^+\pi^-$ decays via $\pi/\mu$ misidentification. 

The rare decay with two neutrinos, $B_s\to \phi\nu\bar{\nu}$, are also analyzed with the full CEPC detector simulation~\cite{Li:2022tov}. The estimated relative uncertainty at the Tera-$Z$ stage reaches $\leq 2\%$. The study also illustrates the accurate kinematic reconstructions and differential measurements involving a large missing momentum by virtue of the advanced tracking and calorimetry techniques.   It is also noteworthy that the above discussions of CEPC may also apply to other FCNC rare decays like (double) radiative decays~\cite{Colangelo:2010wg} or charm decays~\cite{Bause:2020xzj}, although more thorough studies are needed.

\subsection{Low multiplicity and $\tau$ Physics}
The large amount of $\tau$'s produced at the $Z$-pole are clean and energetic ($\gamma_\tau \sim 26$). Including the tagging efficiency, the effective $\tau$ statistics at CEPC will exceed those at Belle II~\cite{Kou:2018nap} or even the future Super Tau Charm Factory~\cite{Shi:2020nrf}. The $Z$-pole phase of CEPC is therefore an ideal facility to study $\tau$ physics, with the physics reaches extensively discussed in~\cite{Dam:2018rfz}. On the other hand, the low final state multiplicity of the $Z\to\tau^+\tau^-$ process is analogous to many distinct physics cases, $e.g.$, exclusive $Z$ hadronic decays~\cite{Grossmann:2015lea} and $e^+e^-$ interactions with significant radiative return~\cite{Karliner:2015tga}.
The phenomenological similarities between the above physics lead to close connections between their signal identification or even shared backgrounds.

The current focus of the $\tau$ physics at the $Z$-pole is the charged-lepton flavor violating (cLFV) interaction in $\tau$ and $Z$ decays. As the cLFV interactions are highly suppressed in the SM, even tiny BSM amplitudes would overwhelm the SM prediction and introduce striking signals~\cite{Calibbi:2017uvl,Qin:2017aju,Li:2018cod}. The latest study also emphasized the complementarity between cLFV $Z$ and $\tau$ decays~\cite{Calibbi:2021pyh}. The projected sensitivities to two benchmarks $\tau$ exotic decays, $\tau\to 3\mu$ and $\tau\to \mu\gamma$, are comparable to the Belle II expectations~\cite{Kou:2018nap}. Additionally, the basic $\tau$ properties ($e.g.$, mass and lifetime) and LFU tests in leptonic $\tau$ decays are also appropriate targets at CEPC. 
Many aspects of $\tau$ physics remain to be fully explored at CEPC, including the $\alpha_s$ determination in hadronic $\tau$ decays~\cite{Pich:2013lsa}, the $\tau$ polarimetry as a measurement of EW physics~\cite{Yumino:2022vqt}, the production of $\tau^+\tau^-$ bound states~\cite{dEnterria:2022ysg}, and $CP$ violating $\tau$ decays~\cite{Grossman:2011zk}.

\begin{table}[h!]
\centering
\resizebox{1 \textwidth}{!}{\begin{tabular}{ccccc}
\hline
Measurement & Current~\cite{ParticleDataGroup:2020ssz} & FCC~\cite{Dam:2018rfz} & Tera-$Z$ Prelim.~\cite{Talk_Dan} &  Comments \\
\hline
Lifetime [sec] & $\pm 5\times 10^{-16}$ & $\pm 1\times 10^{-18}$ &  & from 3-prong decays, stat. limited\\
BR($\tau \to \ell \nu\bar\nu$)   & $\pm 4\times 10^{-4}$ & $\pm 3\times 10^{-5}$  & & $0.1\times$ the ALEPH systematics  \\
m($\tau$) [MeV]  & $\pm 0.12$  & $\pm 0.004 \pm 0.1$ & & $\sigma(p_{\rm track})$ limited \\
BR($\tau \to 3\mu$) &$< 2.1\times 10^{-8}$& $\mathcal{O}(10^{-10})$ & same & bkg free  \\
BR($\tau \to 3e$)  & $<2.7\times 10^{-8}$ &  $\mathcal{O}(10^{-10})$ & & bkg free\\
BR($\tau^\pm \to e\mu\mu $) &  $<2.7\times 10^{-8}$ & $\mathcal{O}(10^{-10})$ & & bkg free\\
BR($\tau^\pm \to \mu ee $) &   $<1.8\times 10^{-8}$ & $\mathcal{O}(10^{-10})$& & bkg free\\
BR($\tau \to \mu\gamma$) & $<4.4\times 10^{-8}$ & $\sim 2\times 10^{-9}$ & $\mathcal{O}(10^{-10})$ & $Z\to \tau\tau \gamma$ bkg , $\sigma(p_{\gamma})$ limited\\
BR($\tau \to e\gamma $) & $<3.3\times 10^{-8}$ &   $\sim 2\times 10^{-9}$ & & $Z\to \tau\tau \gamma$ bkg, $\sigma(p_{\gamma})$ limited\\
\hline
BR($Z \to \tau \mu $) & $<1.2\times 10^{-5}$ &   $\mathcal{O} (10^{-9})$ & same &  $\tau\tau$ bkg,  $\sigma(p_{\rm track})$ \& $\sigma(E_{\rm beam})$ limited\\
BR($Z \to \tau e $) & $<9.8\times 10^{-6}$ &   $\mathcal{O} (10^{-9})$ &  &  $\tau\tau$ bkg,  $\sigma(p_{\rm track})$ \& $\sigma(E_{\rm beam})$ limited\\
BR($Z \to \mu e $) & $<7.5\times 10^{-7}$ &   $10^{-8}-10^{-10}$ & $\mathcal{O}(10^{-9})$ &  PID limited\\
\hline
BR($Z\to \pi^+\pi^-$) & & & $\mathcal{O}(10^{-10})$ & $\sigma(\vec{p}_{\rm track})$ limited, good PID\\
BR($Z\to \pi^+\pi^-\pi^0$) & & & $\mathcal{O}(10^{-9})$ & $\tau\tau$ bkg\\
BR($Z\to J/\psi \gamma $) & $<1.4\times 10^{-6}$ & & $10^{-9}-10^{-10}$ & $\ell\ell\gamma$+$\tau\tau\gamma$ bkg\\
BR($Z\to \rho \gamma $) & $<2.5\times 10^{-5}$ & & $\mathcal{O}(10^{-9})$ & $\tau\tau\gamma$ bkg, $\sigma(p_{\rm track})$ limited \\
\hline
\end{tabular}}
\caption{Projected sensitivities of $\tau$ physics at the $Z$-factory run of FCC-$ee$~\cite{Dam:2018rfz} and recent Tera-$Z$ updates~\cite{Talk_Dan}. All numbers are presented as absolute instead of relative values. For $\tau\to 3e$, $\tau\to \mu ee$, and $\tau\to e\mu\mu$ limits, we assume the sensitivities are similar to that of $\tau \to 3\mu$. The expected reaches for several exclusive hadronic $Z$ decays are also listed.}
\label{tab:tau1}
\end{table}

The low multiplicity of aforementioned $\tau$ physics studies also inspires the search for exclusive $Z$ hadronic decays. In those searches, the hierarchy between $m_Z$ and $\Lambda_{\rm QCD}$ and small backgrounds offer an excellent chance to test the factorization theorem and extract the internal structure of mesons~\cite{Internal_Cheng}. In Table~\ref{tab:tau1}, we list the FCC-$ee$ projections provided in~\cite{Dam:2018rfz} and the comparison with current limits. A preliminary update~\cite{Talk_Dan} from the CEPC full detector simulations and lepton PID performance~\cite{Yu:2021pxc} are also listed for comparison.

\section{Beyond the Standard Model Physics}
CEPC not only serves as a Higgs factory, but also has a tremendous  potential to search for the direct production of new physics states. In particular, with a very clean collision background, CEPC has the discovery advantage in many scenarios which are challenging  at hadron colliders due to sizable backgrounds,  large pile-up, trigger constraints from high energy objects, and difficulties in object reconstruction and identification. The identification of the Higgs bosons via the recoil mass method together with its high reconstruction efficiency makes it extremely sensitive to probe the exotic decays of the Higgs bosons at the CEPC.  Moreover, it can also be sensitive to models various new physics scenarios, including Supersymmetry, Dark Matter, Long-Lived Particles, and more.  The rest of the section is dedicated to describe the reach of the CEPC in probing these new physics scenarios, within an emphasis on updating the results reported in the CDR \cite{CEPCStudyGroup:2018ghi}. 

\subsection{Higgs Exotic Decays}
\label{ssec:Higgsexotic}

New physics could reveal themselves in the Higgs exotic decays, especially
hidden sector dynamics through this generic Higgs portal. The Higgs exotic decay program complements the Higgs coupling precision measurements and is a critical component of future Higgs factories. 

\begin{figure}[!ht]
\centering
\includegraphics[scale=0.45,clip]{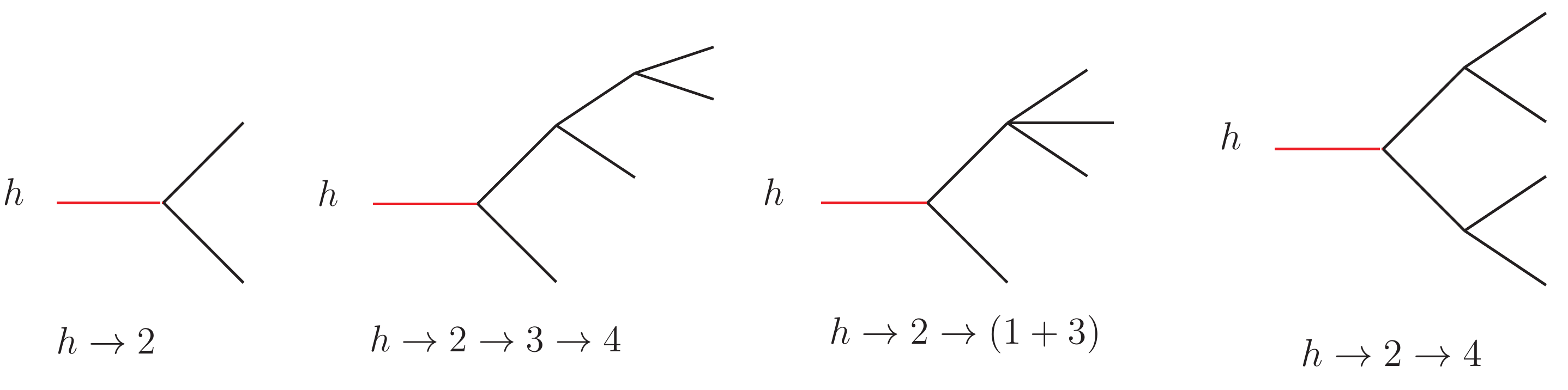}
\caption{\label{fig:topo}Representative topologies of the Higgs exotic decays. }
\end{figure}

A survey of lepton collider sensitivities to Higgs exotic decays into final states was initially carried 
out in Ref.~\cite{Liu:2016zki}, focusing on challenging channels at hadron colliders. The study shows promising sensitivities at lepton colliders.
This study focuses on two-body Higgs decays into BSM particles, dubbed as $X_i$, $h\to X_1 X_2$, which are
allowed to decay further, to up to four-body final states.
The cascade decay modes are classified into four cases, schematically
shown in Fig.~\ref{fig:topo}.
A large class of BSM physics, such as singlet extensions,  two-Higgs-doublet-models, SUSY models, Higgs portals, gauge extensions of the SM~\cite{Curtin:2013fra,Liu:2016zki,deFlorian:2016spz,Cepeda:2021rql}, motivates these exotic decay considerations.

\begin{figure}[!ht]
\centering
\includegraphics[width=16cm]{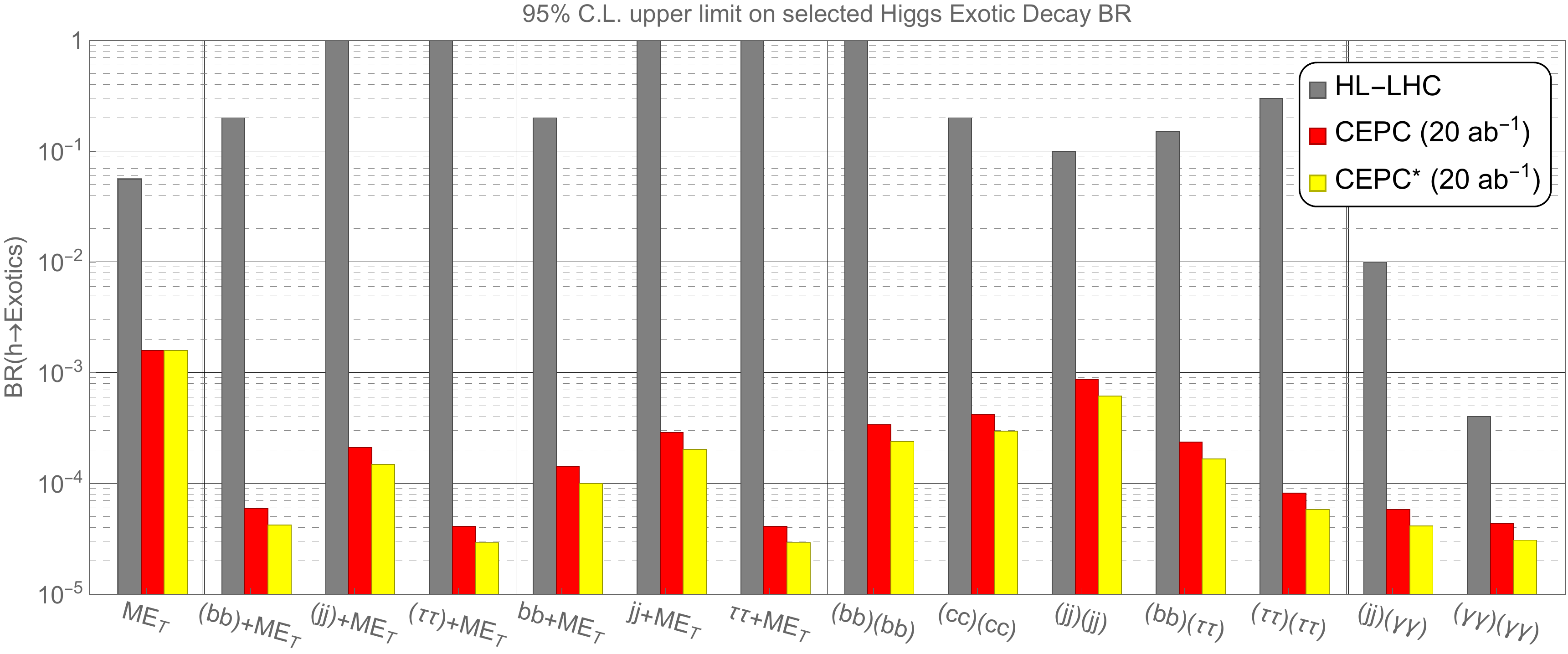}
\caption{\label{fig:ExoticHiggssummary} The 95\% C.L. upper limit on selected Higgs exotic decay
branching fractions at HL-LHC and CEPC, based on
Ref~\cite{Liu:2016zki}. The CEPC curves are derived using results from
Ref~\cite{Liu:2016zki} with leptonic decaying Z boson in the
$e^-e^+\rightarrow ZH$ process. The CEPC$^*$ scenario further utilizes
the hadronically decaying Z boson and includes an estimated
(indicative) improvement of 40\%. Each set of three bars describes a
different topology of exotic Higgs decay. For a recent review on current LHC constraints, see Ref.~\cite{Cepeda:2021rql}.}
\end{figure}

For CEPC $240$~GeV, the dominant Higgs production is $Z$-Higgs associated production process. The $Z$ boson with visible decays enables Higgs tagging using the ``recoil mass'' technique.
A selection cut around the peak of the recoil mass removes most of the  SM background. A large number of analyses are described in \cite{Liu:2016zki}, and we summarize the results in Fig.~\ref{fig:ExoticHiggssummary}, providing the
projection 95\% C.L. limits for the CEPC with 20~ab$^{-1}$ integrated luminosity. We also include the projected LHC sensitivities in gray bars. We use the up-to-date projected sensitivities for the LHC constraints, but
many do not exist or are very conservative. More recent studies, e.g., Ref.~\cite{Shelton:2021xwo} on $h\to 4\tau$, and Ref.~\cite{Kato:2022} on $h\to 4b$, show consistent projection on sensitivities as well. 

The LHC will provide strong constraints on many channels with muons, electrons, and photons.
For the more challenging channels shown here, which rely on signals from jets, heavy
quarks, and taus, the improvements over the LHC expectations
vary from one to four orders of magnitude.
This great advantage benefits from the low QCD background and Higgs tagging from the recoil mass technique at future lepton colliders.
For the Higgs exotic decays without missing energy, the improvement varies between two to
three orders of magnitude, except for the one order of magnitude improvement for the
$(\gamma\gamma)(\gamma\gamma)$ channel.   Here the possibility at the LHC of reconstruction of the Higgs mass from the final state particles provides additional signal-background discrimination power. Channels with electrons, muons, and photons, which 
are relatively clean objects at the LHC, can take advantage of the
higher statistics available from the HL-LHC.

Many new and interesting channels deserve further study.
Higgs exotic decays of $H\to XX \to 4f$ where the intermediate
resonant particle $X$ mass is below 10~GeV is one of these 
channels. This scenario is particularly motivated by the recent discussion
of the connection between  Higgs exotic decay and strongly first-order
electroweak phase
transitions~\cite{Carena:2019une,Kozaczuk:2019pet}. In this
region, the particle $X$ can be long-lived, so the study should be
extended into long-lived particle regime~\cite{Alipour-fard:2018mre}.

Another example is the Higgs decay into a dark shower, that is, a
shower of dark-sector particles.\footnote{These can be bosons or be fermions, for example, composite neutrinos~\cite{Chacko:2020zze}.}  These   can either
decay promptly or be long-lived and their decay back to visible SM
particles 
can be either hadronic or leptonic. The process is motivated by
generic considerations of hidden sector strong dynamics. It also
appears in the discussion of neutral
naturalness~\cite{Craig:2015pha}.   Current
studies have been focusing on the Higgs decays into a pair of twin
glueballs~\cite{Curtin:2015fna,Liu:2018wte,Alipour-fard:2018mre,Liu:2020vur,Carena:2022yvx,Wang:2022dkz},
but this is only a subclass of the generic Higgs decays into these final states. This dark shower channel is also motivated by
the class of models with a large number of light scalars~\cite{Jung:2021tym}, e.g.,
NNaturalness~\cite{Arkani-Hamed:2016rle}, EW scale as a
trigger~\cite{Arkani-Hamed:2020yna}, and delayed or non-restored electroweak 
symmetry~\cite{Meade:2018saz,Baldes:2018nel,Glioti:2018roy,Matsedonskyi:2020mlz}.

\subsection{Supersymmetry}
Supersymmetry (SUSY) provides an intriguing candidate to solve the gauge hierarchy problem in the Standard Model (SM). In Supersymmetric SMs (SSMs) have many appealing features, including gauge coupling unification, dynamical electroweak symmetry breaking. In addition, the Lightest Supersymmetric Particle (LSP) such as neutralino can serve as a viable dark matter (DM) candidate with R-parity conservation. The SUSY searches at the LHC have already set strong constraints on the SSMs~\cite{Aad:2020nyj, Aad:2020sgw,Aad:2019vvi, Aad:2019qnd}. The CEPC will run at much lower energies. At the same time, it can be complementary in covering parameter spaces which are difficult for the LHC to reach. This is particularly important for the search for some of uncolored new physics particles. In this section, as a demonstration, we present recent studies of a couple such examples. In addition, the precision measurements at the CEPC can also probe SUSY even without direct production of the new particles. To give an example of such an approach, we will show a recent study of a global using GAMBIT.

\subsubsection{Light electroweakino and slepton searches} 
In this section, we will present recent studies of the reach of the CEPC on several scenarios with light electroweakino and sleptons. These scenarios can have various physics motivations (for some examples, see ~\cite{Leggett:2014mza, Leggett:2014hha, Du:2015una, Li:2015dil}). 

\begin{figure}[h!]
  \centering
    {\includegraphics[width=.49\textwidth]{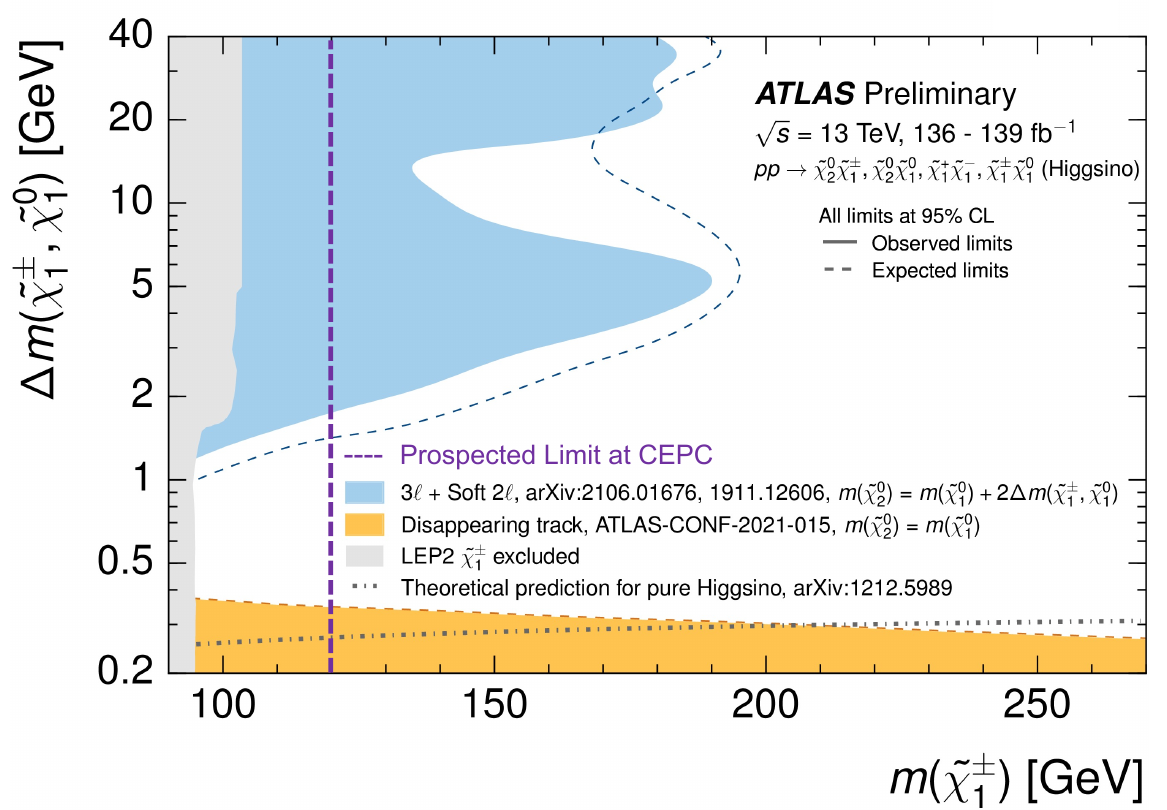}}
    {\includegraphics[width=.46\textwidth]{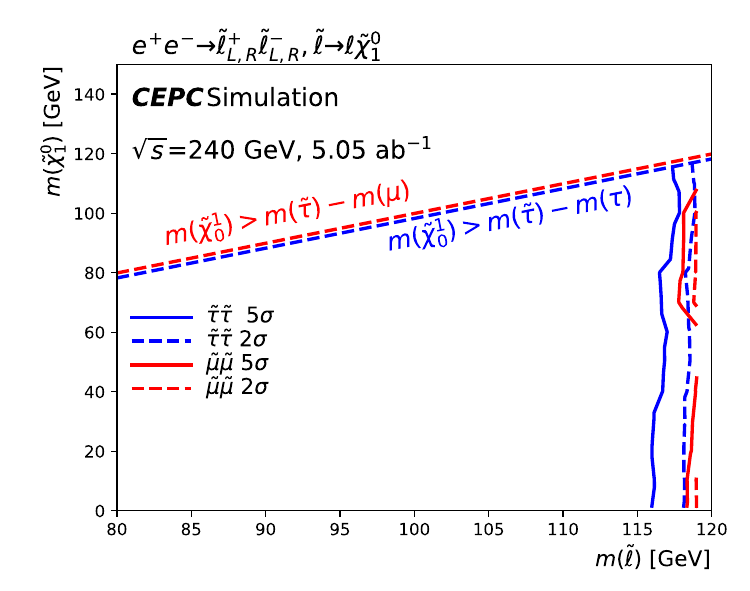}}
    \caption{Left plot shows the observed and expected exclusion limits on simplified SUSY models for chargino-pair production with Higgsino-like LSP obtained by ATLAS. The observed limits obtained by LEP are shown with light grey. The prospected limits at CEPC are also shown in the dotted purple line for rough comparison. Right plot shows the 5$\sigma$ discovery contour (solid line) and 2$\sigma$ exclusion contour (dashed line) for the direct stau production and direct smuon production with 5\% flat systematic uncertainty. }
    \label{fig:ewksl}
  \end{figure}

The light Higgsino particles, well-motivated by naturalness conditions, tend to have small mass splitting among the chargino and neutralino. Therefore, they are quite challenge to be probed in the LHC experiments due to the very soft decay products. The sensitivity studies for chargino pair production by considering scenarios for both a Bino-like and a Higgsino-like neutralino as the LSP have been performed and published at ~\cite{Yuan:2022ewk}. With the cleaner collision environment and better low energy particle reconstruction, CEPC has shown the capability of probing the very compressed region. 


\begin{figure*}[h!]
\centering
\includegraphics[width=1.0\textwidth]{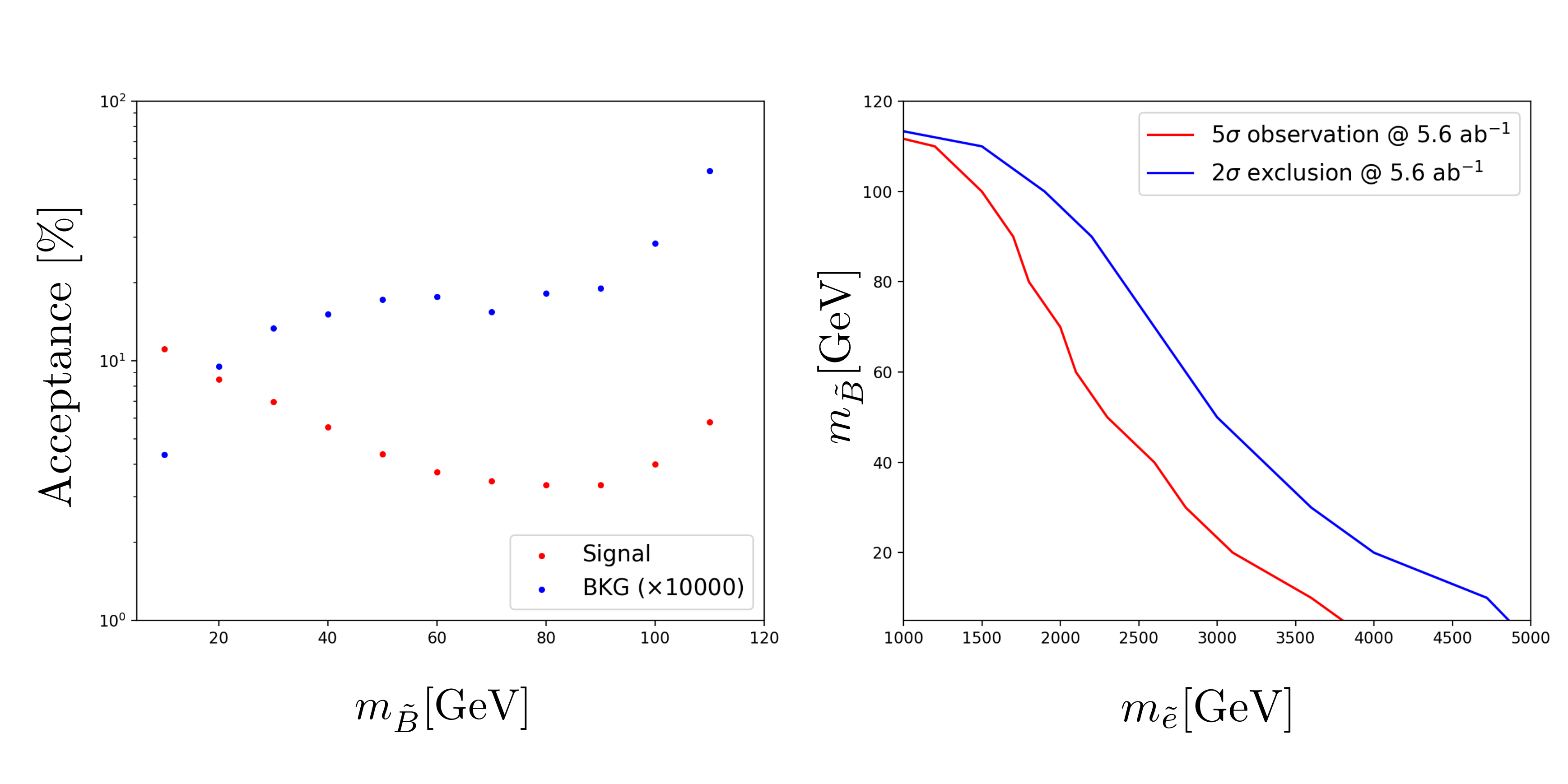}
\vspace*{-1.0cm}
\caption{
\textit{Left}: acceptance of signal and background processes as functions of $m_{\tilde{B}}$. 
Here the acceptance of background process has been multiplied by 10,000. 
\textit{Right}: 2$\sigma$ exclusion and 5$\sigma$ observation limits on the $m_{\tilde{e}} - m_{\tilde{B}}$ plane 
at a future lepton collider running with an integral luminosity 5.6 ab$^{-1}$ and center-of-mass energy 240 GeV. 
Regions below the red (blue) curves are observable (excluded).}
\label{limits} 
\end{figure*}

A light Bino (SUSY partner of $U(1)_Y$ gauge boson, to be labeled as $\tilde{B}$) ($\mathcal{O}(10)$ GeV scale) are still not excluded by the current SUSY searches. A search for light bino scenario (motivated by Gauge-Mediated SUSY Breaking~\cite{Pagels:1981ke}) at CEPC has been performed~\cite{Chen:2021omv}, in which bino is the next-to-lightest supersymmetric particle (NLSP), while the lightest supersymmetric particle (LSP) and dark matter candidate is the sub-GeV gravitino (to be labeled as $\tilde{G}$). 
The process of bino pair production via a t-channel selectron (labeled as $\tilde{e}$), where bino subsequently decay to gravitino and a photon, has been considered, namely $e^+ e^- \to \tilde{B}\tilde{B} \to \gamma\gamma \tilde{G}\tilde{G}$.
The corresponding dominant background process is $e^+ e^- \to \gamma\gamma \nu\bar{\nu}$ (via $Z$ boson invisible decay), which has been suppressed by a dedicated cut-flow using a list of kinematic variables with good signal and background seperation power.
The study shows that CEPC is able to exclude selectron lighter than 4.5 TeV (2 TeV) with bino mass around 10 GeV (100 GeV), see Fig.~\ref{limits}, which is much larger than current LHC bound which exclude selectron mass up to several hundreds GeV.

Light smuon and stau particles are interesting to search for in their own right, and they are also favored by the latest muon g-2 excess. At the same time, it is challenging to search them at the LHC, especially in the region where their masses are close to that of the LSP. Such regions are also favored by dark matter relic density requirements. They have also been explored with CEPC detector~\cite{Yuan:2022slp}. Assuming a flat 5\% systematic uncertainty, the discovery sensitivity can reach up to 117 (116) GeV for smuon (stau) mass via direct smuon (stau) production. The above results can fill a significant region in the gap in the LHC search.

Somewhat heavier selectrons, above the kinematic limit for direct production, can also be searched for in the process ~\cite{Ahmed:2022ude}: $e^{+}_{R} e^{-}_{R}\rightarrow {\tilde \chi_{1}^{0}({\rm bino})}+{\tilde \chi_{1}^{0}({\rm bino})}+{\gamma}$. The reach depends on the model assumptions. For example, if the relic abundance requirement is satisfied by LSP annihilating through the Z-pole,  the right-handed selectron will be excluded up to 180 (210) GeV respectively at 3(2)$\sigma$. On the other hand, if the annilation through the Higgs pole dominates, right-handed selectron will be excluded up to 140 (180) GeV at 3(2)$\sigma$.

For both electroweakino and slepton searches, the discovery potential can reached  up to the kinematic limit of the detector $\sqrt{s}/2$, and cover interesting parameter regions. It is shown that CEPC can have its own role to play in the search for SUSY. The examples shown here  have a minor dependence on the reconstruction model and detector geometry. Moreover, these results can be considered as a reference and benchmark for similar searches at other proposed electron-positron colliders, such as the Future Circular Collider ee (FCC-ee) or the International Linear Collider (ILC), given the similar nature of the facilities, detectors, center-of-mass energies, and target luminosities.

\begin{figure}[h!]
 \centering
 \includegraphics[width=0.49\textwidth]{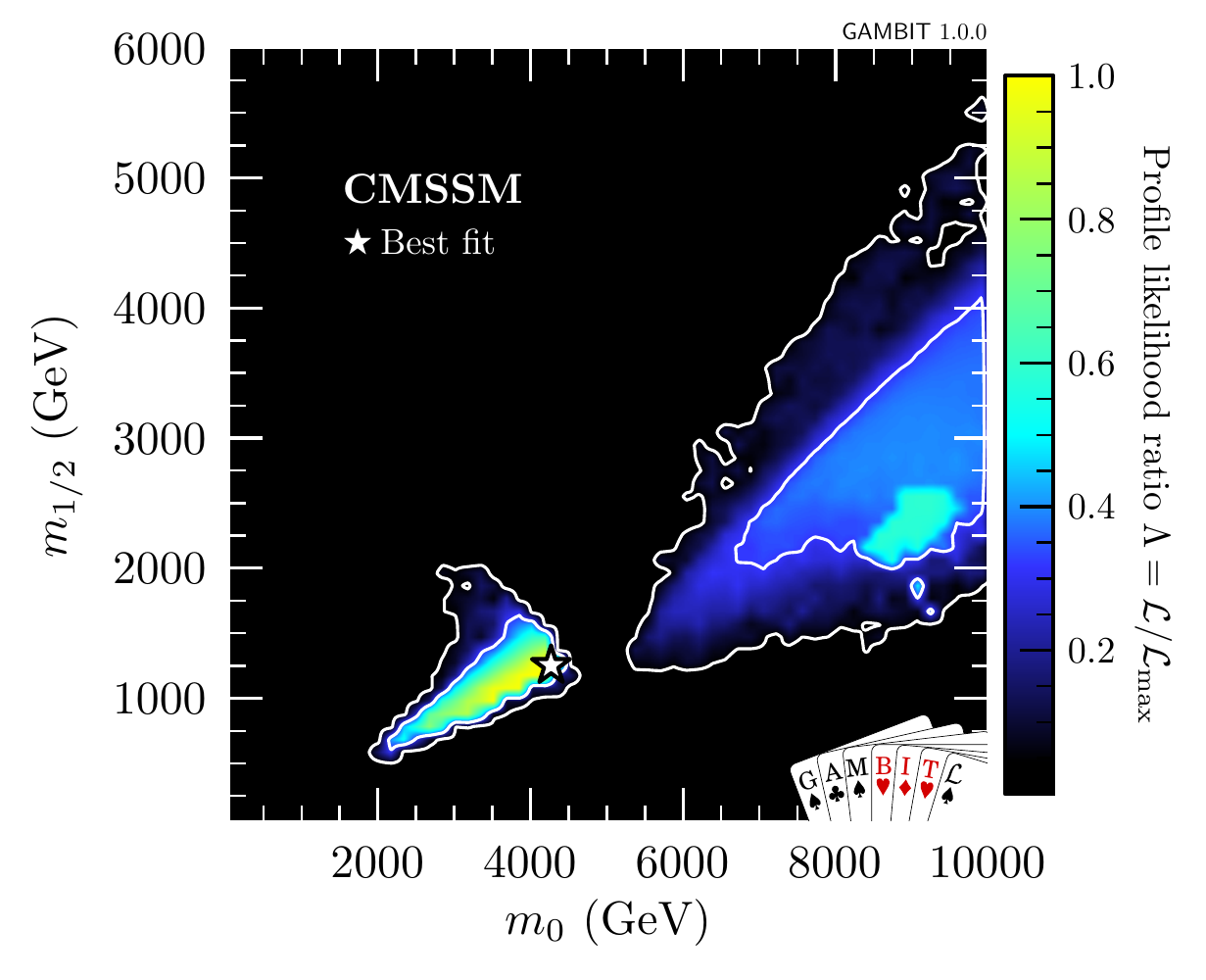}
 \includegraphics[width=0.49\textwidth]{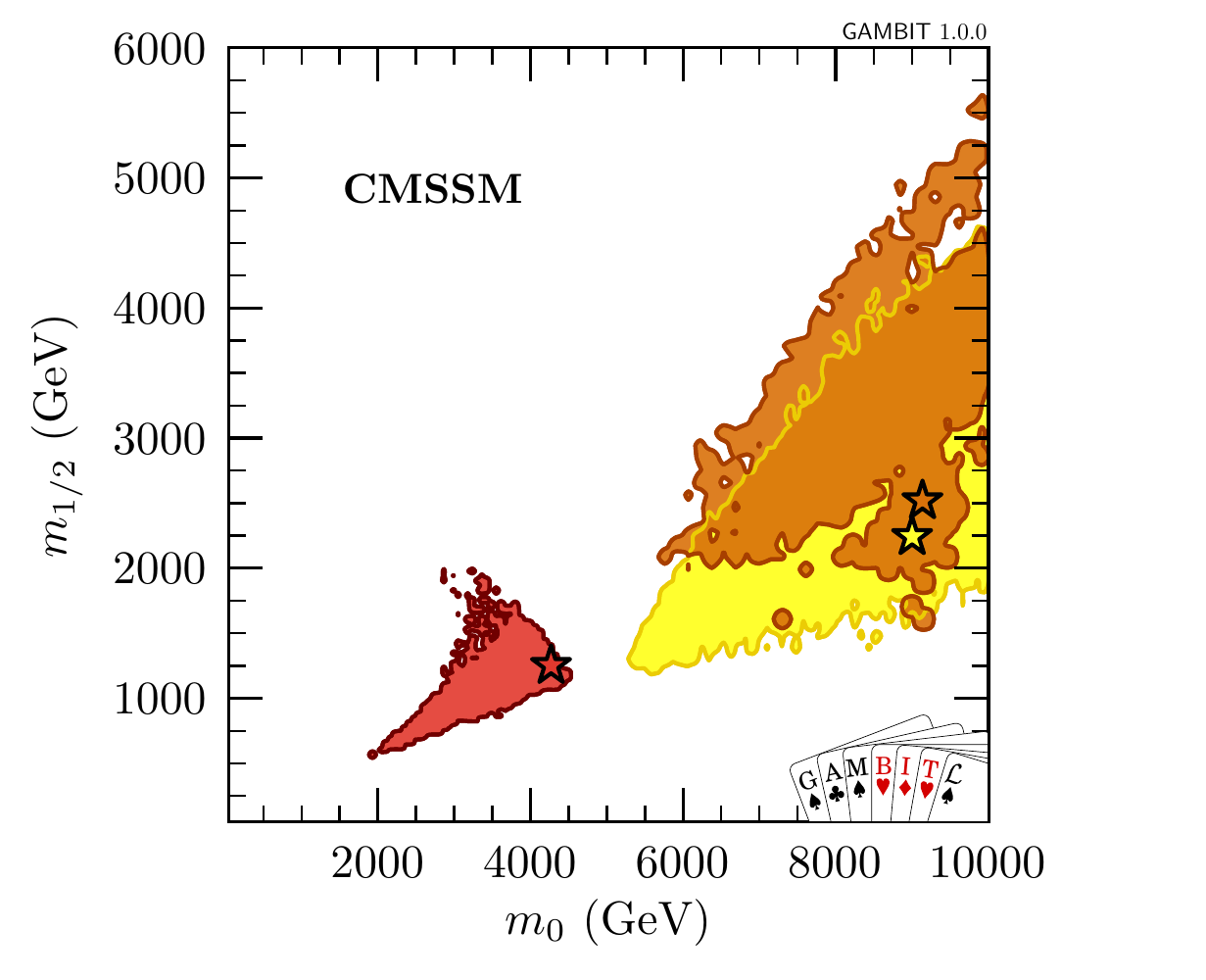}
 \includegraphics[width=0.49\textwidth]{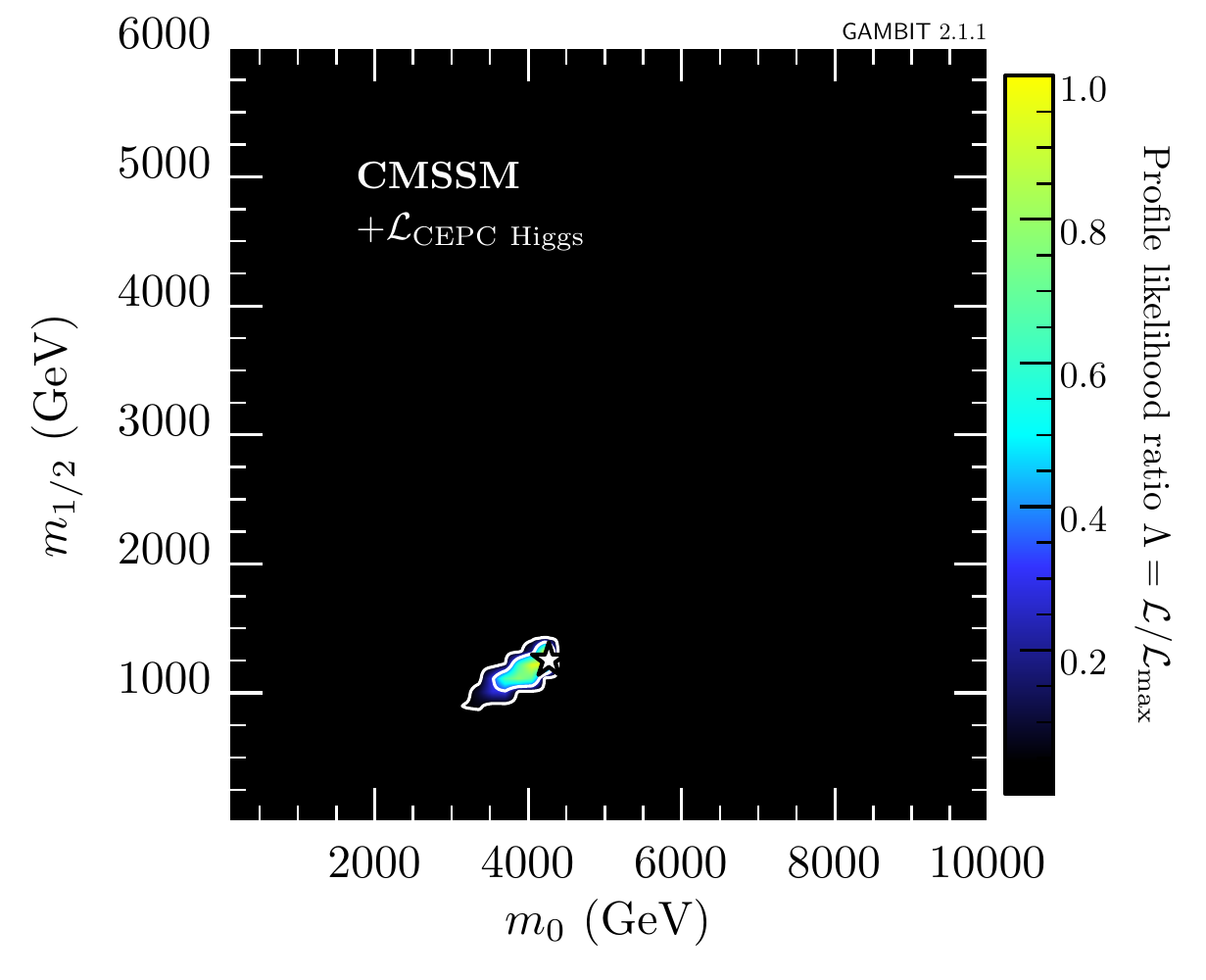}
 \includegraphics[width=0.49\textwidth]{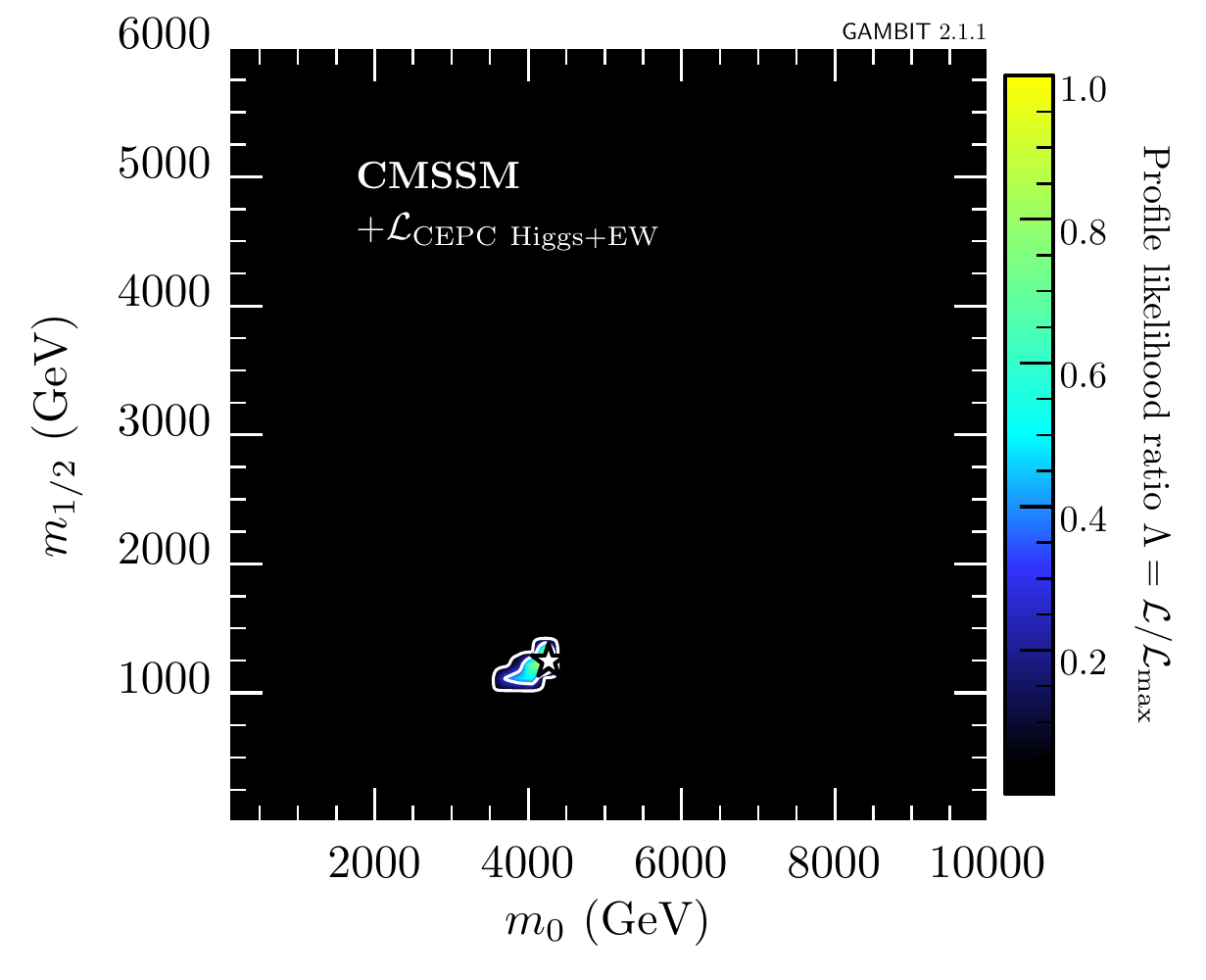}\\
 \caption{The profile likelihood ratio in the CMSSM for the present  
(top left panel) and for the future with additional CEPC measurements (bottom panels), with 68\% and 95\% CL contours drawn in white, and the best-fit point indicated by a star. The top right panel
shows mechanisms for ensuring that the dark matter relic density does not exceed the measured value in 2$\sigma$ contours of present likelihoods. 
 We use values of the best-fit point in CMSSM to set the central values of measurements at CEPC, and the theoretical uncertainties are $k=1/5$ 
times smaller than the current SM Higgs theoretical uncertainties. }
 \label{fig:gambit}
\end{figure}
\subsubsection{SUSY global fits} 
In this section, we present a study of the impact of the Higgs and electroweak precision measurements at the CEPC with GAMBIT global fits of several constrained versions of the MSSM, namely the Constrained Minimal Supersymmetric Standard Model (CMSSM), the Non-Universal Higgs Mass generalisations (NUHM1 and NUHM2) and the seven-dimensional weak scale phenomenological MSSM (MSSM7)~\cite{Athron:2022uzz}. Besides the likelihoods of Higgs and electroweak measurements at the CEPC, the global fits include several direct and indirect dark matter searches, a large collection of electroweak precision and flavor observables, direct searches for SUSY at the LEP, and Runs I and II of the LHC. We showed that the CEPC can further test the currently allowed parameter space of these models, advance our understanding of the mass spectrum, and be complementary to the dark matter searches, as illustrated in Figure~\ref{fig:gambit}. 


\subsection{Dark Matter and Dark Sector}
There are plenty of evidence for the existence of  Dark Matter (DM) from astrophysical and cosmological observations, and detecting DM at the colliders is very important as a complementary of direct and indirect detection experiments. And we have many ways to probe DM at colliders. The case of the SUSY LSP (Bino, Higgsino, Wino, Gravitino, Axino etc.) in the SSMs have  already been discussed in the previous sub-section. More broadly, thermally produced light DM particles can also couple to the SM through various portals, such as the lepton, Higgs, or dark quark portals. Many studies have been carried out in this scenario. In this section, we present several new results which highlight the capability of the CEPC. 

\subsubsection{Lepton portal Dark Matter} 
\begin{figure}
\centering	
\includegraphics[width=0.32 \columnwidth]{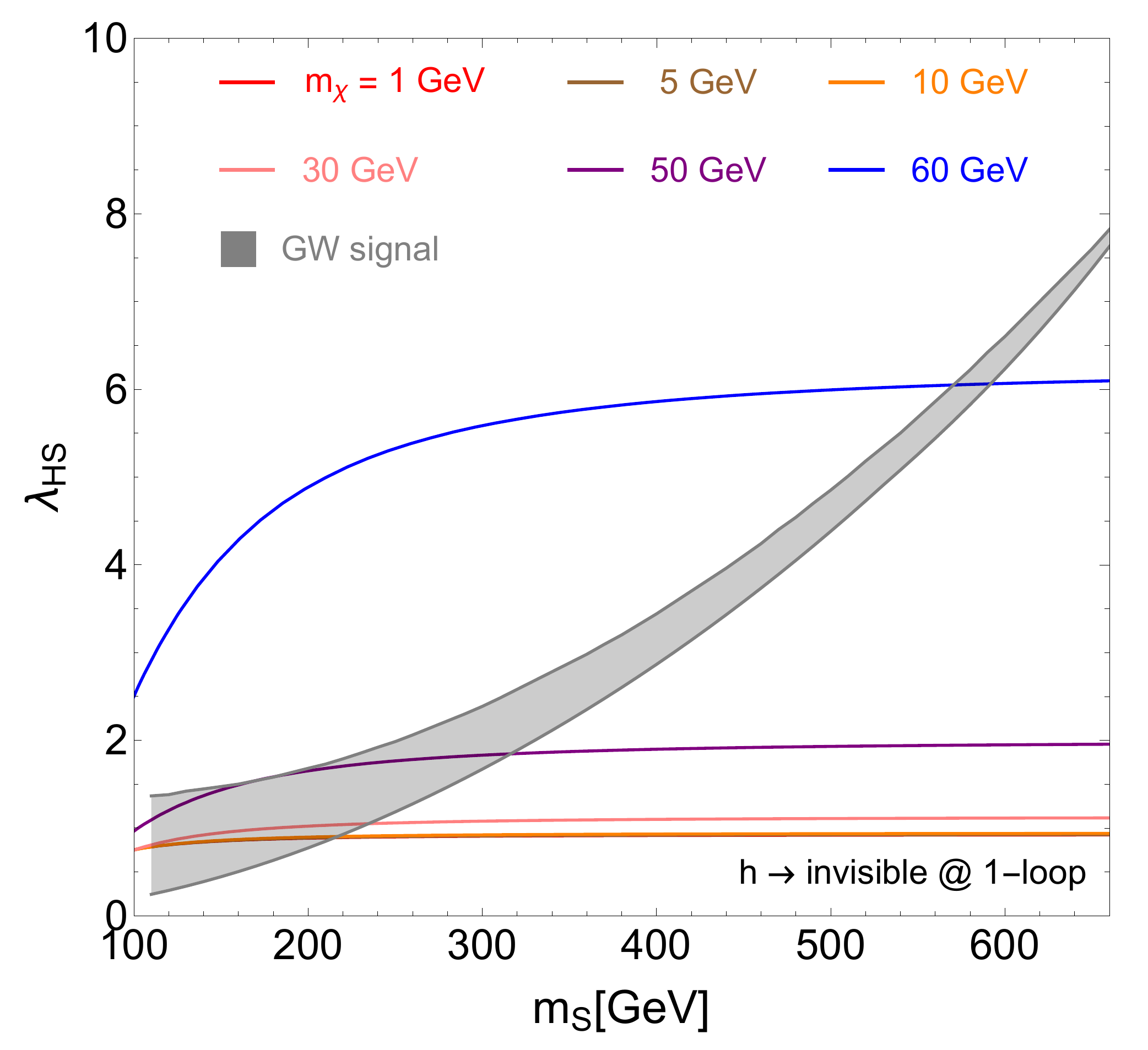}
\includegraphics[width=0.32 \columnwidth]{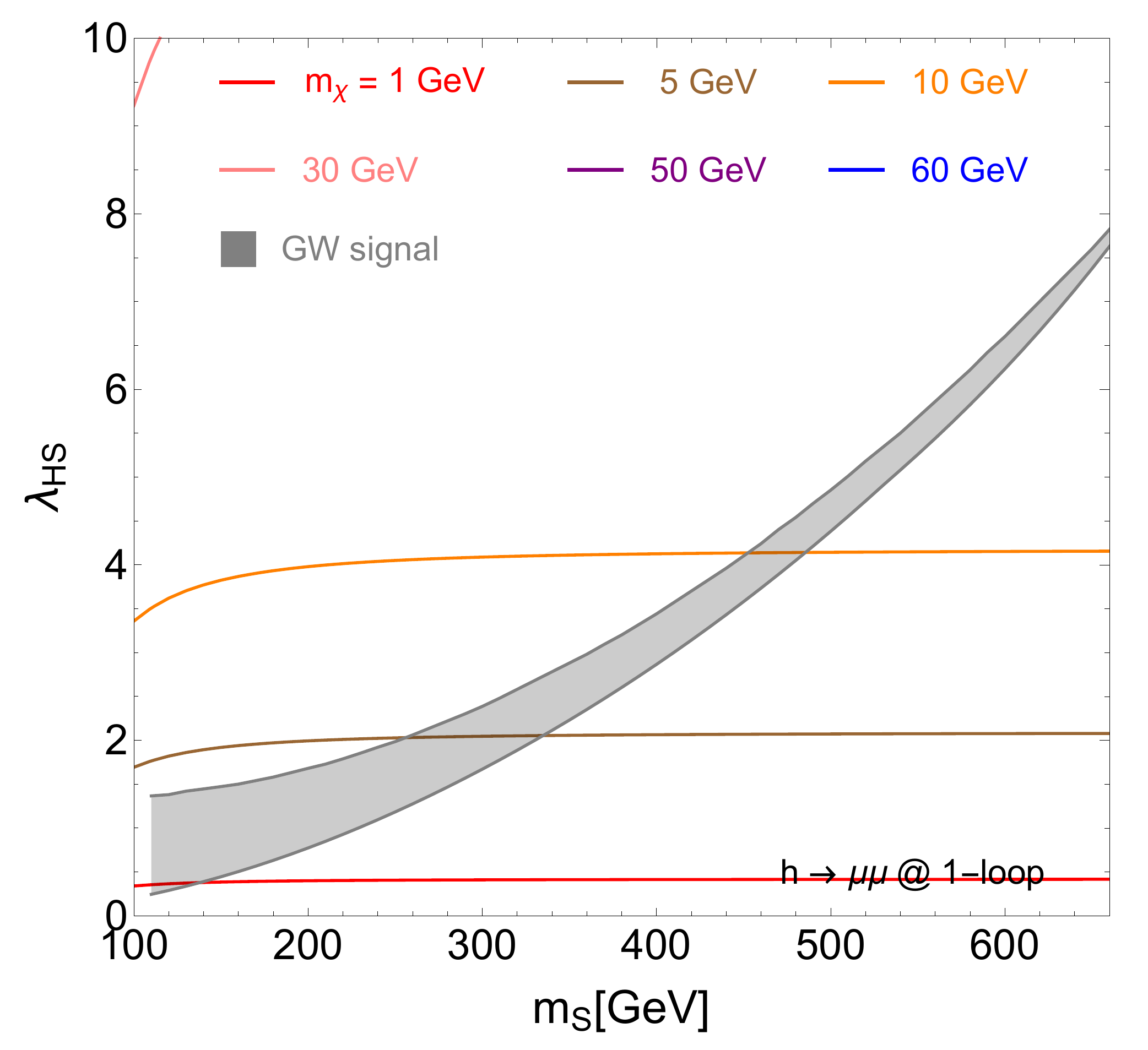}
\includegraphics[width=0.32 \columnwidth]{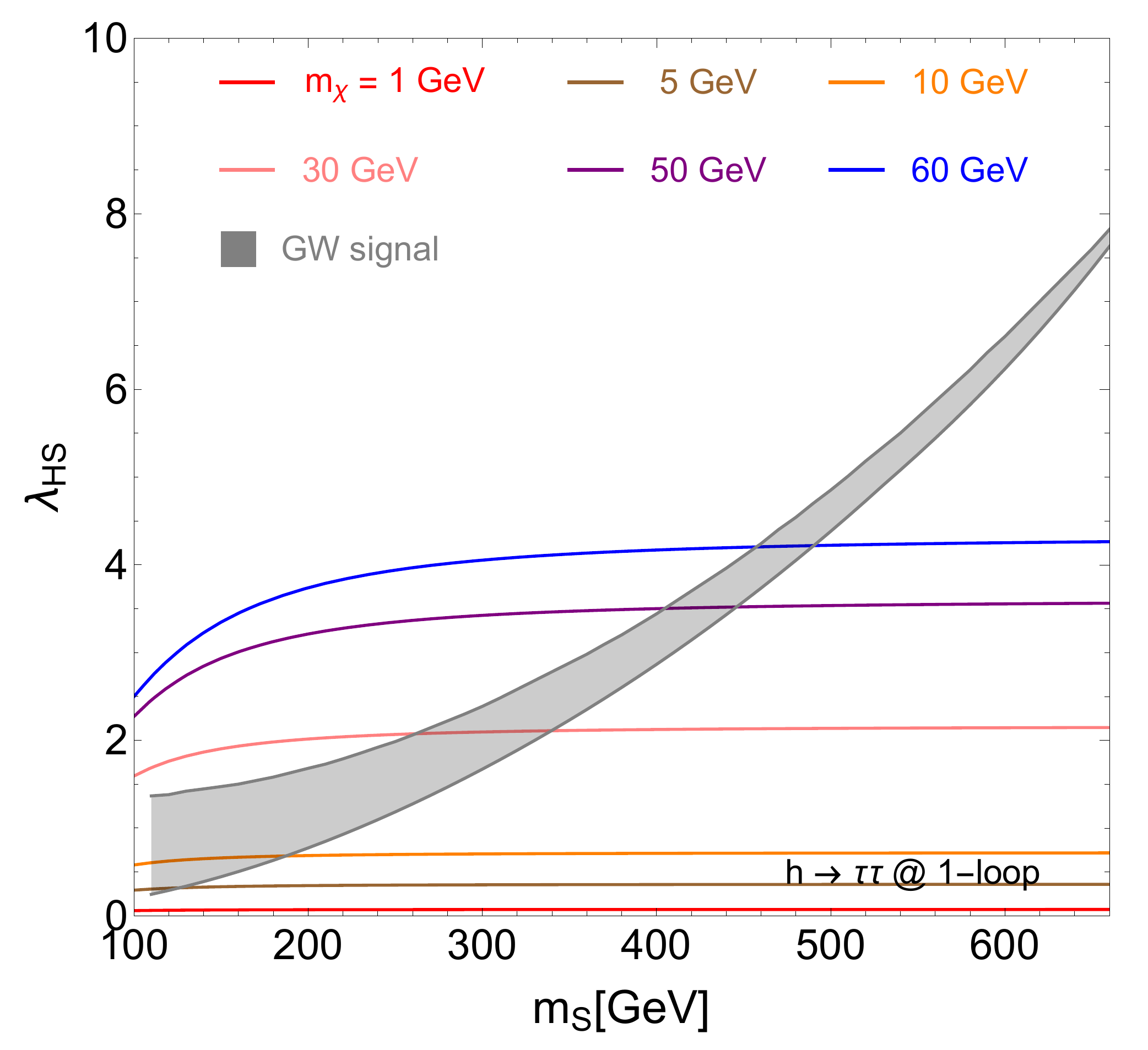}
\caption{Figure from Ref.~\cite{Liu:2021mhn}, the interplay between GW detection and future $e^+e^-$ collider searches. The gray shaded region is the LISA detectable parameter space. From left to right, the sensitivities for $\lambda_{HS}$ are shown from future CEPC precision measurements, in which the region above a given $m_\chi$ (corresponding to a colored line) can be probed.}
\label{fig:h-1loop-GW}
\end{figure}

In many WIMP models, the signals for direct and indirect detection can be  suppressed. In this case, the collider searches are crucial. One such example is the lepton portal model~\cite{Bai:2014osa}, in which a Majorana DM candidate, denoted as $\chi$, couples to the SM right-handed leptons $\ell_R$ via a complex charged scalar mediator $S$ via $y_\ell \bar\chi_L\ell_RS^\dagger$. Reference~\cite{Liu:2021mhn} studied the collider phenomenology of this model and the interplay with the gravitational wave (GW) astronomy. The masses of DM and mediator, as well as the lepton coupling $y_\ell$ and the Higgs portal coupling $|S|^2|H|^2$ can be probed at the CEPC via the pair production of mediators $e^+e^-\to S^{\pm(*)} S^\mp\to\ell^+\chi\ell'^-\chi$, exotic decays of the Higgs or $Z$ boson, $h/Z\to S^{\pm (*)} S^{\mp (*)}\to\ell^+\chi\ell'^-\chi$ and $h\to \chi\chi$, and the Higgs couplings, including $h\ell^+\ell^-$, $h\gamma\gamma$ and $hZZ$. In addition to the collider signals, the model might trigger a first-order phase transition in the early Universe, provided that the mediator mass parameter $\mu_S^2$ is negative, and the portal coupling $\lambda_{HS}$ is large enough. In this case, the phase transition GWs can also be a probe of the model. Figure~\ref{fig:h-1loop-GW} shows the LISA projections and the CEPC Higgs precision measurement sensitivities for comparison, where the overlap of the parameter space reachable by the two probes can be used for crosschecking any potential excess and obtaining more information. For the rest of the parameter space, the two approaches are complementary. The idea of Ref.~\cite{Liu:2021mhn} can be generalized to other WIMP models that are difficult to be probed in the direct and indirect detections, especially the models with scalar DM and/or mediators in which a first-order phase transition may happen.

\subsubsection{Asymmetric Dark Matter} 
In addition to DM, the observed baryon asymmetry of the Universe (BAU) is also a main puzzle in cosmology and particle physics.
Current measurements show that the abundance of baryon and DM are roughly at the same order of magnitude ($\Omega_{\text{DM}} \simeq 5\Omega_\text{B}$)~\cite{Akrami:2018vks,Ade:2015xua}. 
This coincidence provides the motivation to consider the so-called ``asymmetric DM" (ADM) model~\cite{Kaplan:2009ag,Petraki:2013wwa,Zurek:2013wia,Bai:2013xga}. 

A new ADM model has been proposed and studied in~\cite{Zhang:2021orr}. In this model, the dark sector is charged under a dark QCD,  $SU(3)'$, 
and the mass of DM is generated via the dark confinement (so our DM is actually a ``dark baryon'').
Furthermore, to generate dark and visible asymmetry simultaneously, we introduce a scalar mediator (labeled as $\Phi$) that is charged under $SU(3)'$ and standard model (SM) $U(1)_Y$. 
Mediator $\Phi$ couples to dark quark (labeled as $q'$) and SM right-hand leptons, and thus provide a portal for us to search for. 

\begin{figure}[h!]
\centering
{
\includegraphics[width=3.0in]{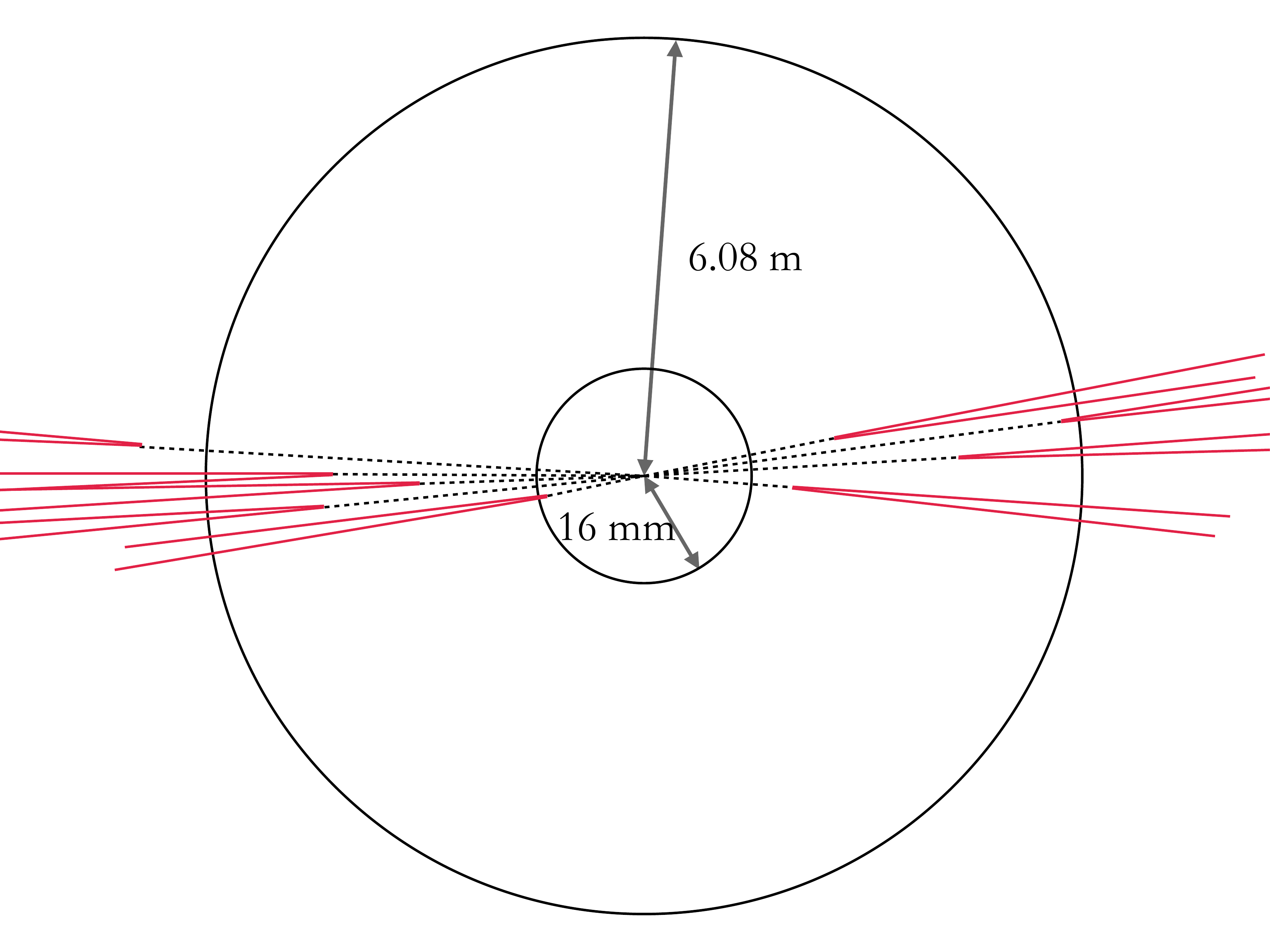}
}
\quad
{
\includegraphics[width=3.0in]{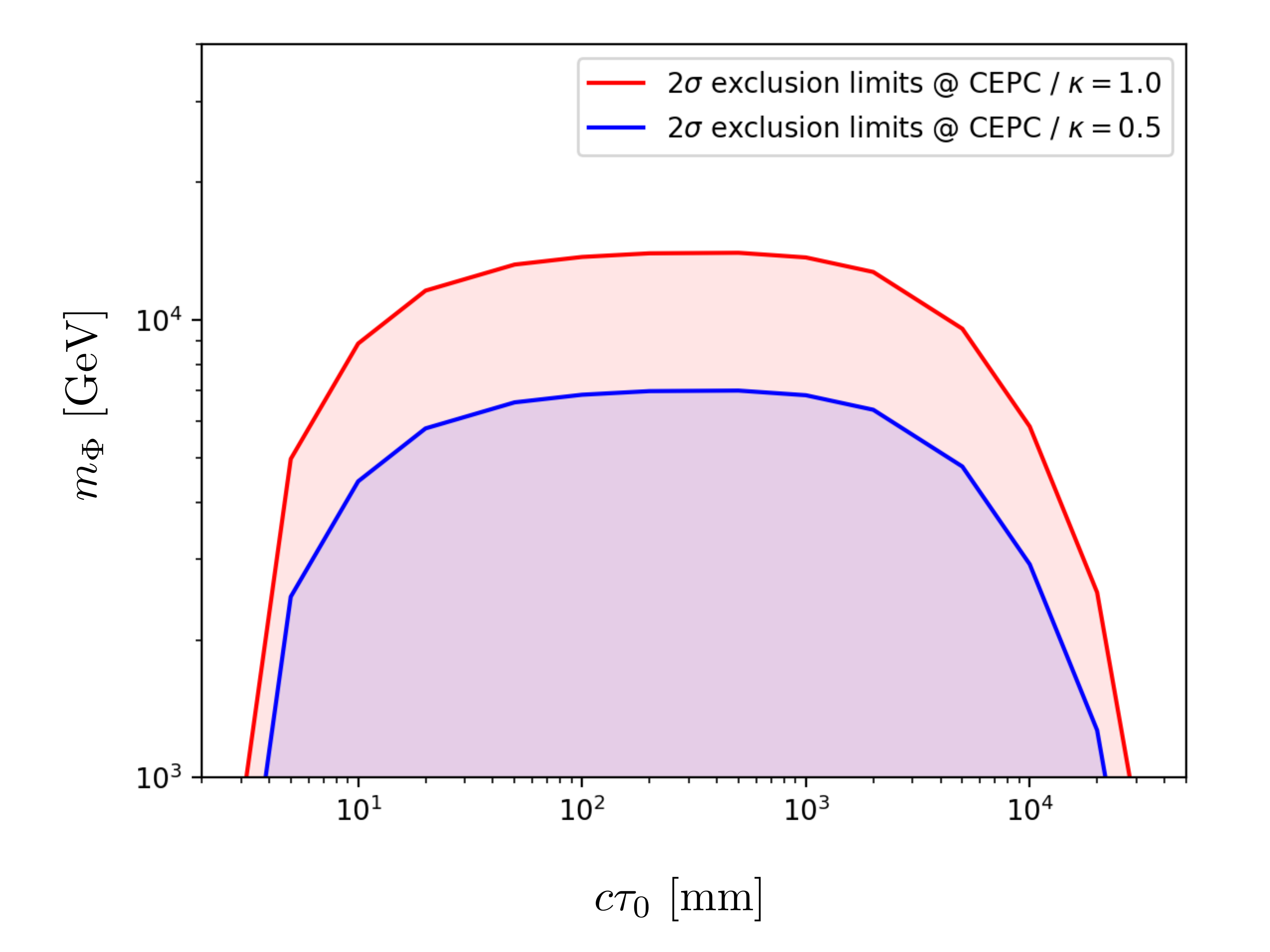}
}
\caption{
Left: An illustration of the signal process at CEPC. Detector is represented by two circles. Black dotted lines and red solid lines are dark pions and muons, respectively.
Right: 2 $\sigma$ exclusion limits on the mediator mass $m_{\Phi}$ as a function of the dark pion proper decay length, with coupling $\kappa$ fixed to 0.5 and 1.0 respectively.}
\label{CEPC2}
\end{figure}
The Lagrangian related to collider search is
\begin{eqnarray}
\mathcal{L} \supset \bar{q'} ( D \!\!\!\!/  - m_{q'} ) q'  + (D_{\mu} \Phi )^{\dagger}(D^{\mu} \Phi ) - m^2_{\Phi} \Phi^{\dagger} \Phi -\frac{1}{4} {G'}^{\mu\nu} {G'}_{\mu\nu} - ( \kappa \Phi \bar{q'}_L  {l}_R  +h.c.)~,~
\label{Lag_collider}
\end{eqnarray}
where $G'^{\mu\nu}$ is the field strength of dark gluon. 
Mediator $\Phi$ can be produced in pairs at LHC via the Hyper charge it carries.  
Then $\Phi$ decays to a SM lepton and a dark quark $q'$. 
While on CEPC, $q'\bar{q}'$ can be produced directly via a t-channel $\Phi$. 
Due to the dark confinement, $q'$ will hadronize to a cluster of dark mesons (labeled as $\pi'$). 
Dark meson $\pi'$ (long-lived) will decay to lepton pair via the $\Phi$ portal, and leave displaced vertex inside detector. 
Fig.~\ref{CEPC2} (left) shows the predicted signal process on CEPC for illustration. 
The study shows that CEPC has the ability to cover a large parameter space of this model, see Fig.~\ref{CEPC2} (right). 
The mass of mediator can be excluded up to $\mathcal{O}(10)$ TeV, if the proper lifetime of dark pion $\pi'$ is between 10 mm and 10 m. 
This bound is stronger than the limit from current ATLAS displaced lepton jet search result~\cite{ATLAS:2016jza}.

\subsubsection{Dark sector from exotic Z decay}

Operating in the Z-factory mode, the CEPC can produce on the order of $10^{12}$ of Zs (Tera-Z). At the same time, CEPC has less QCD background and fixed center of mass energy, these features help us to reconstruct missing energy of final states and easier to distinguish DM signal from the SM background. Hence, comparing to the LHC, the CEPC \cite{CEPCStudyGroup:2018rmc, CEPCStudyGroup:2018ghi} has some unique advantages in the search for the dark sector. 

\begin{figure}[htb]
	\centering
	\includegraphics[width=0.98\textwidth]{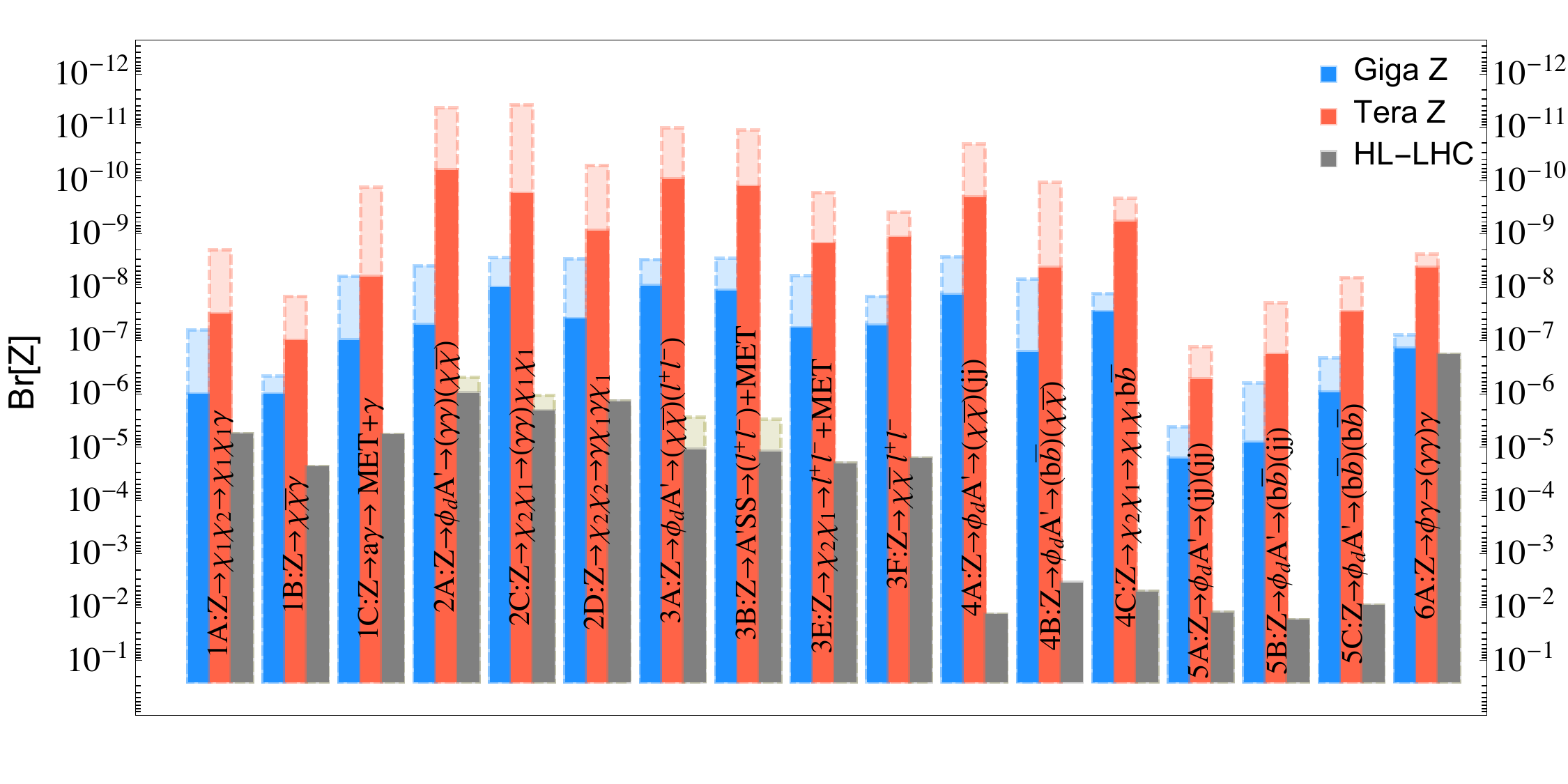}
	\caption{The reach for the branching ratio of various exotic Z decay modes at the future Z-factories (Giga Z and Tera Z) and the HL-LHC at 13 TeV with $\mathcal{L} = 3  ~\rm{ab^{-1}}$. The sensitivities in general generally also depend on model parameter, such as the masses of the  mediator and dark matter (see Fig.~\ref{fig:summary1} for some examples).  The dark colored	regions with solid boundary represent the reach for worst case in the parameter space the, while the lighter regions with dashed boundary indicates reach in the the best case. }
	\label{fig:summary}
\end{figure} 

In \cite{Liu:2017zdh}, we have studied a broad range of dark sector models and model independent exotic Z decay channels at future $e^+ e^-$ colliders with the Giga Z and Tera Z options. Four general categories of dark sector models have been included: Higgs portal dark matter, vector portal dark matter, inelastic dark matter and axion like particles.
Focusing on channels motivated by the dark sector models, a model independent study of 
the sensitivities of Z-factories are also carried out. The results are  compared with the reach of high luminosity LHC (HL-LHC). 
The final states of the exotic decays are categorized according to the number of resonances, and possible topologies. 
The projected reach for those channels is shown in Fig.~\ref{fig:summary}. In comparison with the HL-LHC, the future Z-factories can be more sensitive to many interesting decay modes.

\begin{figure*}[tb!]
\includegraphics[width=0.48\textwidth]{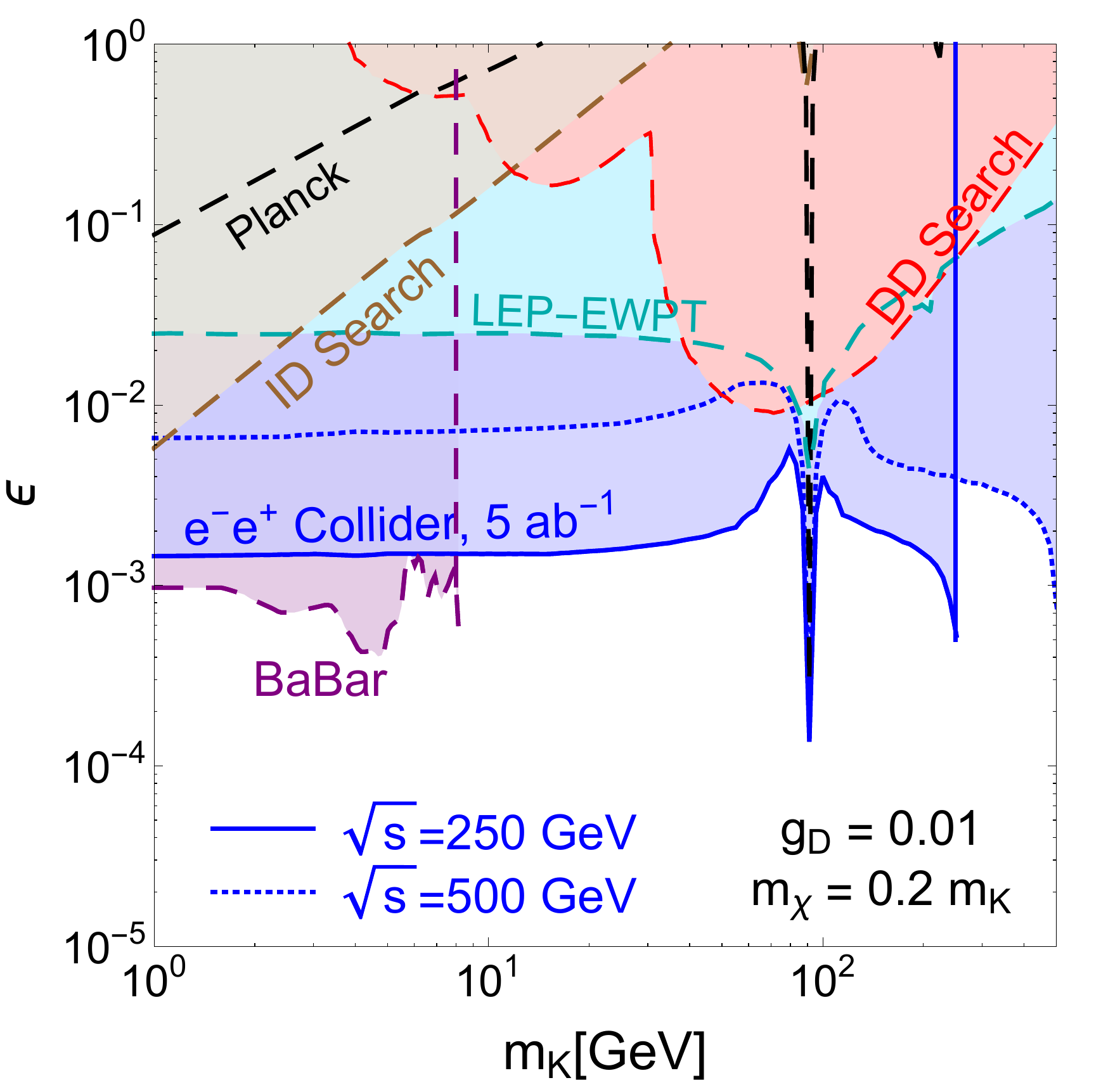}
\includegraphics[width=0.48\textwidth]{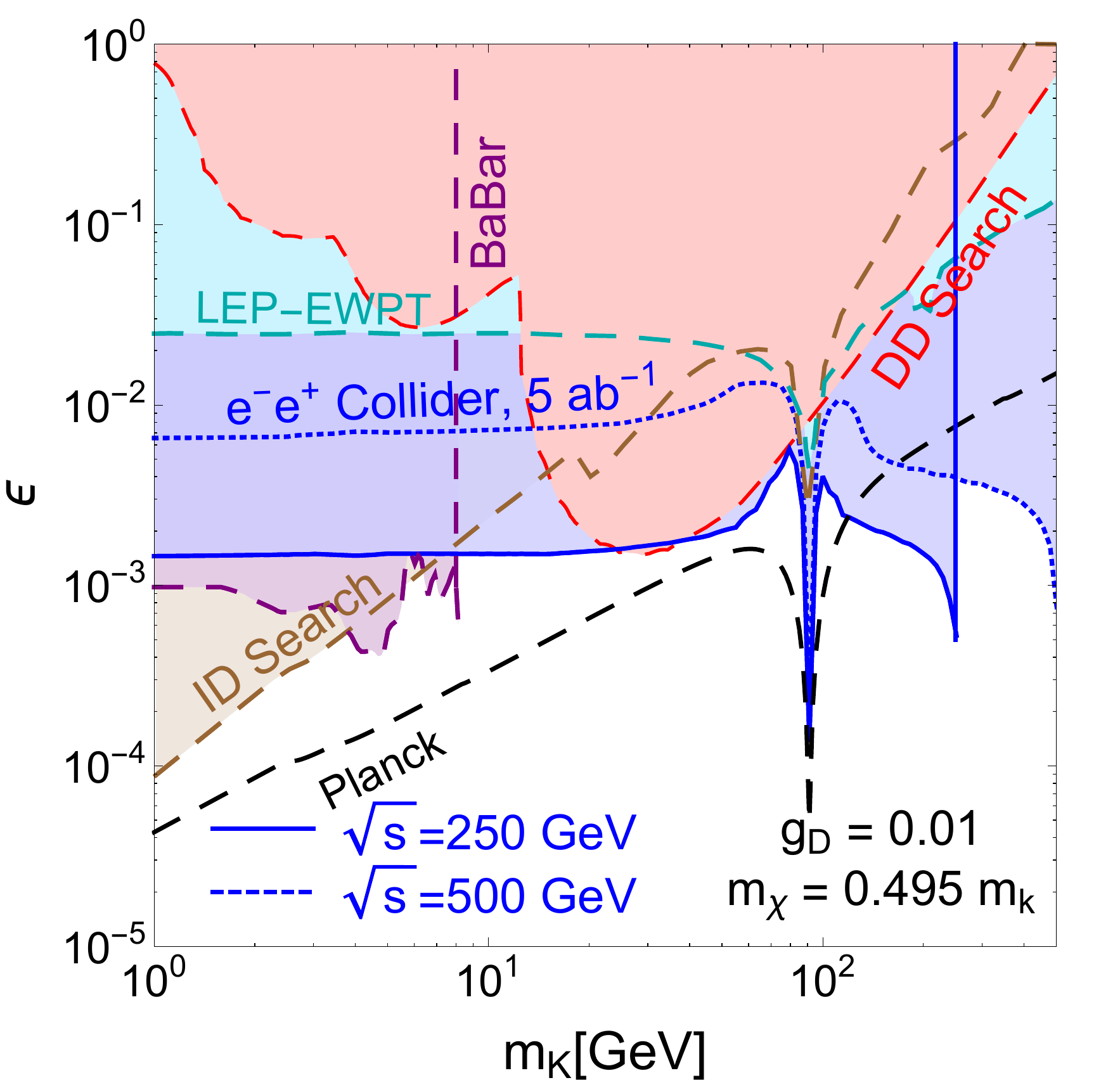}
\caption{The reach of direct detection (red), indirect
  detection (brown), and searches at $e^-e^+$ collider (blue) in
  the $\epsilon$ vs.~$m_K$ plane.  We fix $g_D = 0.01$, $m_\chi=0.2
  m_K$ (left panel) and $m_\chi=0.495 m_K$ (right panel).  We also
  show the contours when $\chi$ satisfies the relic density
  measurement by the Planck collaboration~\cite{Planck:2015fie} as black  dashed lines.  
  The existing constraints from LEP electroweak precision measurements
  (LEP-EWPT) and the BaBar search for the $\tilde{K}$ invisible decay
  (BaBar) are also included.}
\label{fig:summary1}
\end{figure*}

In addition to the Z factory mode, new hidden states can be associated produced with $Z$ and $H$ bosons at the CEPC. In ~\cite{Liu:2017lpo}, we  studied the reach of an ultraviolet (UV) model: Double Dark Portal model from this production mode at CEPC with $\mathcal{L}=5{\rm \ ab}^{-1}$. This model contains both vector and scalar portals. The collider searches do not depend sensitively on the dark matter mass, as long as the dark photon can decay to dark matter. On the other hand,  the constraints from dark matter detection and indirect detection experiments, as well as the relic abundance requirement,  are sensitive to the dark matter mass. We also emphasize that the collider constraint is not sensitive to the coupling between the DM and the mediator, as long as the invisible decay of dark photon dominates. Therefore, the reach from a future $e^+e^-$ collider will complement and supersede that of the dark matter searches.

\subsection{Long-lived Particle Searches}
Many new physics theories predict Long-Lived Particles (LLPs)~\cite{Alekhin_2016, Lee:2018pag,Alimena:2019zri,Lee_2019, Alimena_2020, Alimena:2021mdu}, which have detectable macroscopic decay lengths.
The long lifetime can be due to feeble couplings with the SM particles,  phase space suppression, or heavy mediators.

The LLPs, after being produced at colliders, travel a macroscopic distance before decaying into other SM and/or new particles, which gives interesting signal in detectors.
If LLPs are neutral and they can decay into visible objects, they will lead to the typical ``Displaced Vertex" (DV) signature, which means that the decay vertex of LLP is considerably displaced from its production vertex. On the other hand, they lead to the so called ``disappearing track'' signature for charged LLPs.

The displaced distance depends on the lifetimes and velocities of LLPs.
When their lifetimes  match the size of a usual detector (such as ATLAS, CMS, or a general purpose detector at the CEPC) at an interaction point (IP), which we call ``\textit{near detector}" or abbreviate as ``ND" in this section, the LLPs have a significant probability of decaying inside the detector.
When LLPs have much longer lifetime, since LLPs usually have feeble couplings to the detector material, neutral LLPs are more likely to decay outside the ND, contribute to the missing energy.
A detector far away from the IP, which we call ``\textit{far detector}" or abbreviate as ``FD" in this section, can also be beneficial since it would not have a lot the background coming from the hard collision.

\subsubsection{Results with Near Detectors}

\vspace{0.2 cm}
\textbf{1.1 Light Neutralinos from $Z$-boson Decays}
\vspace{0.2 cm}

\begin{figure}[h]
\centering
\includegraphics[width=0.45\columnwidth]{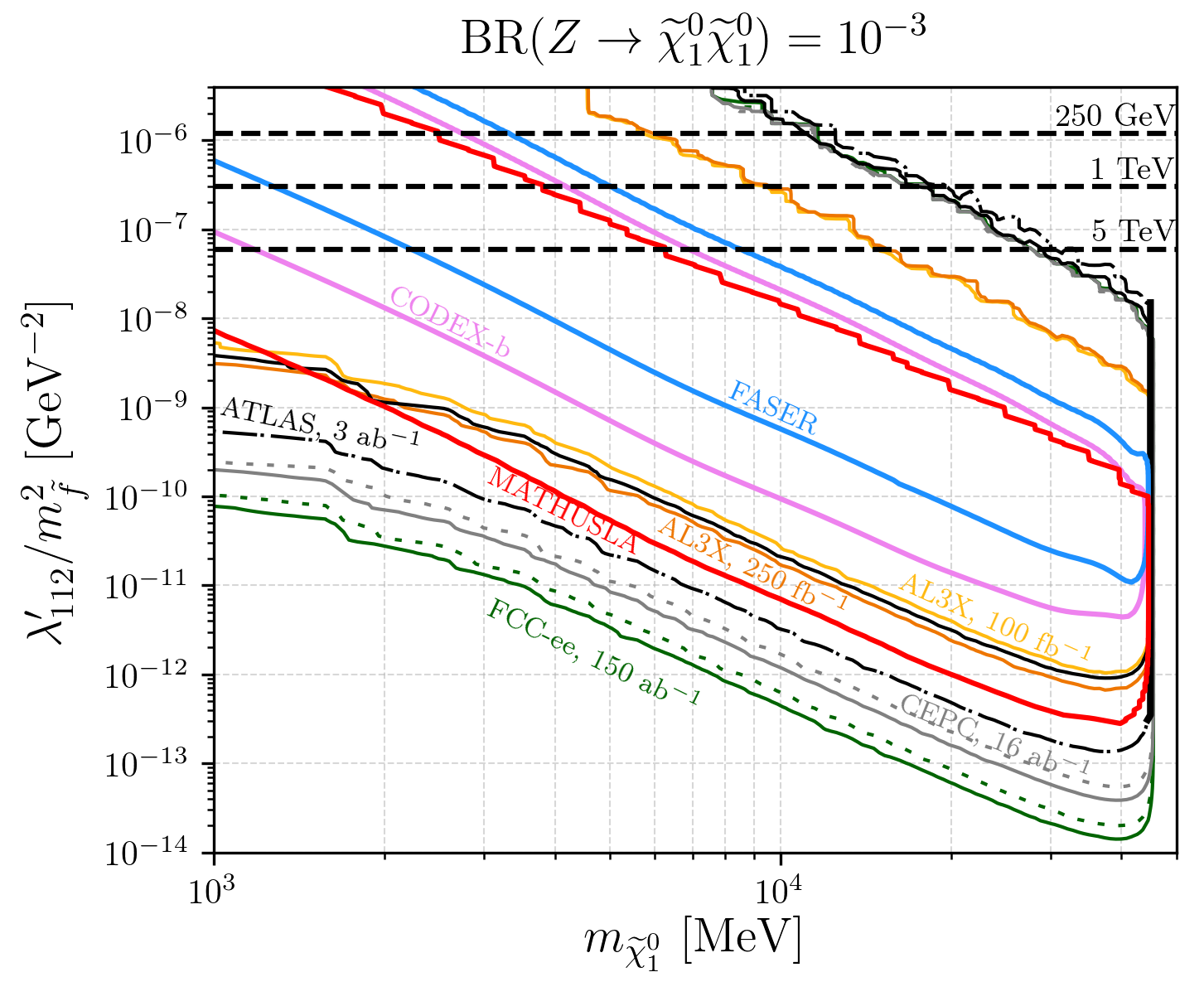}
\caption{
The discovery limits of long-lived neutralinos for the near detector of CEPC/FCC-ee~\cite{Wang:2019orr}.
The solid contour curves correspond to three decay events in the fiducial volume when considering all decay modes of $\tilde{\chi}_1^0$, while the dashed lines include only visible/charged decay modes ($K^{(*)\pm} e^\mp$, $e^-us$ or $e^+\bar{u}\bar{s}$).
The estimates for experiments at the LHC: AL3X, CODEX-b, FASER and MATHUSLA, are reproduced from Refs.~\cite{Helo:2018qej,Dercks:2018wum}.
The ATLAS results correspond to HL-LHC with $\sqrt{s}=14$ TeV and 3 $ab^{-1}$ integrated luminosity.
The black horizontal dashed lines correspond to the current RPV bounds on the single coupling $\lambda^\prime_{112}$ \cite{Kao:2009fg} for three different degenerate sfermion masses $m_{\tilde{f}}=250$ GeV, 1 TeV, and 5 TeV as labelled.
}
\label{fig:zneuneuplots}
\end{figure}

Ref.~\cite{Wang:2019orr} investigates the potential of detecting the LLPs for the near detector of CEPC/FCC-ee.
Long-lived lightest neutralinos pairs $(\tilde{\chi}_1^0\tilde{\chi}_1^0)$, in the context of the R-parity violating supersymmetry (RPV-SUSY), are produced from $Z-$decays.
The analysis indicates that when assuming BR$(Z\rightarrow \tilde{\chi}_1^0\tilde{\chi}_1^0) = 10^{-3}$ and $m_{\tilde{\chi}_1^0} \sim 40$ GeV, the model parameter $\lambda'_{112} / m^2_{\tilde{f}}$ can be discovered down to as low as $\sim 1.5 \times 10^{-14}$ ($3.9 \times 10^{-14}$) GeV$^{-2}$ at the CEPC and FCC-ee  operating in the Z-pole mode with 16 (150) $ab^{-1}$ integrated luminosity.
Fig.~\ref{fig:zneuneuplots} shows sensitivity estimate of the CEPC (grey) and the FCC-ee (green) presented in the 2D plane of $\lambda'_{112}/m_{\tilde f}^2$ vs. $m_{\tilde{\chi}_1^0}$ assuming BR($Z\rightarrow \tilde{\chi}_1^0 \tilde{\chi}_1^0$)$=10^{-3}$.
Limits are compared with other experiments, and results exceed the projected sensitivity reaches of the ATLAS experiment at the HL-LHC and the proposed LHC experiments with far detectors (AL3X, CODEX-b, FASER, and MATHUSLA).
More results can be found in Ref.~\cite{Wang:2019orr}.

\subsubsection{Results with FADEPC}

Inspired by the proposed new experiments MATHUSLA~\cite{Chou:2016lxi,Curtin:2018mvb}, CODEX-b~\cite{Gligorov:2017nwh}, FASER~\cite{Feng:2017uoz}, AL3X~\cite{Gligorov:2017nwh} and ANUBIS~\cite{Bauer:2019vqk}, Ref.~\cite{Wang:2019xvx} has proposed to install Far Detectors at the Electron Positron Collider (FADEPC), which are new detectors at a position far from the interaction point (IP) at generic high energy $e^-e^+$ colliders such as the CEPC, FCC-ee, ILC and CLIC
\footnote{Similar idea was also proposed later in Ref.~\cite{Chrzaszcz:2020emg}}.
Ref.~\cite{Wang:2019xvx} has developed eight different designs of such far detectors (``FD1$-$FD8") by varying the locations, volumes, and geometries.
The design with small geometry size can be placed inside the experiment hall or can be placed in a cavern or shaft near the experiment hall. Other designs have big volume and can be placed on the ground above the IP.

The discovery potential of such FDs for four physics scenarios has been investigated in Refs.~\cite{Wang:2019xvx, Tian:2022rsi}.
It was found that such new experiments with far detectors at future lepton colliders can extend and complement the sensitivity to the LLPs of the experiments at the future lepton colliders with usual near detectors and the present and future experiments at the LHC.

\begin{figure}[h!]
	\centering
	\includegraphics[width=0.45\textwidth]{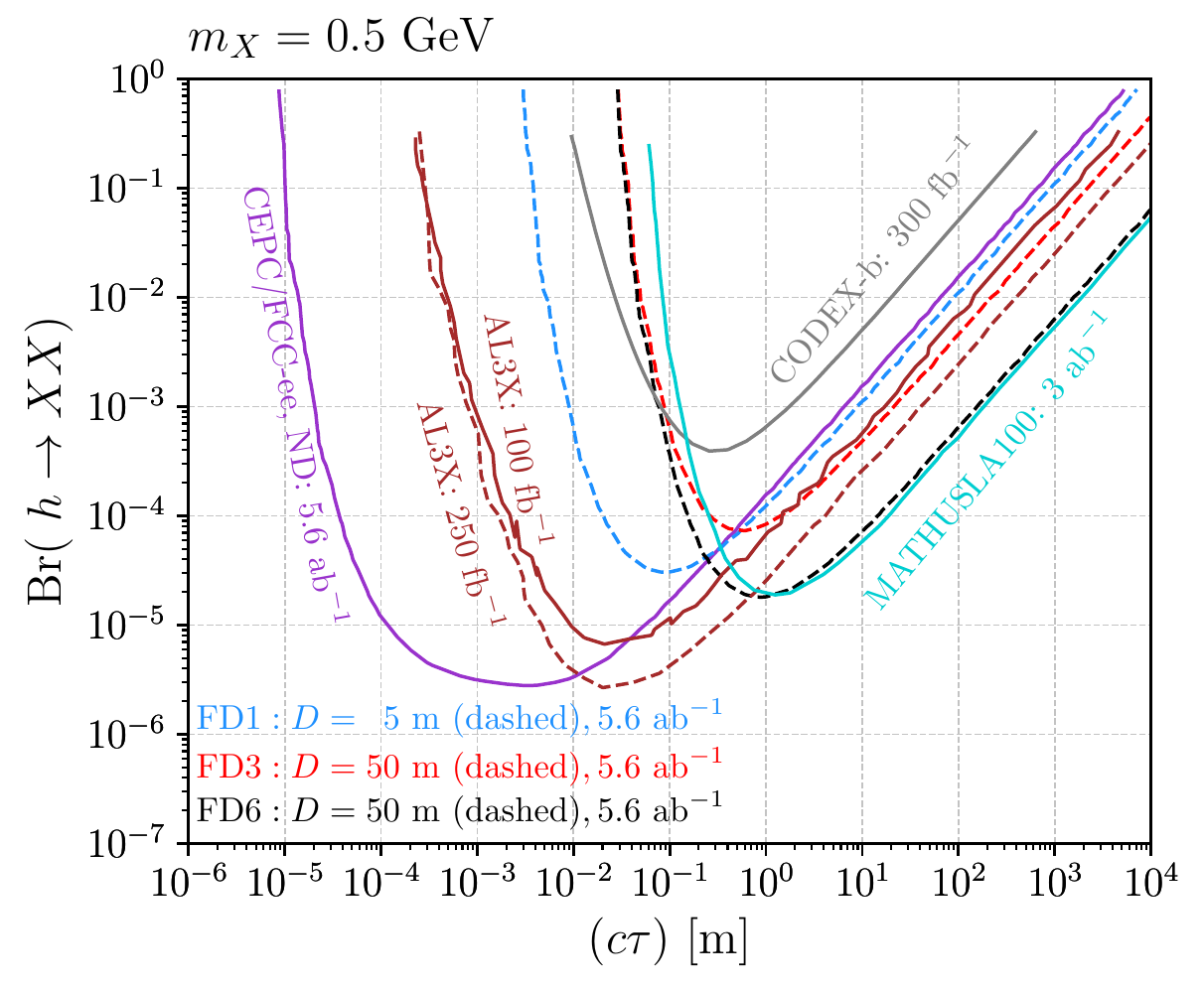}
	\renewcommand{\figurename}{Fig.}
	\caption{The reachs for Br$(h\to XX)$ of the CEPC/FCC-ee's far detectors FD1, FD3, FD6, compared with predictions for the CEPC/FCC-ee's near detector (ND) and for AL3X, CODEX-b and MATHUSLA100~\cite{Wang:2019xvx}.
	 }
	\label{fig:H2XX-0p5}
\end{figure}

\vspace{0.2 cm}
\textbf{2.1 Light  Scalars from Exotic Higgs Decays}
\vspace{0.2 cm}

Ref.~\cite{Wang:2019xvx} considered the Higgs bosons decaying into a pair of light scalars $X$: $h\rightarrow XX$, varying the proper decay length $(c\tau)$ of $X$ and branching ratio of the Higgs boson decaying into a pair of $X$, Br($h\rightarrow XX$), so as to find the sensitive parameter spaces for various far detector designs. 
Fig.~\ref{fig:H2XX-0p5} presents the results in the Br$(h\rightarrow XX)$ vs. $c\tau$ plane for benchmark value of $m_X=0.5$ GeV.
It compares the sensitivity projections of FD1, FD3 and FD6, with those of the CEPC/FCC-ee's near detector, and of other future detectors at the LHC such as CODEX-b \cite{Gligorov:2017nwh}, MATHUSLA \cite{Alpigiani:2018fgd} and AL3X \cite{Gligorov:2018vkc}.
More results can be found in Ref.~\cite{Wang:2019xvx}.

\begin{figure}[h!]
	\centering
	\includegraphics[width=0.45\textwidth]{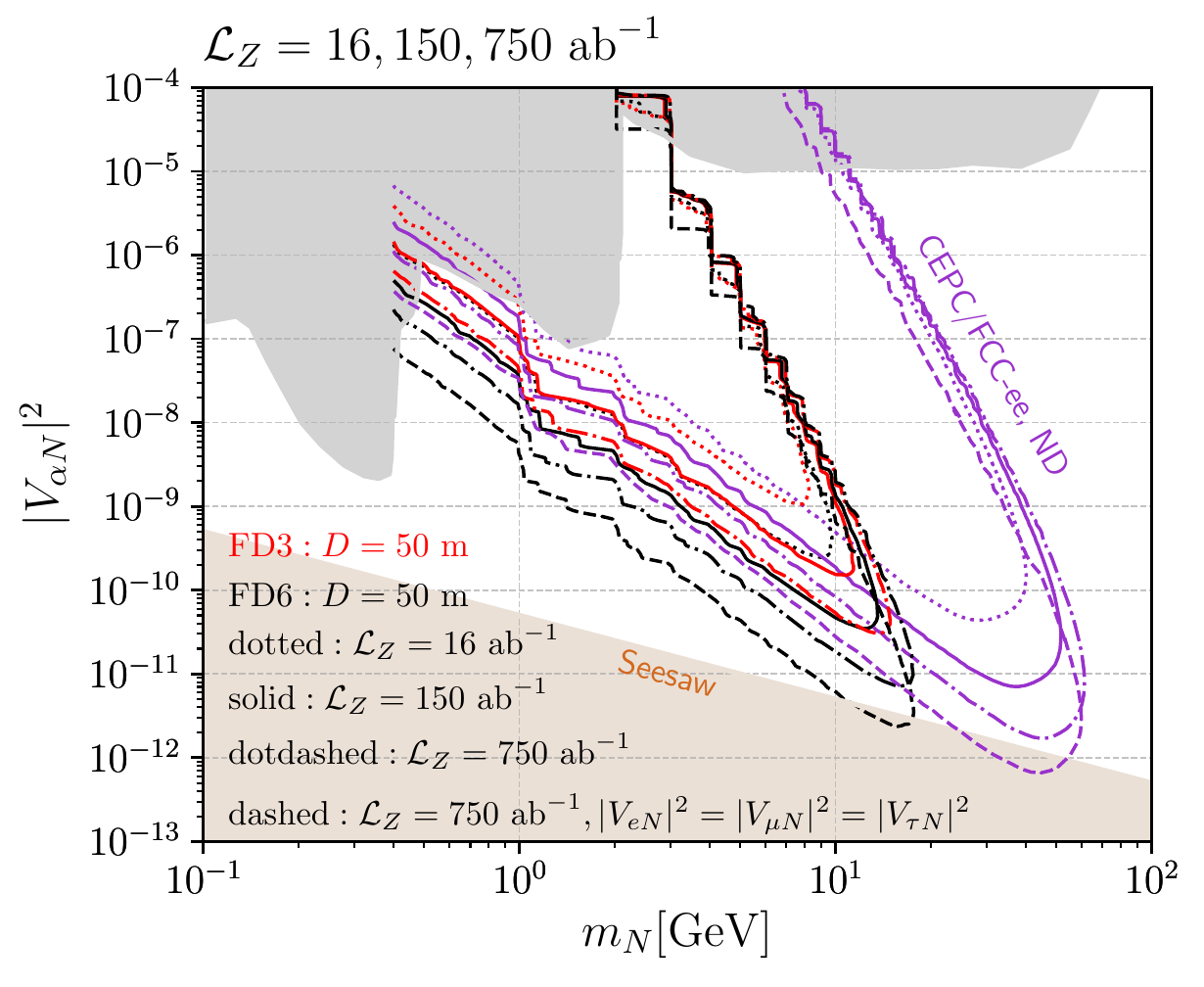}
	\renewcommand{\figurename}{Fig.}
	\caption{
	Reaches for HNLs of both the far and near detectors at the CEPC/FCC-ee with three different integrated luminosities $\mathcal{L}_Z$~\cite{Wang:2019xvx}.
}
\label{fig:HNL}
\end{figure}

\vspace{0.2 cm}
\textbf{2.2 Heavy Neutral Leptons}
\vspace{0.2 cm}

Ref.~\cite{Wang:2019xvx} considered 
the heavy neutral leptos (HNLs) produced from $Z-$decays at an $e^- e^+$ collider running at the $Z-$pole, and made sensitivity predictions for both the far detectors FD1$-$FD8 and near detectors at the CEPC and FCC-ee. 
Fig.~\ref{fig:HNL} compares the performance of the CEPC/FCC-ee's ND, FD3, and FD6 for a variety of integrated luminosities $\mathcal{L}_Z$.
For $\mathcal{L}_Z=750$ ab$^{-1}$, they find that FD6 may reach $\sim 10^{-11}$ for $m_N$ between 10 and 20 GeV.
Furthermore, the previous limits all assume only one single HNL mixes with one single generation of active neutrino generations.
If one HNL has equal mixings with all three active neutrino generations, i.e. $|V_{e N}|^2=|V_{\mu N}|^2=|V_{\tau N}|^2$, with $\mathcal{L}_Z=750$ ab$^{-1}$, the combination of FD6 and the near detector at the CEPC or FCC-ee may probe the Type I seesaw model for $m_N$ between 10 and 60 GeV.
The strategy is thus able to test the Type I seesaw model directly in such case.
Limits are also compared with other experiments and details can be found in Ref.~\cite{Wang:2019xvx}.

\vspace{0.2 cm}
\textbf{2.3 Light Neutralinos from $Z$-boson Decays}
\vspace{0.2 cm}

\begin{figure}[t]
	\centering
	\includegraphics[width=0.45\textwidth]{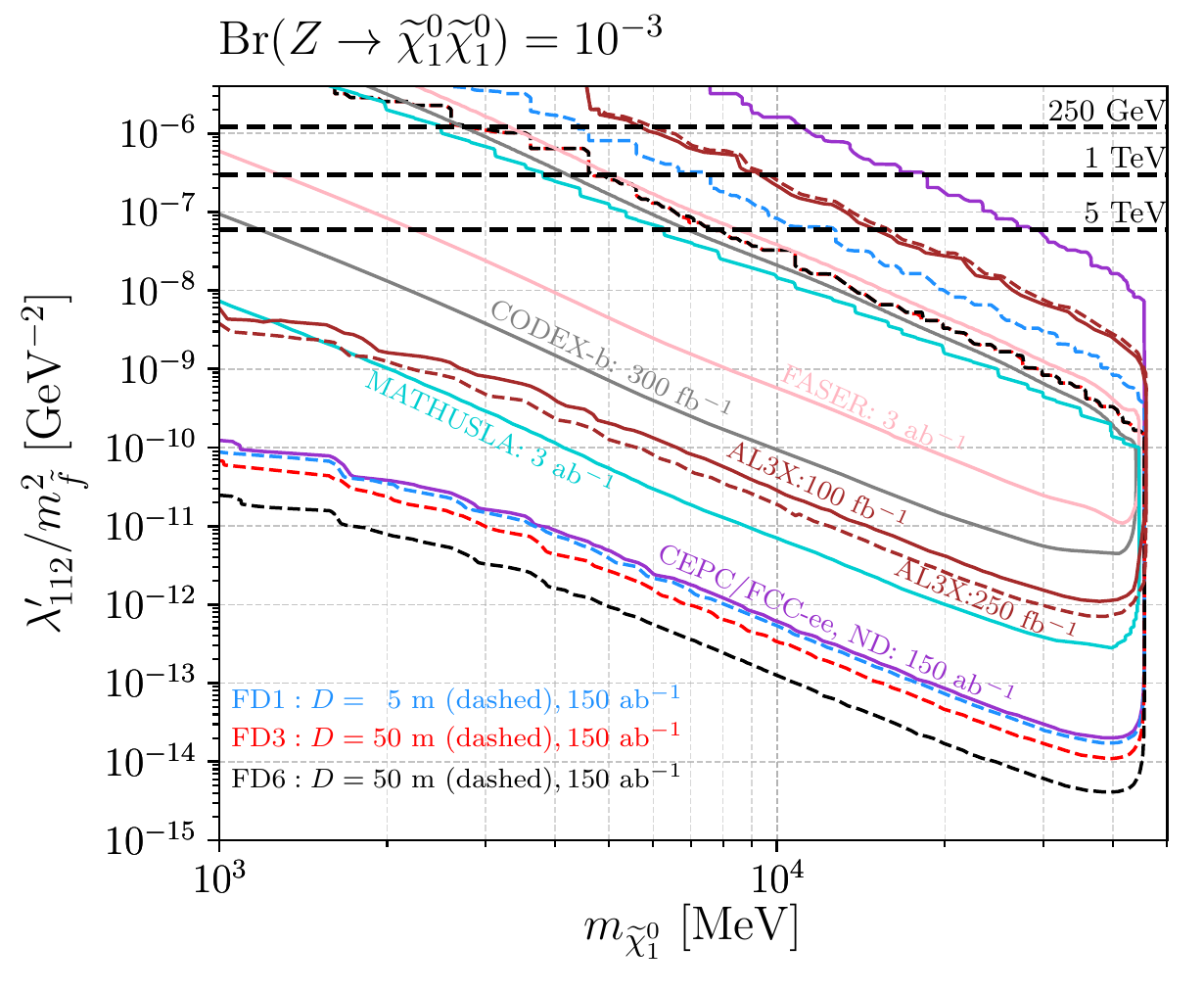}
	\renewcommand{\figurename}{Fig.}
	\caption{
Reaches for RPV neutralino of the CEPC/FCC-ee's far detectors FD1, FD3, FD6, compared with predictions for the CEPC/FCC-ee's near detector (ND) and other experiments~\cite{Wang:2019xvx}.
	}
	\label{fig:Z2n1n1:BrLim}
\end{figure}

Ref.~\cite{Wang:2019xvx} considered the case of a pair of the lightest neutralinos from $Z$ decays, $Z\rightarrow \tilde{\chi}_1^0 \tilde{\chi}_1^0$, in the context of RPV-SUSY. 
Fig.~\ref{fig:Z2n1n1:BrLim} 
compares the sensitivity reaches of representative far detectors with those of the CEPC/FCC-ee's near detector, and future experiments at the LHC.
The black horizontal dashed lines correspond to the current RPV bounds on the single coupling $\lambda^\prime_{112}$~\cite{Kao:2009fg} for three different degenerate sfermion masses $m_{\tilde{f}}=250$ GeV, 1 TeV, and 5 TeV as labelled.
The 3-signal-event isocurves of the near detectors at the CEPC and FCC-ee are reproduced from Ref.~\cite{Wang:2019orr} by adopting  the CEPC's baseline detector, and the predictions for future LHC detectors (CODEX-b, FASER, MATHUSLA and AL3X) are extracted from Refs.~\cite{Helo:2018qej,Dercks:2018wum} for the same physics scenario. 
More results including the sensitivities for the signal scenario at HL-LHC with $\sqrt{s}=14$ TeV and 3 $ab^{-1}$ integrated luminosity can be found in Ref.~\cite{Wang:2019xvx}.

\vspace{0.2 cm}
\textbf{2.4 Axion-like Particles}
\vspace{0.2 cm}

\begin{figure}[h]
\centering
\includegraphics[height=7cm, width=9cm]{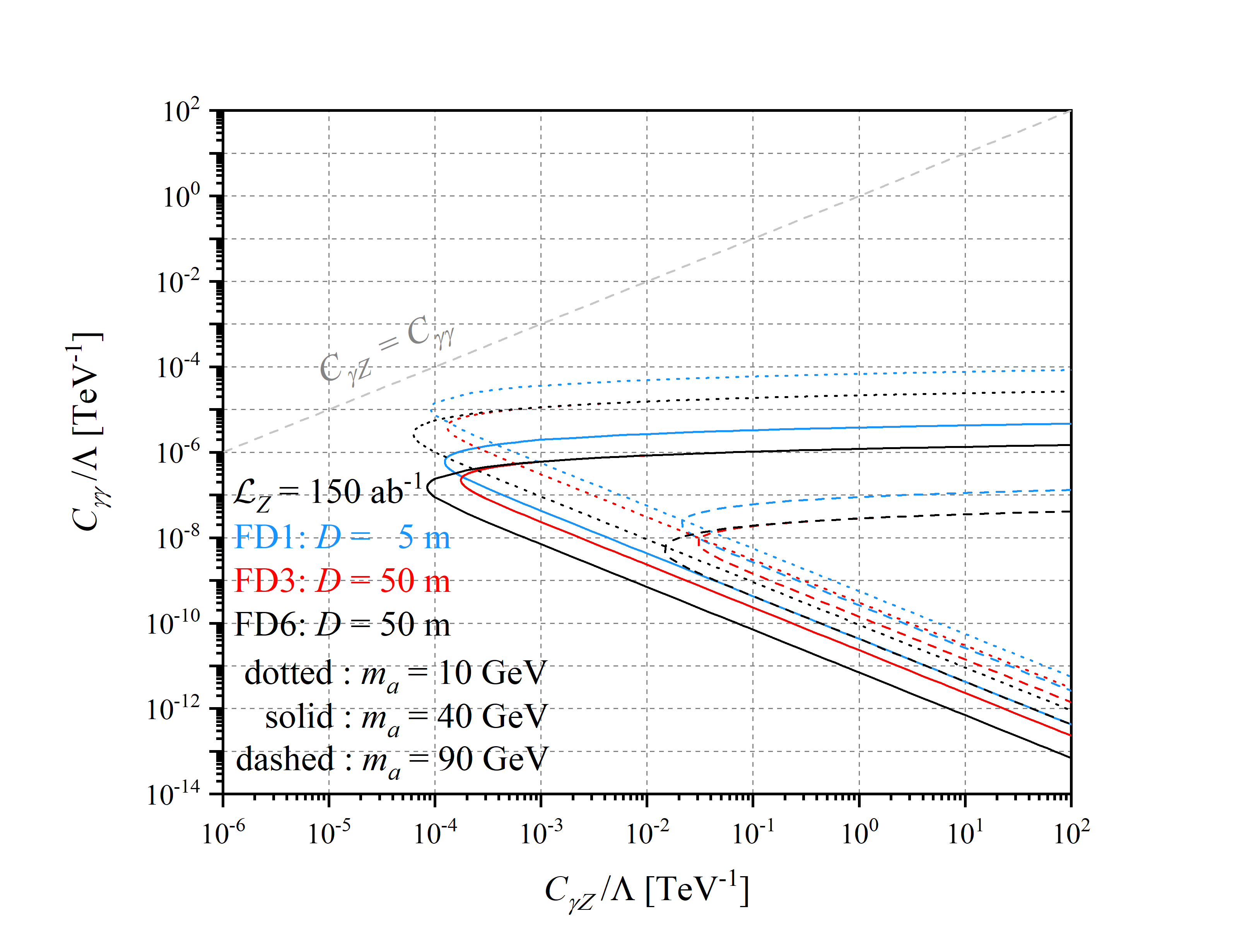}
\caption{
Reaches for ALPs of far detectors with $m_a = 10$ GeV (solid line), 40 GeV (dotted line), and 90 GeV (dashed line)~\cite{Tian:2022rsi}.
}
\label{fig:limitsM10}
\end{figure}

Ref.~\cite{Tian:2022rsi} investigates FDs' potential for discovering long-lived axion-like particles (ALPs) via the process  $e^-e^+ \rightarrow  \gamma \,\, a,~ a \to \gamma\gamma $  at future $e^{-}e^{+}$ colliders running at center-of-mass energy of $\sqrt{s} = 91.2$ GeV and integrated luminosities of 16, 150, and 750 ab$^{-1}$. 
Sensitivities on the model parameters are estimated in terms of the effective ALP-photon-photon coupling $C_{\gamma \gamma} / \Lambda $, the effective ALP-photon-$Z$ coupling $C_{\gamma Z} / \Lambda$, and ALP mass $m_a$.
Fig~\ref{fig:limitsM10} presents the discoverable parameter space  of the FD1, FD3, and FD6 in the $C_{\gamma\gamma}/\Lambda$ vs $C_{\gamma Z}/\Lambda$ plane for various ALP mass values when both $C_{\gamma Z}$ and $C_{\gamma \gamma}$ can freely change.
Results for $C_{\gamma Z} = 0$ and $C_{\gamma Z} = C_{\gamma \gamma}$ cases and more details can be found in Ref.~\cite{Tian:2022rsi}.

\subsection{A couple more examples of exotics}
There are more scenarios in which the new physics couples to the SM electroweak bosons. The search for such new physics can benefit from the high luminosity at CEPC's $Z$-pole and Higgs factory runs. Precision measurement on $Z$, $h$ width and decay products offer powerful test of exotic processes, including but not limited to lepton number/flavor violation, sterile states, axion-like particles. Low hadronic activity level at the CEPC avoids major ISR contamination and offers high identification rate to signals that typically contain comparably soft leptons, photons and jets.

\subsubsection{Heavy neutrinos} 

Motivated by explaining the neutrino mass, testing the seesaw mechanism~\cite{Minkowski:1977sc,Yanagida:1979as,Mohapatra:1979ia,Glashow:1979nm,GellMann:1980vs} includes the search of heavy neutrinos at colliders. In models with massive right-handed neutrino(s), the tiny mass of the active neutrino $\nu$ is generated by mixing the SM left-handed neutrino ($\nu_L$) with massive hypothetical $N_R$, resulting in a heavy mass eigenstate $N$ that has a small SM $\nu_L$ component whiling giving $\nu$ a small mass. The heavy $N$ talks to SM model gauge bosons via its weakly charged $\nu_L$ component~\cite{Atre:2009rg} and has been extensively studied at colliders such as the LHC~\cite{Sirunyan:2018mtv, CMS:2018jxx, SHiP:2018xqw, ATLAS:2019kpx, LHCb:2020wxx, CMS:2021lzm} using production channels via its weak coupling to SM bosons, or new physics interactions in models with a more extended interaction sector.  Massive neutrinos' weak production its proportional to their $\nu_L$ mixing, typically suppressed as $|V_{lN}|\sim {\cal O}(\sqrt{m_{\nu}/m_N})$ in vanilla Type-I like scenarios, while larger mixings are also possible~\cite{Mohapatra:1986aw}. A complementary $N$ production channel at the weak scale is via the Higgs boson, where $N$ couples to BSM scalars that mix with the Higgs, and $N$ acquires an $hNN$ coupling which is not suppressed by the smallness of active neutrino mass. For recent theory reviews, see ~\cite{Deppisch:2015qwa,Cai:2017mow} and references therein.

\begin{table}[h]
\centering
\begin{tabular}{c|c|cccc}
\hline
\hline
\multicolumn{2}{c|}{SM $\ell^\pm\ell^{\pm}$} & Pre-cut &$N_\ell$ and SS dilepton & Jet counting & Missing $E_{\rm T}$  \\
\hline
\multirow{8}{*}{Bkg.}
& $4\tau$    & $1.69\times10^4$   & 870  & $4.6\times10^{-2}$  & $7.7\times10^{-3}$ \\
& $ 2\tau Z$ & $6.80\times10^5$   & $2.91\times10^3$  & 4.6  & 0.93  \\
& $ 2\ell Z$  & $1.74\times10^6$   & $3.98\times10^3$  & - & -  \\
& $4\tau Z$ & 93.0   & 2.0  & 0.19  & $5.9\times10^{-2}$  \\
& $2\tau 2W$ & $4.42\times10^3$  & 63.6  & 0.92  & $8.2\times10^{-2}$  \\
& $ 2\ell 2\tau Z$ &  584  & 13.8 & 2.0 & 0.75 \\
& $ 4\ell Z$ & 862 &  16.5 & 2.2 & 2.1 \\
& $ 2\ell 2W$ & $2.74\times10^4$   & 639  & 11.7  & 1.2  \\
\hline
\hline
\end{tabular}
\caption{
Cut flow table the event numbers of the SM processes in the $\ell^\pm\ell^{\pm}$ channel at $\sqrt{s}=240$ GeV and 5.6 $\text{ab}^{-1}$ integrated luminosity. Selection cuts on the flavor, sign and number of the leptons, and the number of  jets are efficient in removing background. Adapted from~\cite{Gao:2021one}.
}
\label{tab:hNN2l}
\end{table}

CEPC can search for heavy $N$ within the kinematical reach of the center of mass energy. There have been studies on the weak single $N$ production at CEPC in the process $e^- e^+ \to \nu N$ for center-of-mass energy $\sqrt{s} = 240$ GeV~\cite{Liao:2017jiz}, and on high luminosity $Z$-pole running mode~\cite{Ding:2019tqq, Blondel:2021mss}.
As $N$ has a large Majorana mass, lepton number violation occurs in $N$ decay. Same-sign, same flavor dileptons, and a reconstructable $N$ mass peak of final state lepton-jet system are the `smoking gun' signals~\cite{Shoemaker:2010fg} for heavy $N$ search. CEPC is designed to yield $\sim$4M Higgs events. The high identification efficiency for soft leptons and low hadronic background at the CEPC offers a clean search opportunity for $h\rightarrow NN$. The dominant Higgs production channel at CEPC is $e^+e^-\rightarrow Zh$. The associated $Z$ complicates the signal and background analysis, as the $Z$ boson's decay products can be confused with those from heavy $N$ decay. On the other hand, with an extra $Z$ boson, the SM backgrounds can also be suppressed by requiring one more weak vertex. The leading SM backgrounds are from multi-tau production with one or two associated weak vector bosons ($V=W,Z$), e.g. $4\tau, 4\tau V, 2\tau 2\ell V$, etc., in which non-isolated and missing leptons can lead to same-sign same-flavor lepton pairs.

\begin{figure}[h]
\centering
\includegraphics[width=9.5cm,height=5.8cm]{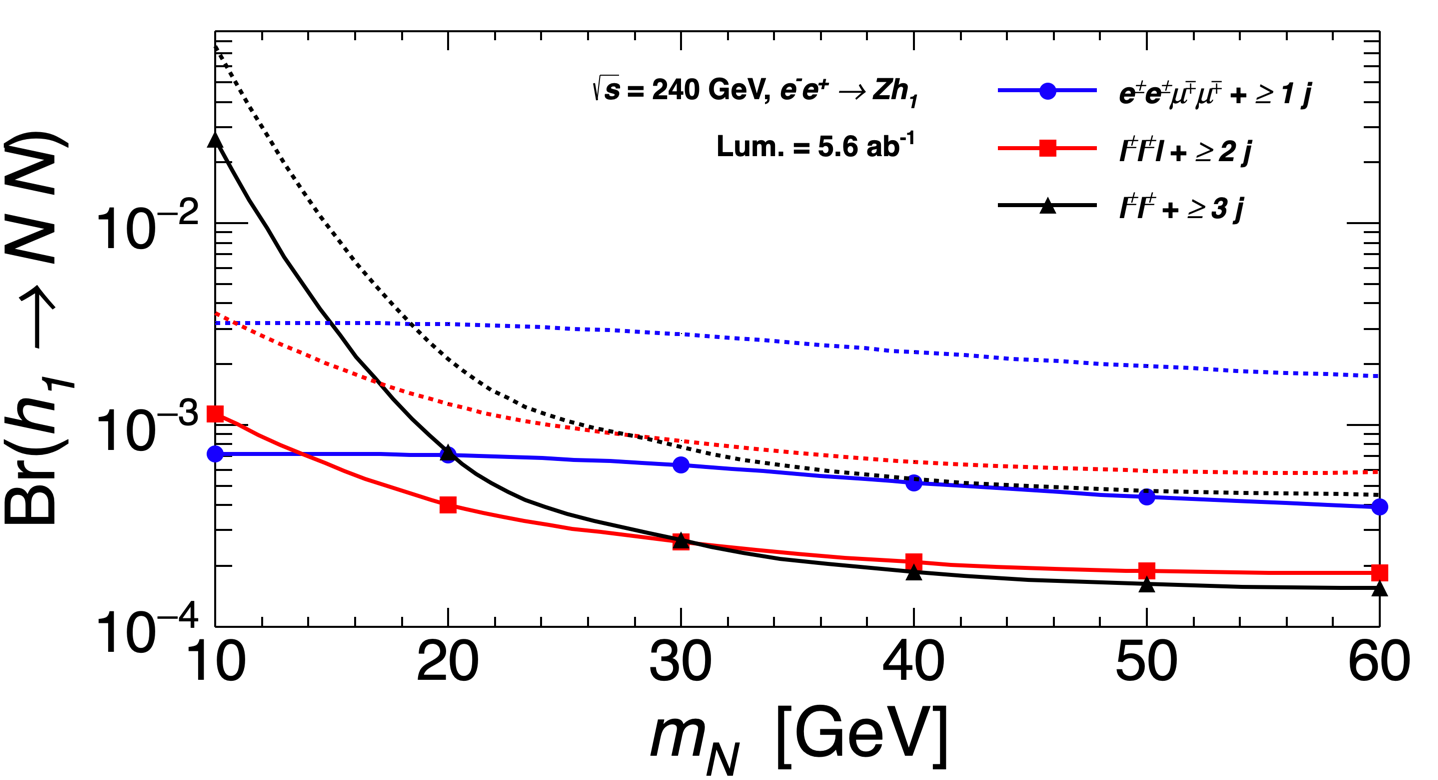}
\caption{
The projected CEPC sensitivities, $2\sigma$ (solid) and $5\sigma$ (dashed), to the decay branching ratio of $h\rightarrow NN$ for 2-4$\ell$ channels. Sensitivities at 240 GeV and 5.6 ab$^{-1}$ are comparable to HL-LHC projections~\cite{Gao:2019tio}. Figure adapted from~\cite{Gao:2021one}.
}
\label{fig:hNNlimits}
\end{figure}

SM background analysis of the $n$-lepton ($n\ge 2$) channels with at least one set of same-sign dileptons~\cite{Gao:2021one} shows that the semileptonic heavy $N$ decay, requiring only one same-sign lepton pair, gives higher sensitivity than fully leptonic $N$ decay channels. Jet and lepton number counting plays an essential role (see Table~\ref{tab:hNN2l}) in removing the SM background contamination. Leptonic decay of the associated $Z$ boson also leads to a same-sign same-flavor trilepton signal. For CEPC 240 GeV @ 5.6 ab$^{-1}$ luminosity, multi-lepton rare decay search for $h\rightarrow NN$ will be  sensitive to Higgs-BSM scalar mixing angle up to around $|\sin\alpha|^2\le 10^{-4}$. The reaches on multi-lepton Higgs rare decay branching ratios are shown in Fig.~\ref{fig:hNNlimits}. 
\subsubsection{Axion-like particles} 
To solve ``strong-CP" problem, one solution is called Peccei-Quinn mechanism proposed by Peccei and Quinn, which predicts the existence of QCD axion~\cite{Peccei:1997,Weinberg:1978,Wilczek:1978}. The axion-like particle (ALP) is a generalization of the QCD axion, which is predicted by new physics models with the breaking of a global U(1) symmetry~\cite{Svrcek:2006,Arvanitaki:2010,Cicoli:2012,Arias:2012}. The prospects for discovering ALPs via a light-by-light (LBL) scattering at two colliders, the future circular collider (FCC-ee) and circular electron-positron collider (CEPC), have been investigated. The promising sensitivities to the effective ALP-photon coupling $g_{a\gamma\gamma}$ are obtained. The study shows that the FCC-ee and CEPC can be more sensitive to the ALPs with mass 2 GeV - 10 GeV than the LHC and CLIC~\cite{Zhang:2019alp}.


\section{Detector requirements and R\&D activities}

The CEPC is a multi-purpose, large collider facility. 
It has a highest physics event rate of the order of $10^{5}$ Hz and is expected to be operated for decades. 
It is very challenging to maintain and monitor the stability of the detector system for such a long time. Similar long-term difficulties arise in beam energy calibration and luminosity measurements. 

The extremely high statistics of physics events and rich physics objectives at the CEPC impose stringent and multi-fold physics requirements for its detector system.
The CEPC detector shall have a very compact machine detector interface, because the final focus system at the circular collider requires its last quadrupole magnet to be placed close enough to the interaction point. 
The CEPC detector shall also provide large solid angle coverage, 
high accuracies on final state particle energy and momentum measurements, highly efficient reconstruction of secondary vertexes, and excellent reconstruction on the jet: its energy, momentum, flavor type, and Charge. 
The CEPC detector designs are based on the strict performance criteria required to offer a precision physics program that tests the Standard Model and explores new physics over a wide range of center-of-mass energies and luminosities. 

A primary requirement for the CEPC detector is to efficiently reconstruct individual final state particles, especially those from hadronically decayed W, Z, and Higgs bosons. 

The efficient reconstruction of individual final state particles provides a solid basis for a high efficiency-purity identification and reconstruction of all physics objects.
These physics objects include single particle physics object like leptons and photons, and composite physics objects such as $\pi_{0}$, $\Lambda$, $K_{S}^{0}$, $\tau$ lepton, and jets. 

The efficient reconstruction of individual final state particles could also significantly improve the accuracy of energy and momentum measurements of composite physics objects: once individual final state particles are identified, their energy and momentum could be measured in the optimal sub-detector systems. 

The identification of signal event, and the consequent accuracy of the corresponding physics measurements, strongly rely on the identification and reconstruction of the key physics objects, as well as on the accuracies of their energy/momentum reconstruction. 
Therefore, the efficient reconstruction of individual final state particle becomes a critical requirement for the CEPC detector. 

On top of the efficient reconstruction of individual final state particle, we would like to emphasize two sets of physics requirements. 

The first set of requirements aims at a successful program of flavor physics measurements. 
Being a Tera-Z factory, the CEPC can perform multiple flavor physics measurements at its Z pole operation, including the time-dependent CKM measurement, the rare decays, search for lepton flavor violation signal, and test of the lepton flavor universality, etc. 
Many measurements need to identify the objective heavy hadrons in $Z\to q\bar{q}$ events, while these objective hadrons decay rapidly inside the fiducial volume of the detector. 
Therefore, it is critical to identify the objective heavy hadron decay final state – from all the surrounding final state particles in the same jet. 

Clearly, these physics measurements appreciate an efficient reconstruction of individual final state particles. 
A precise reconstruction of the energy-momentum of these individual final state particles can also significantly boost the signal-to-noise ratio in relevant physics measurements. 
An highly efficient reconstruction of the jet flavor type (from what type of quark or gluon the jet actually originates from) could also strongly suppress the background. 
In addition, multiple time-dependent CKM measurements requires a highly efficient determination of the jet charge. 

A precise identification of the particle type, especially the identification of charged kaons, is critical to reduce the combinatorial background. 
Looking into the benchmark reconstruction performance of $K^\pm$, $\phi$, $\Lambda$, and $b-$ or $c-$ hadrons decaying into the final state with charged kaons, we believe that the CEPC detector shall provide a pion-kaon separation better than 3-$\sigma$.
Note that a decent PID not only benefits the flavor measurements at the Z pole, but can also enhance the performance of jet flavor tagging and jet charge measurements.

The second set of requirements aims to maximize the scientific output from physics measurements with hadronic final states. 
The CEPC is a high-yield heavy Standard Model particle factory. 
The majority of W, Z, and Higgs bosons decay into hadronic final states, 97\% of the Higgs events at CEPC decay into hadronic or semi-leptonic final states. 
Because the electron-positron collider is free of QCD background, the physics measurements with hadronic final states are strong comparative advantages of the electron-positron collider compared to the hadron collider, especially those to reveal the coupling behaviors betwen the Higgs boson and its hadronic decay final states.  
The reconstruction performance of the hadronic final state can be characterized by Boson Mass Resolution (BMR), which is defined as the relevant mass resolution of the hadronic system, especially the hadronically decayed Higgs boson.

For a ZH event decaying into semi-leptonic final state, the identification of its hadronic system can be straightforward. 
Meanwhile, the semi-leptonic ZZ events can be the irreducible background for the ZH signal, for example, the $qqH, H\to invisible$ signal versus the $ZZ\to\nu$ background, and the $qqH, H\to\tau\tau$ signal versus the $ZZ\to qq\tau\tau$ background. 
For these measurements, the recoil mass of the hadronic system can be used to distinguish the ZH signal from the ZZ background. 
Quantitative analyses show that a BMR better than 4\% is therefore required~\cite{CEPCStudyGroup:2018ghi}. 
A detector that provides a BMR better than 4\% can also separate the hadronically decayed W, Z, and Higgs bosons through their invariant mass distribution. 
The CEPC can perform intensive flavor physics measurements with neutrino final states.
These measurement will certainly benefit from the decent BMR performance. 

From the view point of sub-detectors, 
the CEPC detectors must be able to discriminate $b$-jets, $c$-jets, and light jets from each other in order to quantify the coupling of the Higgs boson to the charm quark, which relies on the high-accuracy, high-efficiency, and low-material vertex detector placed close enough to the interaction point. 
For the tracking system, per mille level relative precision on track momentum resolution is essential to obtain decent sensitivity for the $H \to \mu^+\mu^-$ measurement, and is highly appreciated for multiple flavor physics measurements at the Z pole. 
A good ECAL energy resolution is beneficial to the  $H \to \gamma\gamma$ measurement, as well as multiple flavor measurements with photon or $\pi_{0}$ in the final state.
The measurement of inclusive Higgsstrahlung cross section from $Z(\to q\bar{q})H$ events requires not only an excellent reconstruction of the hadronic system, but also a clear identification of the hadronically decayed Z boson.  
Hunting for the dark matter via the Higgs invisible decay requires a good reconstruction of the missing energy and momentum. 
The latter two drive actually the performance requirements on the BMR, and the identification of Color Singlet in full hadronic physics events.

To fulfill these physics requirements, multiple detector concepts are proposed, and an intensive R\&D program on several sub-detector systems has been performed~\cite{CEPCStudyGroup:2018ghi}. 
At the CEPC CDR, two basic detector designs are investigated: a baseline detector concept with two approaches of tracking systems and an alternative detector concept to meet the requirement on jet energy resolution (which is highly correlated with BMR). 
The baseline detector design combines the particle flow principle with a precision vertex detector, a time projection chamber, a silicon tracker, a 3-Tesla solenoid, a high granularity calorimeter, and a muon detector. 
A full silicon tracker is also an option for the baseline detector concept. Another detector design is based on the dual-readout calorimetry and consists of a precision vertex detector, a drift chamber tracker, a 2-Tesla solenoid, and a muon detector. The various technologies for each detector subsystem are being actively explored through R\&D projects.

\begin{figure}[htbp!]
    \centering
    \includegraphics[width=0.6\textwidth]{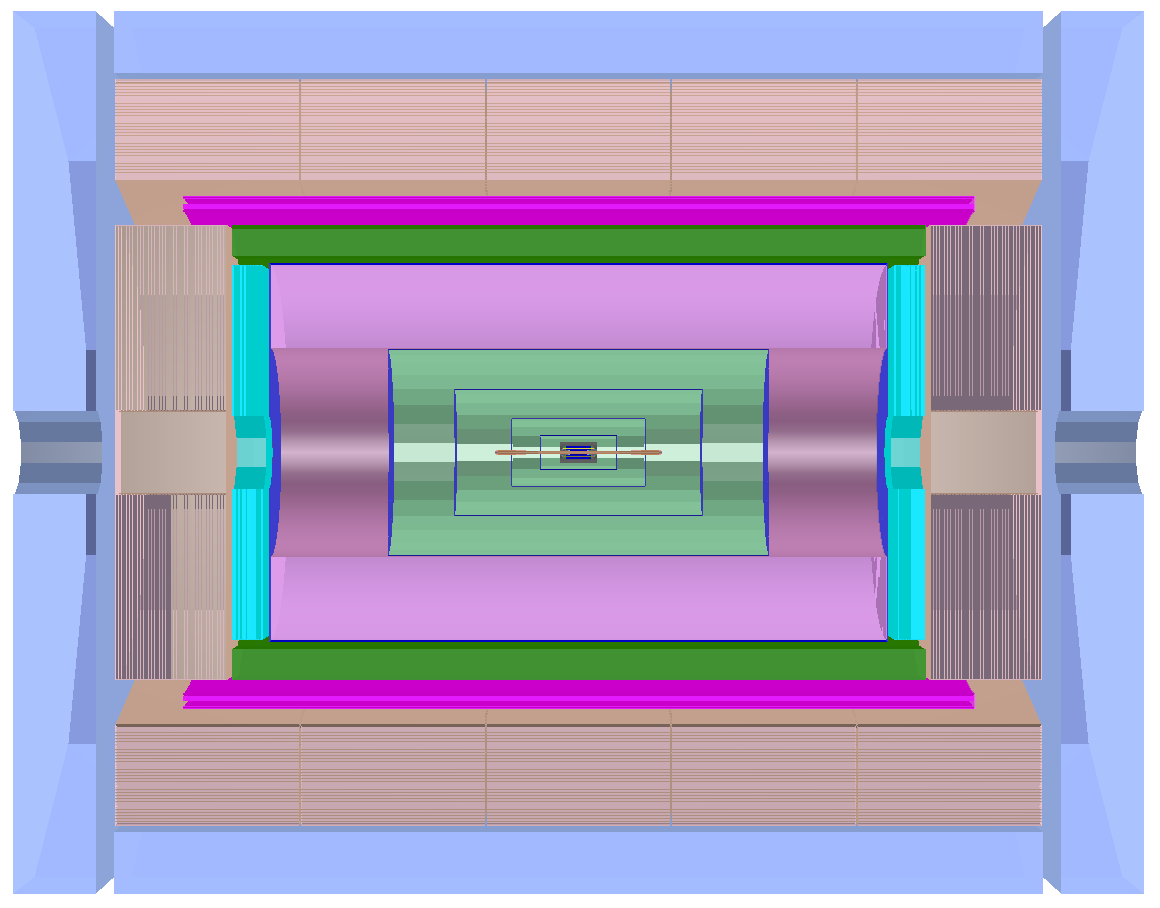}
    \caption{Layout of the new conceptual detector.}
    \label{fig3:det4}
\end{figure}

Recently, a progressive new detector concept has been proposed and is under development and exploration~\cite{jianchun-4th-det}. The new concept has a very thin beam pipe with a small radius and a novel cooling system design. It also integrates several cutting-edge technologies, $e.g.$, a silicon tracker with the HV-COMS technique, a wire chamber optimized for PID and TOF resolution. A transverse crystal bar ECAL will provide high angular and energy resolution. The detector concept will also adopt a thin superconducting magnet to reduce the overall material budget. As a result, the proposed design shall satisfy the requirements of Higgs, EW, flavor, and BSM physics studies. The layout of the new detector design is shown in Fig~\ref{fig3:det4}. 

\section{Message to the Snowmass}

Since completing the CEPC CDR, the CEPC study group has made significant progress on the CEPC accelerator design and key technology R\&D. 
Those collective activities have converged into a new set of nominal beam parameters and operation scenarios. 
Compared to the CDR, the particle yields of the CEPC increase significantly and should be taken into account for the physics potential evaluation. 

Historically, precision measurements have been a great way of making progress in particle physics, having solved some of its most important questions.  
The CEPC project has the potential to characterize the Higgs boson in the same way LEP did to the Z boson and search for possible deviations from the Standard Model. 
It can also make EW, flavor physics, and QCD measurements unprecedented precision and many unique approaches to searching for new physics signals. Given such a comprehensive physics program, the CEPC will 
significantly enhance our knowledge of many aspects of the physics around the TeV scale, with a great prospect of finding new physics principles underlying the SM.

The CEPC project fits well in the strategy of the international particle physics community. 
We would like to seek support from Snowmass~2021 for active 
research programs in the US and worldwide and work towards the realization of the CEPC or at least one electron-positron Higgs factory in the future.

\bibliographystyle{utphys}
\bibliography{references}

\end{document}